Electronic properties of graphene: a perspective from scanning tunneling microscopy and magneto-transport.

# Eva Y. Andrei[1], Guohong Li[1] and Xu Du[2]

[1]Department of Physics and Astronomy, Rutgers University, Piscataway, NJ 08855, USA

[2]Department of Physics, SUNY at Stony Brook, NY, USA

## Abstract

This review covers recent experimental progress in probing the electronic properties of graphene and how they are influenced by various substrates, by the presence of a magnetic field and by the proximity to a superconductor. The focus is on results obtained using scanning tunneling microscopy, spectroscopy, transport and magneto-transport techniques.

A.  Introduction

In 2004 a Manchester University team led by Andre Geim demonstrated a simple mechanical exfoliation process[1, 2] by which graphene, a one-atom thick 2 dimensional (2D) crystal of Carbon atoms arranged in a honeycomb lattice [3-8], could be isolated from graphite. The isolation of graphene and the subsequent measurements which revealed its extraordinary electronic properties [9, 10] unleashed a frenzy of scientific activity the magnitude of which was never seen. It quickly crossed disciplinary boundaries and in May of 2010 the Nobel symposium on graphene in Stockholm was brimming with palpable excitement. At this historic event graphene was the centerpiece for lively interactions between players who rarely share common ground: physicists, chemists, biologists, engineers and field- theorists. The excitement about graphene extends beyond its unusual electronic properties. Everything about graphene – its chemical, mechanical, thermal and optical properties - is different in interesting ways.

This review focuses on the electronic properties of single layer graphene that are accessible with scanning tunneling microscopy and spectroscopy and with transport measurements. Part A gives an overview starting with a brief history in section A1 followed by methods of producing and characterizing graphene in sections A2 and A3. In section A4 the physical properties are discussed followed by a review of the electronic properties in section A5 and a discussion of effects due to substrate interference in section A6. Part B is devoted to STM (scanning tunneling microscopy) and STS (scanning tunneling spectroscopy) measurements which allow access to the atomic structure and to the electronic density of states. Sections B1 and B2 focus on STM/STS measurements on graphene supported on standard $SiO_2$ and on metallic substrates. B3 is devoted to graphene supported above a graphite substrate and the observation of the intrinsic electronic properties including the linear density of states, Landau levels, the Fermi velocity, and the quasiparticle lifetime. This section discusses the effects of electron-phonon interactions and of interlayer coupling. B4 is dedicated to STM/STS studies of twisted graphene layers. B5 focuses on graphene on chlorinated $SiO_2$ substrates and the transition between extended and localized electronic states as the carrier density is swept across Landau levels. A brief description of STM/STS work on epitaxial graphene on SiC and on h-BN substrates is given in B6. Part C is devoted to transport measurements. Following a discussion on graphene devices and substrate-induced scattering sources for $SiO_2$ substrates, section C.1 discusses superconductor--graphene--superconductor Josephson junctions. Sections C.2, C.3 and C.4 discuss suspended graphene devices, the fractional quantum Hall effect and the magnetically induced insulating phase. C1 discusses substrate-induced scattering sources in graphene deposited on $SiO_2$. Superconductor/Graphene/superconductor (SGS) Josephson junctions are the focus of C2. C3 and C4 discuss suspended graphene devices, the observation of ballistic transport the fractional quantum Hall effect and the magnetically induced insulating phase.

List of abbreviations: AFM (atomic force microscopy); ARPES (angular resolved photoemission); CNP (charge neutrality point); CVD (chemical vapor deposition); DOS (density of states); DP (Dirac point); e-ph (electron-phonon); HOPG (highly oriented pyrolitic graphite); LL (Landau levels); ΛL (lambda levels); MAR (multiple Andreev reflections); NSG (non-suspended graphene); QHE (quantum Hall effect); FQHE (fractional QHE); SG (suspended graphene); SEM (scanning electron microscopy); STM (scanning tunneling microscopy); STS (scanning tunneling spectroscopy); TEM (transmission electron microscopy0.

## 1. *Historical note*

The story of graphene is both old and new. First postulated in 1947 by J. C. Wallace [11] as a purely theoretical construct to help tackle the problem of calculating the band structure of graphite, this model of a 2D crystal arranged in a honeycomb lattice, was now and again dusted off and reused over the years [12-15]. In 1984 G. Semenoff [12] resurrected it as a model for a condensed matter realization of a three dimensional anomaly and in 1988 D. Haldane [14] invoked it as model for a Quantum Hall Effect (QHE) without Landau Levels. In the 90's the model was used as a starting point for calculating the band structure of Carbon nanotubes [16]. But nobody at the time thought that one day it would be possible to fabricate a free standing material realization of this model. This skepticism stemmed from the influential Mermin-Wagner theorem [17] which during the latter part of the last century was loosely interpreted to mean that 2D crystals cannot exist in nature. Indeed one does not find naturally occurring free standing 2D crystals, and computer simulations show that they do not form spontaneously because they are thermodynamically unstable against out of plane fluctuations and roll-up [18]. It is on this backdrop that the realization of free standing graphene came as a huge surprise. But on closer scrutiny it should not have been. The Mermin-Wagner theorem does not preclude the existence of finite size 2D crystals: its validity is limited to infinite systems with short range interactions in the ground state. While a finite size 2D crystal will be prone to develop topological defects at finite temperatures, in line with the theorem, it is possible to prepare such a crystal in a long-lived metastable state which is perfectly ordered provided that the temperature is kept well below the core energy of a topological defect. How to achieve such a metastable state? It is clear that even though 2D crystals do not form spontaneously they can exist and are perfectly stable when stacked and held together by Van der Waals forces as part of a 3D structure such as graphite. The Manchester group discovered that a single graphene layer can be dislodged from its graphite cocoon by mechanical exfoliation with Scotch tape. This was possible because the Van der Waals force between the layers in graphite is many times weaker than the covalent bonds within the layer which help maintain the integrity of the 2D crystal during the exfoliation.

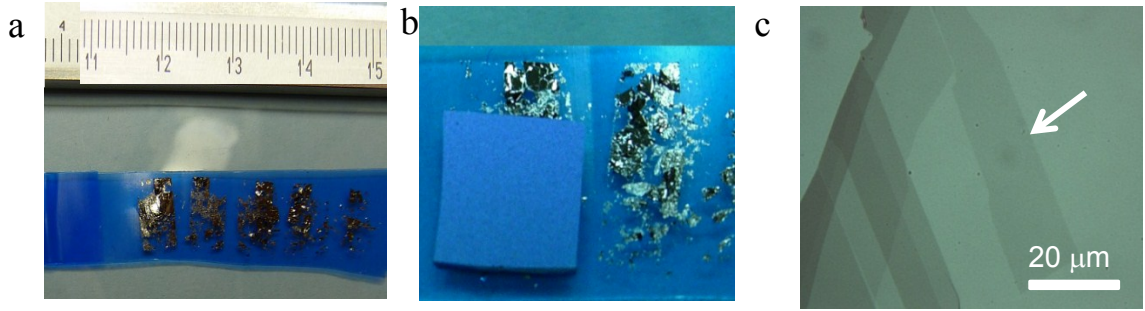

**Figure A-1. Making exfoliated graphene. a)** HOPG graphite flakes are deposited on Scotch tape shown with cm ruler. **b)** A Si/SiO$_2$ substrate is pressed onto flakes on the tape. **c)** Optical micrograph of graphene deposited on SiO$_2$ showing flakes with various number of layers. A large flake of single layer graphene, corresponding to the faintest contrast, is indicated by the arrow. Image credits: A. Luican-Mayer.

The exfoliated graphene layer can be supported on a substrate or suspended from a supporting structure[19] [20-23]. Although the question of whether free-standing graphene is truly 2D or contains tiny out-of-plane ripples [18] (as was observed in suspended graphene membranes at room temperature [20]) is still under debate, there is no doubt about its having brought countless opportunities to explore new physical phenomena and to implement novel devices.

## 2. Making graphene

We briefly describe some of the most widely used methods to produce graphene, together with their range of applicability.

### Exfoliation from graphite.

Exfoliation from graphite, illustrated in Fig. A-1, is inexpensive and can yield small (up to 0.1 mm) high quality research grade samples[1, 2]. In this method, which resembles writing with pencil on paper, the starting material is a graphite crystal such as natural graphite, Kish or HOPG (highly oriented pyrolitic graphite). Natural and Kish graphite tend to yield large graphene flakes while HOPG is more likely to be chemically pure. A thin layer of graphite is removed from the crystal with Scotch tape or tweezers. The layer is subsequently pressed by mechanical pressure (or dry $N_2$ jet for cleaner processing) unto a substrate, typically a highly doped Si substrate capped with 300nm of $SiO_2$, which enables detection under an optical microscope [1] as described in detail in the next section on optical characterization [24-26]. Often one follows up this step with an AFM (atomic force microscope) measurement of the height profile to determine the thickness (~ 0.3nm /layer) and/or Raman spectroscopy to confirm the number of layers and check the sample quality. Typical exfoliated graphene flakes are several microns in size, but occasionally one can find larger flakes that can reach several hundred μm. Since exfoliation is facilitated by stacking defects, yields tend to be larger when starting with imperfect or turbostratic graphite but at the same time the sample size tends to be smaller. The small size and labor intensive production of samples using exfoliated graphene render them impractical for large scale commercial applications. Nevertheless, exfoliated graphene holds its own niche as a new platform for basic research. The high quality and large single crystal domains, so far not achieved with other methods of fabrication, have given access to the intrinsic properties of the unusual charge carriers in graphene, including ballistic transport and the fractional QHE, and opened a new arena of investigation into relativistic chiral quasiparticles[21, 27-30].

### Chemical vapor deposition (CVD) on metallic substrates.

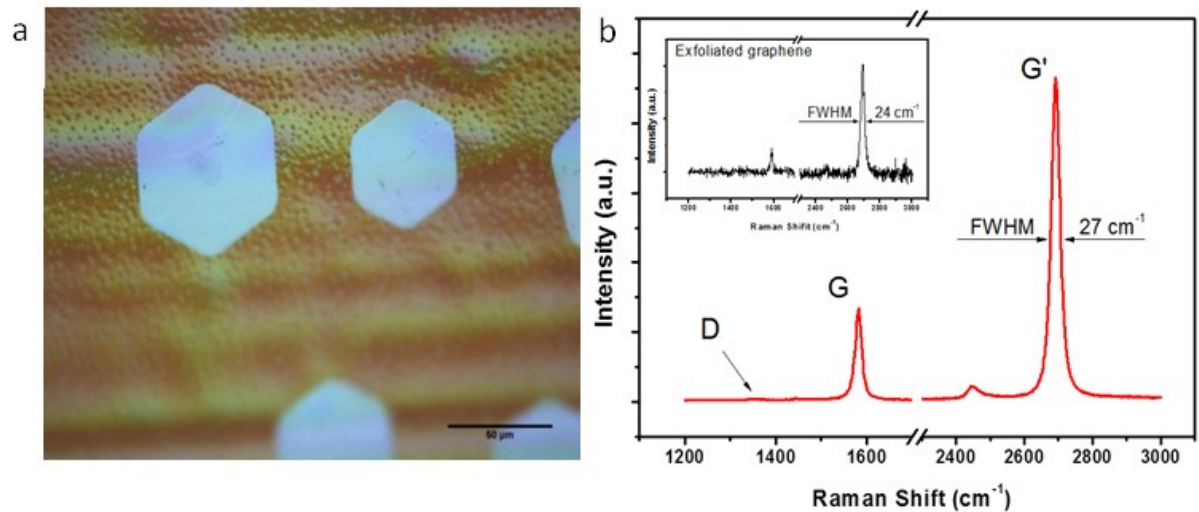

A

quick and relatively simple method to make graphene is CVD by hydrocarbon decomposition on a metallic substrate [31]. This method (Figure A-2a) can produce large areas of graphene suitable, after transfer to an insulating substrate, for large scale commercial applications. In this method a metallic substrate, which plays the role of catalyst, is placed in a heated furnace and is attached to a gas delivery system that flows a gaseous carbon source downstream to the substrate. Carbon is adsorbed and absorbed into the metal surface at high temperatures, where it is then precipitated out to form graphene, typically at around 500-800 $^0$C during the cool down to room temperature. The first examples of graphitic layers on metallic substrates were obtained simply by segregation of carbon impurities when the metallic single crystals were heated during the surface preparation. Applications of this method using the decomposition of ethylene on Ni surfaces [32] were demonstrated in the 70's. More recently graphene growth was demonstrated on various metallic substrates including Rh[33], Pt[34-36], Ir [37], Ru [38-41], Pd [42] and Cu foil [43-46]. The latter yields, at relatively low cost, single layer graphene of essentially unlimited size and excellent transport qualities characterized by mobility in excess of 7000 cm$^2$/V s [47]. The hydrocarbon source is typically a gas such as methane and ethylene but interestingly solid sources also seem to work, such as poly(methyl methacrylate) (PMMA) and even table sugar was recently demonstrated as a viable Carbon source[48].

**Surface graphitization and epitaxial growth on SiC crystals.**

Heating of 6H-SiC or 4H-SiC crystals to temperatures in excess of 1200 °C causes sublimation of the Silicon atoms from the surface[49-51] and the remaining Carbon atoms reconstruct into graphene sheets[52]. The number of layers and quality of the graphene depends on whether it grows on the Si or C terminated face and on the annealing temperature[53]. The first Carbon layer undergoes reconstruction due to its interaction with the substrate forming an insulating buffer layer while the next layers resemble graphene. C face graphene consists of many layers, the first few being highly doped due the field effect from the substrate. Growth on the Si face is more controlled and can yield single or bilayers. By using hydrogen intercalation or thermal release tape[54, 55] one can transfer these graphene layers to other substrates. Epitaxial graphene can cover large areas, up to 4", depending on the size of the SiC crystal. Due to the lattice mismatch these layers form terraces separated by grain boundaries which limit the size of crystal domains to several micrometers[56] as shown in Fig. A-3a, and the electronic mobility to less than 3000 cm$^2$/V s which is significantly lower than in exfoliated graphene. The relatively large size and ease of fabrication of epitaxial graphene make it possible to fabricate high-speed integrated circuits [57], but the high cost of the SiC crystal starting material renders it impractical for large-scale commercial applications.

**Other methods.**

The success and commercial viability of future graphene-based devices rests on the ability to synthesize it efficiently, reliably and economically. CVD graphene is one of the promising directions. Yet, in spite of the fast moving pace of innovation, CVD growth of graphene over large areas remains challenging due to the need to operate at reduced pressures or in controlled environments. The recent demonstration of graphene by open flame synthesis [58] offers the potential for high-volume continuous production at reduced cost. Many other avenues are being explored in the race toward low cost, efficient and large scale synthesis of graphene. Solution-based exfoliation of graphite with organic solvents [59] or non-covalent functionalization [60] followed by sonication can be used in mass production of flakes for conducting coatings or composites. Another promising approach is the use of colloidal suspensions [61]. The starting material is typically a graphite oxide film which is then dispersed in a solvent and reduced. For example the reduction by hydrazine annealing in argon/hydrogen [62] produces large areas of graphene films for use as transparent conducting coating, graphene paper or filters.

## 3. Characterization.

### Optical.

For flakes supported on $SiO_2$ a fast and efficient way to find and identify graphene is by using optical microscopy as illustrated in Figure A-1c. Graphene is detected as a faint but clearly visible shadow in the optical image whose contrast increases with the number of layers in the flake. The shadow is produced by the interference between light-beams reflected from the graphene and the $Si/SiO_2$ interface [24-26]. The quality of the contrast depends on the wavelength of the light and thickness of the oxide. For a ~300 nm thick $SiO_2$ oxide the visibility is optimal for green light. Other "sweet spots" occur at ~90 nm and ~500nm. This method allows to visualize micron-size flakes, and to distinguish between single-layer, bilayer and multilayer flakes. Optical microscopy is also effective for identifying single layer graphene flakes grown by CVD on Copper as illustrated in Figure A-2a.

### Raman spectroscopy.

Raman spectroscopy is a relatively quick way to identify graphene and to determine the number of layers[63, 64]. In order to be effective the spatial resolution has to be better than the sample size; for small samples this requires a companion high resolution optical microscope to find the flakes. The Raman spectrum of graphene, Figure A-2b, exhibits three main features: the G peak ~1580 cm$^{-1}$ which is due to a first order process involving the degenerate zone center E2g optical phonon; the 2D (G') peak at ~2700 is a second order peak involving two $A'_1$ zone-boundary optical phonons; and the D-peak, centered at ~1330 cm$^{-1}$, involving one $A'_1$ phonon, which is attributed to disorder-induced first-order scattering. In pure single layer graphene the 2D peak is typically ~ 3 times larger than the G peak and the D peak is absent. With increasing number of layers, the 2D peak becomes broader and loses its characteristic Lorenzian line-shape. Since the G-peak is attributed to intralayer effects, one finds that its intensity scales with the number of layers.

### Atomic force microscopy (AFM).

The AFM is a non-invasive and non-contaminating probe for characterizing the topography of insulating as well as conducting surfaces. This makes it convenient to identify graphene flakes on any surface and to determine the number of layers in the flake without damage, allowing the flake to be used in further processing or measurement. High-end commercial AFM machines can produce topographical images of surfaces with height resolution of 0.03nm. State of the art machines have even demonstrated atomic resolution images of graphene. The AFM image of epitaxial graphene on SiC shown Figure A-3a clearly illustrates the terraces in these samples. Figure A-3 shows an AFM image of a graphene flake on an h-BN substrate obtained with the Integra Prima AFM by NT-MD.

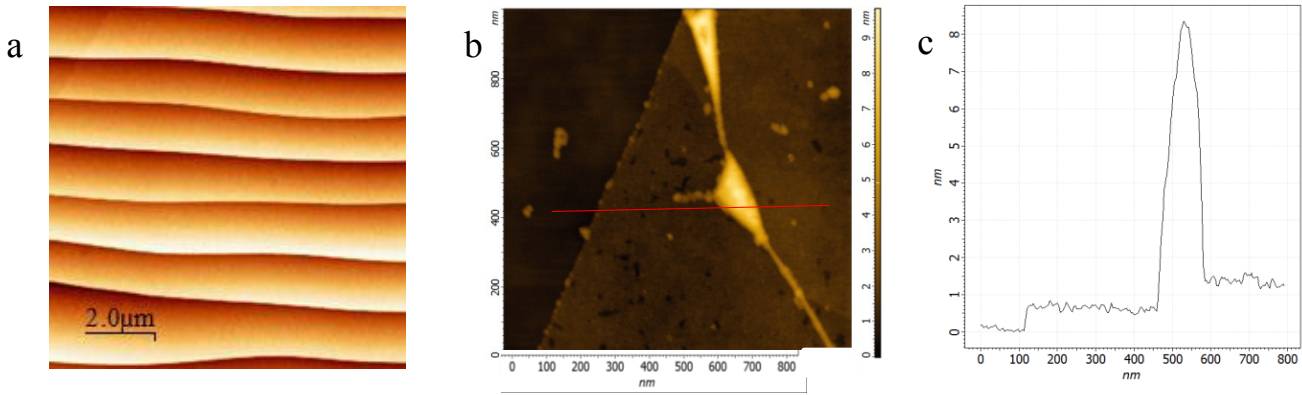

**Figure A-3. a)** AFM image of epitaxial graphene grown on SiC shows micron size terraces . (K.V. Emtsev *et al*. Nature Materials 8 (2009) 203). **b)** AFM scan (NT-MDT Integra prime) of single layer graphene flake on an h-BN substrate. **c)** The height profile shows a 0.7nm step between the substrate and the flake surface. The bubble under the flake is 7nm at its peak height. Image credits: B. Kim NT-MDT.

### Scanning tunneling microscopy and spectroscopy (STM/STS)

STM, the technique of choice for atomic resolution images, employs the tunneling current between a sharp metallic tip and a conducting sample combined with a feedback loop to a piezoelectric motor. It provides access to the topography with sub-atomic resolution, as illustrated in Figure A-4a. STS can give access to the electronic density of states (DOS) with energy resolution as low as ~0.1 meV. The DOS obtained with STM is not limited by the position of the Fermi energy – both occupied and empty states are accessible. In addition measurements are not impeded by the presence of a magnetic field which made it possible to directly observe the unique sequence of Landau levels in graphene resulting from its ultra-relativistic charge carriers [65, 66].

The high spatial resolution of the STM necessarily limits the field of view so, unless optical access is available, it is usually quite difficult to locate small micron size samples with an STM. A recently developed technique [67] which uses the STM tip as a capacitive antenna allows locating sub-micron size samples rapidly and efficiently without the need for additional probes. A more detailed discussion of STM/STS measurements on graphene is presented in part B of this review.

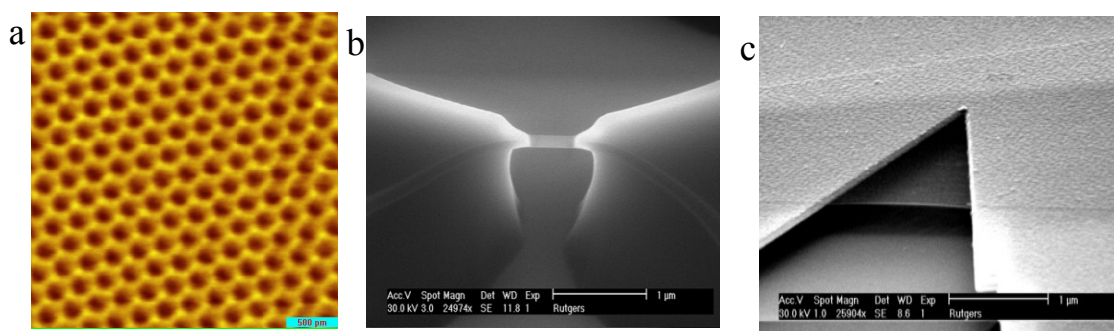

**Figure A-4. STM and SEM on graphene. a)** Atomic resolution STM of graphene on a graphite substrate. **(b,c)** SEM images on suspended graphene (FEI Sirion equipped with JC Nabity Lithography Systems). **b)** Suspended graphene flake supported on LOR polymer. Scale bar 1µm. Image credits: J. Meyerson. **c)** Suspended graphene flake (central area) held in place by an Ti/Au support. Scale bar 1µm. Image credits A. Luican-Mayer.

### Scanning electron microscope (SEM) and transmission electron microscope (TEM)

SEM is convenient for imaging large areas of conducting samples. The electron beam directed at the sample typically has an energy ranging from 0.5 keV to 40 keV, and a spot size of about 0.4 nm to 5 nm in diameter. The image, which is formed by the detection of backscattered electrons or radiation, can achieve a resolution of ~ 10nm in the best machines. Due to the very narrow beam, SEM micrographs have a large depth of field yielding a characteristic three-dimensional appearance. Examples of SEM images of suspended graphene devices are shown in Figure A-4b,c. A very useful feature available with SEM is the possibility to write sub-micron size patterns by exposing an e-beam resist on the surface of a sample. The disadvantage of using the SEM for imaging is electron beam induced contamination due to the deposition of carbonaceous material on the sample surface. This contamination is almost always present after viewing by SEM, its extent depending on the accelerating voltage and exposure. Contaminant deposition rates can be as high as a few tens of nanometers per second.

In TEM the image is formed by detecting the transmitted electrons that pass through an ultra-thin sample. Owing to the small de Broglie wavelength of the electrons, TEMs are capable of imaging at a significantly higher resolution than optical microscopes or SEM, and can achieve atomic resolution. Just as with SEM imaging with TEM suffers from electron beam induced contamination.

### Low energy electron diffraction (LEED) and angular resolved photoemission (ARPES).

These techniques provide reciprocal space information. LEED measures the diffraction pattern obtained by bombarding a clean crystalline surface with a collimated beam of low energy electrons, from which one can determine the surface structure of crystalline materials. The technique requires the use of very clean samples in ultra-high vacuum. It is useful for monitoring the thickness of materials during growth. For example LEED is used for in-situ monitoring of the formation of epitaxial graphene [68].

ARPES is used to obtain the band structure in zero magnetic field as a function of both energy and momentum. Since only occupied states can be accessed one is limited to probing states below the Fermi energy. Typical energy resolution of ARPES machines is ~ 0.2eV for toroidal analyzers. Recently 0.025eV resolution was demonstrated with a low temperature hemispherical analyzer at the Advanced Light Source.

### Other techniques

In situ formation of graphitic layers on metal surfaces was monitored in the early work by Auger electron spectroscopy which shows a carbon peak [69] that displays the characteristic fingerprint of graphite[70]. In X-ray photoemission spectroscopy, which can also be used during the deposition, graphitic carbon is identified by a carbon species with a C1s energy close to the bulk graphite value of 284.5 eV[70].

## 4. *Structure and physical properties*

Structurally, graphene is defined as a one-atom-thick planar sheet of $sp^2$-bonded carbon atoms that are arranged in a honeycomb crystal lattice[3] as illustrated in Figure A-5a. Each Carbon atom in graphene is bound to its three nearest neighbors by strong planar σ bonds that involve three of its valence electrons occupying the sp$^2$ hybridized orbitals. In equilibrium the Carbon-Carbon σ bonds are 0.142 nm long and are 120$^0$ apart. These bonds are responsible for the planar structure of graphene and for its mechanical and thermal properties. The fourth valence electron which remains in the half-filled 2p$_z$ orbital orthogonal to the graphene plane forms a weak π bond by overlapping with other 2p$_z$ orbitals. These delocalized π electrons determine the transport properties of graphene.

**Mechanical properties.**

The covalent σ bonds which hold graphene together and give it the planar structure are the strongest chemical bonds known. This makes graphene one of the strongest materials: its breaking strength is 200 times greater than steel, and its tensile strength, 130 GPa [19, 71, 72], is larger than any measured so far. Bunch *et al*. [72] were able to inflate a graphene balloon and found that it is impermeable to gases[72], even to helium. They suggest that this property may be utilized in membrane sensors for pressure changes in small volumes, as selective barriers for filtration of gases, as a platform for imaging of graphene-fluid interfaces, and for providing a physical barrier between two phases of matter.

**Chemical properties.**

The strictly two dimensional structure together with the unusual massless Dirac spectrum of the low energy electronic excitations in graphene (discussed below) give rise to exquisite chemical sensitivity. Shedin *et al*.[73] demonstrated that the Hall resistivity of a micrometer-sized graphene flake is sensitive to the absorption or desorption of a single gas molecule, producing step-like changes in the resistance. This single molecule sensitivity, which was attributed to the exceptionally low electronic noise in graphene and to its linear electronic DOS, makes graphene a promising candidate for chemical detectors and for other applications where local probes sensitive to external charge, magnetic field or mechanical strain are required.

**Thermal properties.**

The strong covalent bonds between the carbon atoms in graphene are also responsible for its exceptionally high thermal conductivity. For suspended graphene samples the thermal conductivity reaches values as high as 5,000 W/m K [74] at room temperature which is 2.5 times greater than that of diamond, the record holder among naturally occurring materials. For graphene supported on a substrate, a configuration that is more likely to be found in useful applications and devices, the thermal conductivity (near room temperature) of single-layer graphene is about 600 Wm$^{-1}$K$^{-1}$ [48]. Although this value is one order of magnitude lower than for suspended graphene, it is still about twice that of Copper and 50 times larger than for Silicon.

**Optical properties.**

The optical properties of graphene follow directly from its 2D structure and gapless electronic spectrum (discussed below). For photon energies larger than the temperature and Fermi energy the optical conductivity is a universal constant independent of frequency: $G = \dfrac{e^2}{4\hbar}$ where *e* is the electron charge and $\hbar$ the reduced Plank constant[15, 75]. As a result all other measurable quantities - transmittance T, reflectance R, and absorptance (or opacity) P - are also universal constants. In particular the ratio of absorbed to incident light intensity for suspended graphene is simply proportional to the fine structure constant $\alpha = \dfrac{e^2}{\hbar c} = \dfrac{1}{137} : P = (1-T) \approx \pi\alpha = 2.3\%$. Here c is the speed of light. This is one of the rare instances in which the properties of a condensed matter system are independent of material parameters and can be expressed in terms of fundamental constants alone. Because the transmittance in graphene is readily accessible by shining light on a suspended graphene membrane [76], it gives direct access in a simple bench-top experiment to a fundamental constant, a quantity whose measurement usually requires much more sophisticated techniques. The 2.3% opacity of graphene, which is a significant fraction of the incident light despite being only one atom thick, makes it possible to see graphene with bare eyes by looking through a glass slide covered with graphene. For a few layers of graphene stacked on top of each other the opacity increases in multiples of 2.3% for the first few layers.

The combination of many desirable properties in graphene: transparency, large conductivity, flexibility, high chemical and thermal stability, make it[77, 78] a natural candidate for solar cells and other optoelectronic devices.

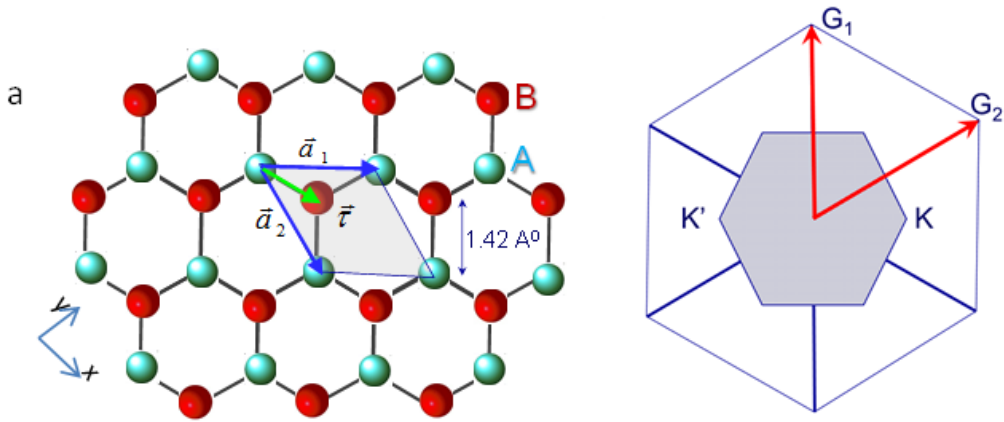

**Figure A-5. Graphene structure. a)Hexagonal lattice. Red and green colors indicate the two triangular sublattices, labeled A and B. The grey area subtended by the primitive translation vectors $\vec{a}_1$ and $\vec{a}_2$ marks the primitive unit cell and the vector marked $\vec{\tau}$ connects two adjacent A and B atoms. b) Brillouin zone showing the reciprocal lattice vectors $G_1$ and $G_2$. Each zone corner coincides with a Dirac point found at the apex of the Dirac cone excitation spectrum shown in Figure A-6. Only two of these are inequivalent (any two which are not connected by a reciprocal lattice vector) and are usually referred to as K and K'.**

## 5. *Electronic properties.*

Three ingredients go into producing the unusual electronic properties of graphene: its 2D structure, the honeycomb lattice and the fact that all the sites on its honeycomb lattice are occupied by the same atoms, which introduces inversion symmetry. We note that the honeycomb lattice is not a Bravais lattice. Instead, it can be viewed as a bipartite lattice composed of two interpenetrating triangular sublattices, A and B with each atom in the A sublattice having only B sublattice nearest neighbors and vice versa. In the case of graphene the atoms occupying the two sub-lattices are identical and as we shall see this has important implications to its electronic band structure. As shown in Figure A-5a, the Carbon atoms in sublattice A are located at positions $\vec{R} = m\vec{a}_1 + n\vec{a}_2$, where $m,n$ are integers and $\vec{a}_1 = \frac{a}{2}(3,\sqrt{3})$, $\vec{a}_2 = \frac{a}{2}(3,-\sqrt{3})$ are the lattice translation vectors for sublattice A. Atoms in sublattice B are at $\vec{R} + \vec{\tau}$, where $\vec{\tau} = (\vec{a}_2 + \vec{a}_1)/3$. The reciprocal lattice vectors, $\vec{G}_1 = \frac{2\pi}{3a}(1,\sqrt{3})$, $\vec{G}_2 = \frac{2\pi}{3a}(1,-\sqrt{3})$ and the first Brillouin zone, a hexagon with the corners at the so-called K points, are shown in Figure A-5b. Only two of the K points are inequivalent, the others being connected by reciprocal lattice vectors. The electronic properties of graphene are controlled by the low energy conical dispersion around these K points.

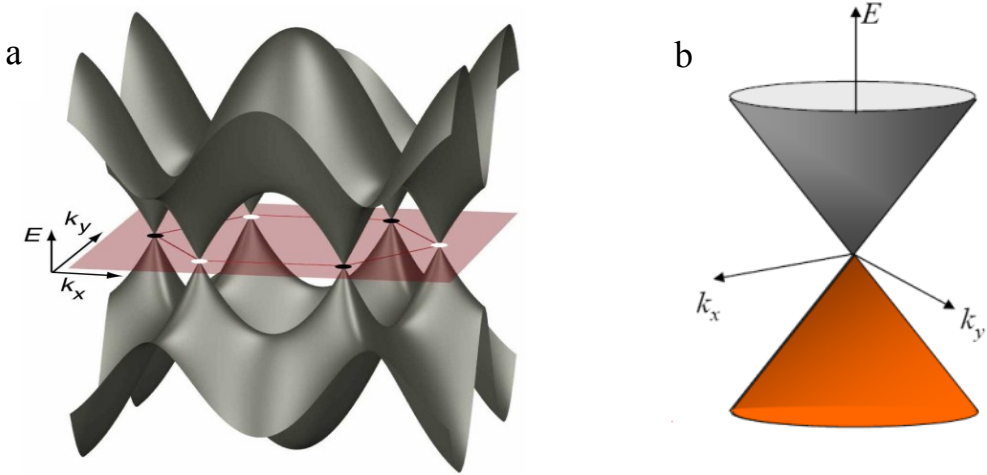

**Figure A-6. Graphene band structure. a) Three dimensional band structure. Adapted from C.W.J. Beenakker, Rev.Mod.Phys., 80 (2008) 1337. b) Zoom into low energy dispersion at one of the K points shows the electron-hole symmetric Dirac cone structure .**

**Tight binding Hamiltonian and band structure.**

The low energy electronic states, which are determined by electrons occupying the $p_z$ orbitals, can be derived from the tight binding Hamiltonian[11] in the Huckel model for nearest neighbor interactions:

1. $$H = -t \sum_{|\vec{R}|} \left( |\vec{R}\rangle\langle\vec{R}+\vec{\tau}| + |\vec{R}\rangle\langle\vec{R}-\vec{a}_1+\vec{\tau}| + |\vec{R}\rangle\langle\vec{R}-\vec{a}_2+\vec{\tau}| + h.c. \right)$$

Here $\langle\vec{r}|\vec{R}\rangle = \Psi_{p_z}(\vec{R}-\vec{r})$ is a wave function of the $p_z$ orbital on an atom in sublattice A, $\langle\vec{r}|\vec{R}+\vec{\tau}\rangle$ is a similar state on a B sublattice atom, and $t$ is the hopping integral from a state on an A atom to a state on an adjacent B atom. The hopping matrix element couples states on the A sublattice to states on the B sublattice and vice versa. It is chosen as t ~ 2.7 eV so as to match the band structure near the K points obtained from first principle computations. Since there are two Bravais sublattices two sets of Bloch orbitals are needed, one for each sublattice, to construct Bloch eigenstates of the Hamiltonian: $|\vec{k}A\rangle = \frac{1}{\sqrt{N}}\sum_{\vec{R}} e^{i\vec{k}\cdot\vec{R}}|\vec{R}\rangle$ and $|\vec{k}B\rangle = \frac{1}{\sqrt{N}}\sum_{\vec{R}} e^{i\vec{k}\cdot\vec{R}}|\vec{R}+\vec{\tau}\rangle$. These functions block-diagonalize the one-electron Hamiltonian into 2 x 2 sub-blocks, with vanishing diagonal elements and with off-diagonal elements given by: $\langle\vec{k}A|H|\vec{k}B\rangle = -te^{i\vec{k}\cdot\vec{\tau}}(1 + e^{-i\vec{k}\cdot\vec{a}_1} + e^{-i\vec{k}\cdot\vec{a}_2}) \equiv e(\vec{k})$. The single particle Bloch energies $\varepsilon(\vec{k}) = \pm|e(\vec{k})|$ give the band structure plotted in Figure A-6a, with $\varepsilon(\vec{k}) = |e(\vec{k})|$ corresponding to the conduction band $\pi^*$ and $\varepsilon(\vec{k}) = -|e(\vec{k})|$ to the valence band $\pi$. It is easy to see that $\varepsilon(\vec{k})$ vanishes when $\vec{k}$ lies at a K point. For example at $\vec{K}_1 = (\vec{G}_1 + 2\vec{G}_2)/3$, $e(\vec{K}) = -te^{i\vec{k}\cdot\vec{\tau}}\left(1 + e^{-i\vec{G}_1\cdot\vec{a}_1/3} + e^{-i2\vec{G}_2\cdot\vec{a}_2/3}\right) = 0$ where we used: $\vec{G}_i \cdot \vec{a}_j = 2\pi\delta_{ij}$. For reasons that will become clear, these points are called "Dirac points" (DP). Everywhere else in k-space, the energy is finite and the splitting between the two bands is $2|e(\vec{k})|$.

### Linear dispersion and spinor wavefunction.

We now discuss the energy spectrum and eigenfunctions for k close to a DP. Since only two of the K points - also known as "valleys" - are inequivalent we need to focus only on those two. Following convention we label them K and K'. For the K valley, it is convenient to define the (2D) vector $\vec{q} = \vec{K} - \vec{k}$. Expanding around $\vec{q} = 0$, and substituting $\vec{q} \to -i\hbar(\partial_x, \partial_y)$ the eigenvalue equation becomes [3-5]:

2.  $$H_K \Psi_K = -i\hbar v_F \begin{pmatrix} 0 & \partial_x - i\partial_y \\ \partial_x + i\partial_y & 0 \end{pmatrix} \begin{pmatrix} \psi_{KA} \\ \psi_{KB} \end{pmatrix} = \varepsilon \begin{pmatrix} \psi_{KA} \\ \psi_{KB} \end{pmatrix}$$

Where $v_F = \frac{\sqrt{3}}{2}\frac{at}{\hbar} \approx 10^6 m/s$ is the Fermi velocity of the quasiparticles. The two components $\Psi_{KA}$ and $\Psi_{KB}$ give the amplitude of the wave function on the A and B sublattices. The operator couples $\Psi_{KA}$ to $\Psi_{KB}$ but not to itself, since nearest-neighbor hopping on the honeycomb lattice couples only A-sites with B- sites. The eigenvalues are linear in the magnitude of q and do not depend on its direction, $\varepsilon(q) = \pm\hbar v_F |\vec{q}|$ producing the electron-hole symmetric conical band shown in Figure A-6b. The electron hole symmetry in the low energy dispersion of graphene is slightly modified when second order and higher neighbor overlaps are included. But the degeneracy at the DP remains unchanged even when the higher order corrections are added as discussed in the next section. The linear dispersion implies an energy independent group velocity $v_{group} = |\partial E / \hbar\partial k| = |\partial E / \hbar\partial q| = v_F$ for low-energy excitations ($|E| \ll t$).

The eigenfunctions describing the low energy excitations near point K are:

3.  $$\Psi_K^\pm(\theta_q) = \begin{pmatrix} \psi_{KA} \\ \psi_{KB} \end{pmatrix} = \frac{1}{\sqrt{2}}\begin{pmatrix} e^{i\theta_q/2} \\ \pm e^{-i\theta_q/2} \end{pmatrix}, \qquad \theta_q \equiv \tan^{-1}(q_x/q_y)$$

This two component representation, which formally resembles that of a spin, corresponds to the projection of the electron wavefunction on each sublattice.

### How robust is the Dirac Point?

A perfect undoped sheet of graphene has one electron per carbon in the $\pi$ band and, taking spin into account, this gives a half filled band at charge neutrality. Therefore, the Fermi level lies between the two symmetrical bands, with zero excitation energy needed to excite an electron from just below the Fermi energy (hole sector) to just above it (electron sector) at the DPs. The Fermi "surface" in graphene thus consists of the two K and K' points in the Brillouin zone where the π and π * bands cross. We note that in the absence of the degeneracy at the two K points graphene would be an insulator! Usually such degeneracies are prevented by level repulsion opening a gap at crossing points. But in graphene the crossing points are protected by discrete symmetries[79]: $C_3$, inversion and time reversal, so unless one of these symmetries is broken the DP will remain intact. Density functional theory calculations[80] show that adding next-nearest neighbor terms to the Hamiltonian removes the electron hole symmetry but leaves the degeneracy of the DPs. On the other hand the breaking of the symmetry between the A and B sublattices, such as for example by a corrugated substrate, is bound to lift the degeneracy at the DPs. The effect of breaking the (A,B) symmetry is directly seen in graphene's sister compound, h-BN. Just like graphene h-BN is 2-dimensional crystal with a honeycomb lattice, but the two sublattices in h-BN are occupied by different atoms and the resulting broken symmetry leaves the DP unprotected. Consequently h-BN is a band insulator with a gap of ~ 6eV.

**Dirac-Weyl Hamiltonian, masssles Dirac fermions and chirality**

A concise form of writing the Hamiltonian in equation 2 is

$$H_K = \hbar v_F \vec{\sigma} \cdot \vec{p}$$

where $\vec{p} = \hbar \vec{q}$ and the components of the operator $\vec{\sigma} = (\sigma_x, \sigma_y)$ are the usual Pauli matrices, which now operate on the sublattice degrees of freedom instead of spin, hence the term pseudospin. Formally, this is exactly the Dirac-Weyl equation in 2D, so the low energy excitations are described not by the Schrödinger equation, but instead by an equation which would normally be used to describe an ultra-relativistic (or massless) particle of spin 1/2 (such as a massless neutrino), with the velocity of light c replaced by the Fermi velocity $v_F$, which is 300 times smaller. Therefore the low energy quasiparticles in graphene are often referred to as "massless Dirac fermions".

The Dirac-Weyl equation in quantum electrodynamics (QED) follows from the Dirac equation by setting the rest mass of the particle to zero. This results in two equations describing particles of opposite helicity or chilarity (for massless particles the two are identical and the terms are used interchangeably). The chiral (helical) nature of the Dirac-Weyl equation is a direct consequence of the Hamiltonian being proportional to the helicity operator: $\hat{h} = \frac{1}{2}\vec{\sigma} \cdot \hat{p}$ where $\hat{p}$ is a unit vector in the direction of the momentum. Since $\hat{h}$ commutes with the Hamiltonian, the projection of the spin is a well-defined conserved quantity which can be either positive or negative, corresponding to spin and momentum being parallel or antiparallel to each other.

In condensed matter physics hole excitations are often viewed as a condensed matter equivalent of positrons. However, electrons and holes are normally described by separate Schrödinger equations, which are not in any way connected. In contrast, electron and hole states in graphene are interconnected, exhibiting properties analogous to the charge-conjugation symmetry in QED. This is a consequence of the crystal symmetry which requires two-component wave functions to define the relative contributions of the A and B sublattices in the quasiparticle make-up. The two-component description for graphene is very similar to the spinor wave functions in QED, but the 'spin' index for graphene indicates the sublattice rather than the real spin of the electrons. This allows one to introduce chirality in this problem as the projection of pseudospin in the direction of the momentum – which, in the K valley, is positive for electrons and negative for holes. So, just as in the case of neutrinos, each quasipartcle excitation in graphene has its "antiparticle". These particle-antiparticle pairs correspond to electron-hole pairs with the same momentum but with opposite signs of the energy and with opposite chirality. In the K' the chirality of electrons and holes is reversed, as we show below.

**Suppression of backscattering**

The backscattering probability can be obtained from the projection of the wavefunction corresponding to a forward moving particle $\Psi_K^+(\vec{q}(\theta))$ on the wavefunction of the back-scattered particle $\Psi_K^+(\vec{q}(\theta+\pi))$. Within the same valley we have $\Psi_K^+(\vec{q}(\theta)) \to \Psi_K^+(\vec{q}(\theta+\pi)) = i\Psi_K^-(\vec{q}(\theta))$ which gives $\langle \Psi_K^+(\vec{q}(\theta))| \Psi_K^-(\vec{q}(\theta)) \rangle = 0$. In other words backscattering within a valley is suppressed. This selection rule follows from the fact that backscattering within the same valley reverses the direction of the pseudospin.

We next consider backscattering between the two valleys. Expanding in $\vec{q}' = \vec{K}' - \vec{k}$ near the second DP yields $H_{K'} = -\hbar v_F \vec{\sigma}^* \cdot \vec{p}$ (* indicates complex conjugation) which is related to $H_K(\vec{q})$ by the time reversal symmetry operator, $\sigma_z C^*$ [5]. The solution in the K' valley is obtained by taking $p_x \to -p_x$ in equation 2 resulting in $\Psi_{K'}^\pm(\theta_q) = \frac{1}{\sqrt{2}}\begin{pmatrix} e^{-i\theta_q/2} \\ \pm e^{i\theta_q/2} \end{pmatrix}$.

Backscattering between valleys is also disallowed because it entails the transformation $\Psi_K^+(\theta_q) \to \Psi_{K'}^+(\theta_q + \pi) = i\Psi_K^-(\theta_q)$ which puts the particle in a state that is orthogonal to its original one. This selection rule follows from the fact that backscattering between valleys reverses the chirality of the quasiparticle.

The selection rules against backscattering in graphene have important experimental consequences including ballistic transport at low temperature [21, 22], extremely large room temperature conductivity [81] and weak anti-localization [82].

**Berry Phase**

Considering the quasiparticle wavefunction in equation 3, we note that it changes sign under a $2\pi$ rotation in reciprocal space: $\Psi_K^\pm(\theta_q) = -\Psi_K^\pm(\theta_q + 2\pi)$. This sign change is often used to argue that the wavefunctions in graphene have a Berry phase, of $\pi$. A non-zero Berry phase [83] which can arise in systems that undergo a slow cyclic evolution in parameter space, can have far reaching physical consequences that can be found in diverse fields including atomic, condensed matter, nuclear and elementary particle physics, and optics. In graphene the Berry phase of $\pi$ is responsible for the zero energy Landau level and the anomalous QHE discussed below.

On closer inspection however the definition of the Berry phase in terms of the wavefunction alone is ambiguous because the sign change discussed above can be made to disappear simply by multiplying the wavefunction by an overall phase factor, $e^{i\theta_q/2}\Psi_K^\pm(\theta_q) = \frac{1}{\sqrt{2}}\begin{pmatrix} e^{i\theta_q} \\ \pm 1 \end{pmatrix}$. For a less ambiguous result one should use a gauge invariant definition for the Berry phase[84] $\gamma = \oint_C d\lambda \left\langle \psi(\lambda) \left| i \frac{\partial}{\partial \lambda} \right| \psi(\lambda) \right\rangle$ where the integration is over a closed path in parameter space and the wavefunction $\psi(\lambda)$ has to be single valued. Applying this definition to the single valued form of the wavefunction, ie $\psi(\theta) = \frac{1}{\sqrt{2}} \begin{pmatrix} e^{i\theta_q} \\ \pm 1 \end{pmatrix}$ and taking $\lambda \rightarrow \theta;\ \oint_C \rightarrow \int_0^{2\pi} d\theta$ over a contour that encloses one of the DPs we find that the gauge invariant Berry phase in graphene is $\gamma = \pi$.

**Density of states and ambipolar gating.**

The linear DOS in graphene is a direct consequence of the conical dispersion and the electron-hole symmetry. It can be obtained by considering $n_K(q) = q^2/2\pi$, the number of states in reciprocal space within a circle of radius $|q| = |\varepsilon|/\hbar v_F$ around one of the DPs, say K, and taking into account the spin degeneracy. The DOS associated with this point is $\frac{1}{\hbar v_F} \frac{dn_K}{dq}$. Since there are 2 DPs the total DOS per unit area is:

4. $\rho(\varepsilon) = \frac{2}{\hbar v_F} \frac{dn_K}{dq} = \frac{2}{\pi} \frac{1}{(\hbar v_F)^2} |\varepsilon|$

The DOS per unit cell is then $\rho(\varepsilon) A_c$ where $A_c = 3\sqrt{3}a^2/2$ is the unit cell area. The DOS in graphene differs qualitatively from that in non-relativistic 2D electron systems leading to important experimental consequences. It is linear in energy, electron-hole symmetric and vanishes at the DP - as opposed to a constant value in the non-relativistic case where the energy dispersion is quadratic. This makes it quite easy to dope graphene with an externally applied gate. At zero doping, the lower half of the band is filled exactly up to the DPs. Applying a gate voltage induces a nonzero charge, which is equivalent to injecting (depending on the sign of the voltage) electrons in the upper half of Dirac cones or holes in the lower half. Due to the electron-hole symmetry, the gating is ambipolar with the gate induced charge changing sign at the DP. This is why the DP is commonly labeled as the charge neutrality point (CDP).

**Cyclotron mass and Landau levels**

Considering such a doped graphene device with carrier density per unit area, $n_s$, at a low enough temperature so that the electrons form a degenerate Fermi sea, one can then define a "Fermi surface" (in 2D a line). After taking into account the spin and valley degeneracies, the corresponding Fermi wave vector $q_F$ is $q_F = (\pi n_s)^{1/2}/2\pi$. One can now define an "effective mass" m* in the usual way, $m^* = \hbar q_F/v_F = \frac{\pi^{1/2}\hbar}{v_F} n_s^{1/2}$. In a 3D solid, the most direct way of measuring m* is through the specific heat, but in a 2D system such as graphene this is not practical. Instead one can use the fact that for an isotropic system the mass measured in a cyclotron resonance experiment, $m_c^*$, is identical to m* defined above. This is because in the semi-classical limit $m_c^* = \frac{1}{2\pi} \left[ \frac{\partial S}{\partial \varepsilon} \right]_{\varepsilon_F}$, where $S(\varepsilon) = \pi q^2(\varepsilon) = \pi \frac{\varepsilon^2}{\hbar^2 v_F^2}$, is the k space area enclosed by an orbit of energy $\varepsilon$, so $m_c^* = \hbar q_F / v_F = m^*$. Cyclotron resonance experiments on graphene verify that m* is indeed proportional to $n^{1/2}$ [9].

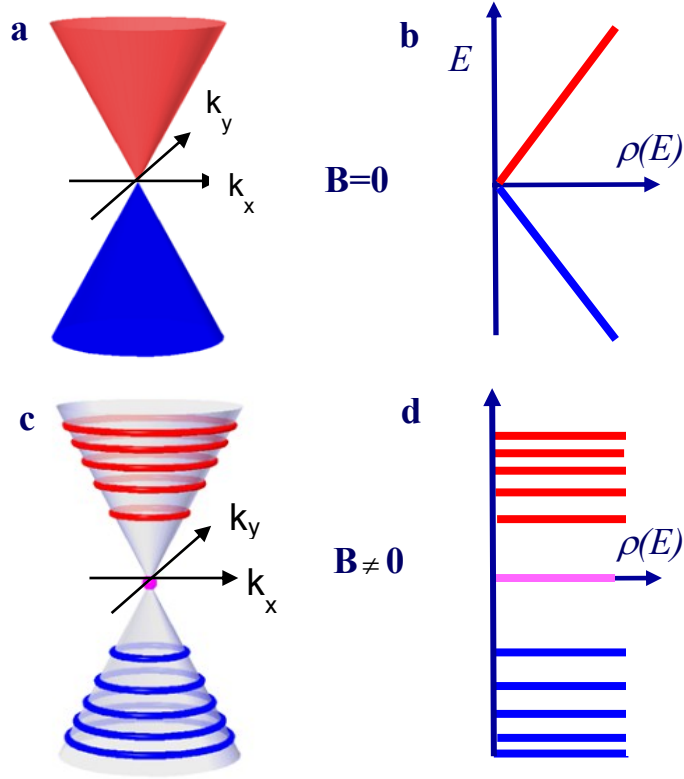

**Figure A-7. Low energy dispersion and DOS. a)** Zero-field energy dispersion of low energy excitations illustrating the electron (red) hole (blue) symmetry. **b)** The zero-field DOS is linear in energy and vanishes at the Dirac point. **c)** Finite-field energy dispersion exhibits a discrete series of unevenly spaced Landau levels symmetrically arranged about the zero-energy level, N=0, at the Dirac point. **d)** DOS in finite magnetic field consists of a sequence of $\delta(E-E_N)$ functions with gaps in between, All peaks have the same height, proportional to the level degeneracy $4B/\phi_0$.

The energy spectrum of 2D electron systems in the presence of a magnetic field, B, normal to the plane breaks up into a sequence of discrete Landau levels. For the nonrelativistic case realized in 2D electron system on helium[85] or in semiconductor heterostructures [86] the Landau level sequence consists of a series of equally spaced levels similar to that of a harmonic oscillator $E_N = \hbar\omega_c(N+1/2)$ with $\omega_c = eB/m^*$ the cyclotron frequency and a finite energy offset of 1/2 $\hbar\omega_c$. This spectrum follows directly from the semi-classical Onsager quantization condition [87] for closed orbits in reciprocal space: $S(\varepsilon) = \left(\dfrac{2\pi|e|B}{\hbar}\right)(N+\lambda); \quad N = 0,1,..$ and $\lambda = 1/2 - \gamma/2\pi$, where $\gamma$ is the Berry phase. The magnetic field introduces a new length scale, the magnetic length $l_B = \sqrt{\dfrac{\hbar}{eB}}$, which is roughly the distance between the flux quanta $\phi_0 = \dfrac{h}{e}$. The Onsager relation is equivalent to requiring that the cyclotron orbit encloses an integer number of flux quanta.

For the case of non-relativistic electrons $\gamma = 0$, resulting in the ½ sequence offset. In graphene, as a result of the linear dispersion and Berry phase of $\pi$ which gives $\lambda = 0$, the Landau level spectrum is qualitatively different. Using the same semiclassical approximation, the quantization of the reciprocal space orbit area, $\pi q_F^2$ gives $S(\varepsilon) = \pi q_F^2 = \left(\dfrac{2\pi |e| B}{\hbar}\right) N$, which produces the Landau level energy sequence:

5. $\qquad E_N = \hbar v_F q_F = \pm\sqrt{2e\hbar v_F^2 B |N|}\;;\quad N = 0, \pm 1, \ldots.$

Here the energy origin is taken to be the DP and +/- refer to electron and hole sectors respectively.

Compared to the non-relativistic case the energy levels are no longer equally spaced, the field dependence is no longer linear and the sequence contains a level exactly at zero energy which is a direct manifestation of the Berry phase in graphene[12].

We note that the Landau levels are highly degenerate, the degeneracy/per unit area being equal to 4 times (for spin and valley) the orbital degeneracy (the density of flux lines): $4\dfrac{B}{\phi_0}$.

The exact finite field solutions to this problem can be obtained [88-91] from the Hamiltonian in equation 2, by replacing $-i\vec{\nabla} \to -i\vec{\nabla} + e\vec{A}$, where in the Landau gauge, the vector potential is $\vec{A} = B(-y, 0)$ and $\vec{B} = \vec{\nabla} \times \vec{A}$. The energy sequence obtained in this approach is the same as above, but now one can also obtain the explicit functional form of the eigenstates.

### From bench-top quantum relativity to nano-electronics

Owing to the ultra-relativistic nature of its quasiparticles, graphene provides a platform which for the first time allows testing in bench-top experiments some of the strange and counterintuitive effects predicted by quantum relativity, but often not yet seen experimentally, in a solid-state context. One example is the so called "Klein paradox" which predicts unimpeded penetration of relativistic particles through high [92] potential barriers. In graphene the transmission probability for scattering through a high potential barrier [93, 94] of width D at an angle $\theta$, is $T = \dfrac{\cos^2(\theta)}{1 - \cos^2(q_x D)\sin^2(\theta)}$. In the forward direction the transmission probability is 1 corresponding to perfect tunneling. Klein tunneling is one of the most exotic and counterintuitive phenomena. It was discussed in many contexts including in particle, nuclear and astro-physics, but direct observation in these systems has so far proved impossible. In graphene on the other hand it may be observed [95]. Other examples of unusual phenomena expected due to the massless Dirac-like spectrum of the quasiparticles in graphene include electronic negative index of refraction[96], zitterbewegung and atomic collapse[97].

Beyond these intriguing single-particle phenomena electron-electron interactions and correlation are expected to play an important role in graphene [98-104] because of its weak screening and large effective "fine structure constant" $\alpha = \dfrac{e^2}{\hbar v_F} \approx 2$ [3] In addition, the interplay between spin and valley degrees of freedom is expected to show SU(4) fractional QH physics in the presence of a strong magnetic field which is qualitatively different from that in the conventional 2D semiconductor structures[104, 105].

The excellent transport and thermal characteristics of graphene make it a promising material for nanoelectronics applications. Its high intrinsic carrier mobility[106], which enables low operating power and fast time response, is particularly attractive for high speed electronics[57]. In addition, the fact that graphene does not lose its electronic properties down to nanometer length scales, is an invaluable asset in the quest to downscale devices for advanced integration. These qualities have won graphene a prime spot in the race towards finding a material that can be used to resolve the bottleneck problems currently encountered by Si-based VLSI electronics.

Amongst the most exciting recent developments is the use of graphene in biological applications. The strong affinity of bio-matter to graphene makes it an ideal interface for guiding and controlling biological processes. For example graphene was found to be an excellent bio-sensor capable of differentiating between single and double stranded DNA [107]. New experiments report that graphene can enhance the differentiation of human neural stem cells for brain repair [108] and that it accelerates the differentiation of bone cell from stem cells[109]. Furthermore, graphene is a promising material for building efficient DNA sequencing machines based on nanopores, or functionalized nano-channels [110].

**Is graphene special?**

The presence of electron-hole symmetric Dirac cones in the band structure of graphene endows it with extraordinary properties, such as ultra-high carrier mobility which is extremely valuable for high speed electronics, highly efficient ambipolar gating and exquisite chemical sensitivity.
One may ask why graphene is special. After all there are many systems with Dirac cones in their band structure. Examples include transition metal dichalcogenites below the charge density wave transition[111], cuprates below the superconducting transition [112] and pnictides below the spin density wave transition[113]. However in all the other cases the effect of the DP on the electronic properties is drowned by states from other parts of the Brillouin zone which, not having a conical dispersion, make a much larger contribution to the DOS at the Fermi energy. In graphene on the other hand the effect of the DPs on the electronic properties is unmasked because they alone contribute to the DOS at the Fermi energy. In fact, as discussed above, had it not been for the DPs, graphene would be a band insulator.

## 6. *Effect of the substrate on the electronic properties of graphene.*

The isolation of single layer graphene by mechanical exfoliation was soon followed by the experimental confirmation of the Dirac-like nature of the low energy excitations [9, 81]. Measurements of the conductivity and Hall coefficient on graphene FET devices demonstrated ambipolar gating and a smooth transition from electron doping at positive gate voltages to hole doping on the negative side. At the same time the conductivity remained finite even at nominally zero doping, consistent with the suppression of backscattering expected for massless Dirac fermions. Furthermore, magneto-transport measurements in high magnetic field which revealed the QHE confirmed that the system is 2 dimensional and provided evidence for the chiral nature of the charge carriers through the absence of a plateau at zero filling (anomalous QHE). Following these remarkable initial results, further attempts to probe deeper into the physics of the DP by measuring graphene deposited on $SiO_2$ substrates, seemed to hit a hard wall. Despite the fact that the QHE was readily observed, it was not possible in these devices to approach the DP and to probe its unique properties such as ballistic transport [56, 114], specular Andreev reflections expected [63, 115] at graphene/superconducting junctions [116, 117] or correlated phenomena such as the fractional QHE [118]. Furthermore, STS measurements did not show the expected linear DOS or its vanishing at the charge neutrality point (CNP)[119, 120].

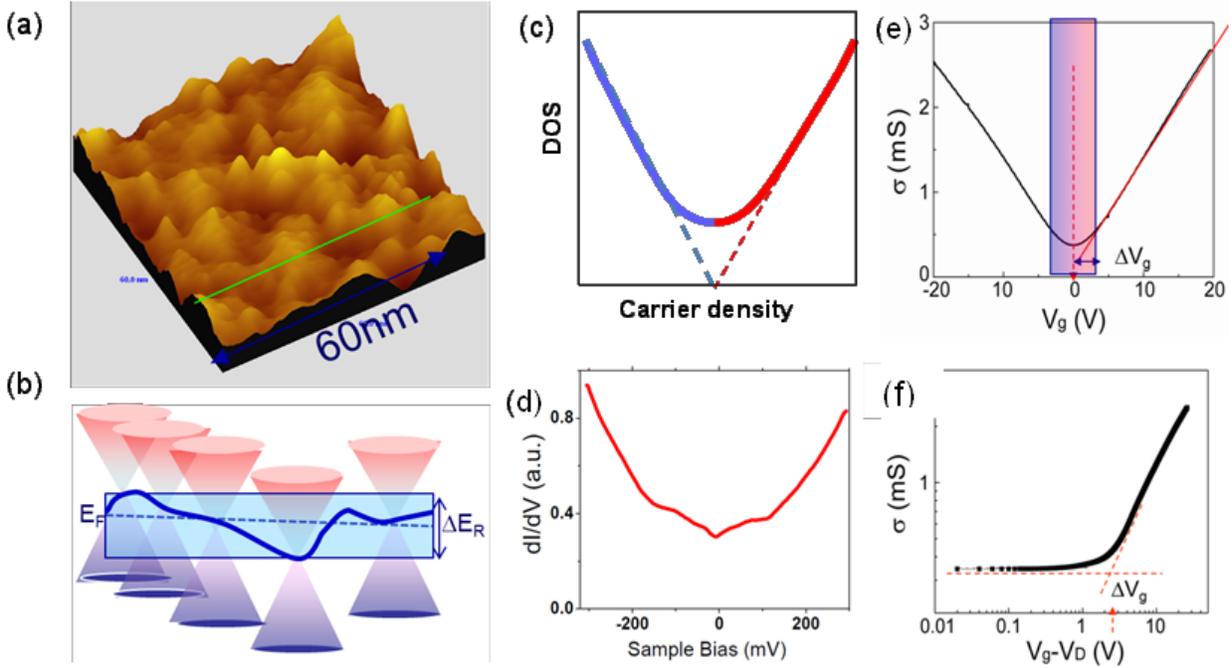

The failure to probe the DP physics in graphene deposited on $SiO_2$ substrates was understood later, after applying sensitive local probes such as STM [119-124] and SET (single electron transistor) microscopy[125], and attributed to the presence of a random distribution of charge impurities associated with the substrate. The electronic properties of graphene are extremely sensitive to electrostatic potential fluctuations because the carriers are at the surface and because of the low carrier density at the DP. It is well known that insulating substrates such as $SiO_2$ host randomly distributed charged impurities, so that graphene deposited on their surface is subject to spatially random gating and the DP energy (relative to the Fermi level) displays random fluctuations, as illustrated in Figure A-8b. The random potential causes the charge to break up into electron-hole puddles: electron puddles when the local potential is below the Fermi energy and hole puddles when it is above. These puddles fill out the DOS near the DP (Figure A-8c,d) making it impossible to attain the zero carrier density condition at the DP for any applied gate voltage as seen in the STS image shown in Figure A-8e. Typically for graphene deposited onto $SiO_2$ the random potential causes DP smearing over an energy range $\Delta E_R \approx 30-100 meV$. When the Fermi energy is within $\Delta E_R$ of the DP, a gate voltage change transforms electrons into holes and vice versa but it leaves the net carrier density almost unchanged. As a result, close to the DP the gate voltage cannot affect significant changes in the net carrier density. This is directly seen as a broadening of order 1-10V in the conductivity versus gate voltage curves, Figure A-8e,f, which corresponds to a minimum total carrier density in these samples of $n_s \sim 10^{11}$ cm$^{-2}$. The energy scale defined by the random potential also defines a temperature $k_B T \sim \Delta E_R$ below which the electronic properties such as the conductivity are independent of temperature.

**Integer and fractional quantum Hall effect.**

The substrate induced random potential which makes the DP inaccessible in graphene deposited on SiO$_2$, explains the inability to observe in these samples the linear DOS and its vanishing at the CNP with STS measurements. As we show below this also helps understand why the integer QHE is readily observed in such samples but the fractional QHE is not.

To observe the QHE in a 2D electron system one measures the Hall and longitudinal resistance while the Fermi energy is swept through the Landau levels (LL), by changing either carrier density or magnetic field [126]. The Fermi energy remains within a LL until all the available states, $4B/\phi_0$ per unit area, are filled and then jumps across the gap to the next level unless, as is usually the case, there are localized impurity states available within the gap which are populated first. In homogeneous samples the LL energy is uniform in the bulk and diverges upwards (downwards) for electrons (holes) near the edges. As a result, when the Fermi energy is placed within a bulk gap between two LLs, it must intersect all the filled LLs at the edge. This produces one dimensional ballistic edge channels, in which the quasiparticles on opposite sides of the sample move in opposite directions, as shown in Figure A-9a. These ballistic channels lead to a vanishing longitudinal resistance and to a quantized Hall conductance: $\sigma_{xy} = \nu \frac{e^2}{h}$ where $\nu$ is the "filling factor". $\nu$ counts the number of occupied ballistic channels which is the number of filled LLs multiplied by the (non-orbital) degeneracy, 4 in the case of graphene. The N=0 level is special because half of its states are electron like (K valley) and half hole like (K' valley) so that its contribution consists of only 2 states for each species. Therefore when the Fermi energy is in between levels N and N+1, the number of occupied states is 4N+2, corresponding to $\nu = 4(N+1/2)$. The ½ offset, absent in the case of non-relativistic electrons, is a direct consequence of the chiral symmetry of the low energy quasiparticles in graphene. As a result the series of QH plateaus in graphene:

6. $\sigma_{xy} = 4(N+1/2)\frac{e^2}{h}$    $N = 0,\pm 1,...$

lacks the plateau at zero Hall conductance which in the non-relativistic case is associated with a gap at zero energy.

The ballistic edge channels which are necessary to observe the QHE are destroyed by excessive disorder. This is because large random potential fluctuations may prevent the formation of a contiguous gap across the entire sample and then the Fermi energy cannot be placed in a gap between two LL as illustrated in Figure A-9b. This could allow the creation of a conducting path that connects the two edges resulting in back-scattering, the destruction of the ballistic channels and the loss of the quantized plateaus. In graphene, the condition for to placing the Fermi energy $\Delta E_R \geq 30 meV$ between the N=0 and N=1 LLs, and thus to observe at least one QH plateau, is: $E_1 - E_0 = 35 meV \cdot (B[T])^{1/2} > \Delta E_R, k_B T$. For a typical graphene sample on SiO$_2$, where this implies that the integer QHE can already be seen in fields $B \geq 1T$, consistent with the early experiments.

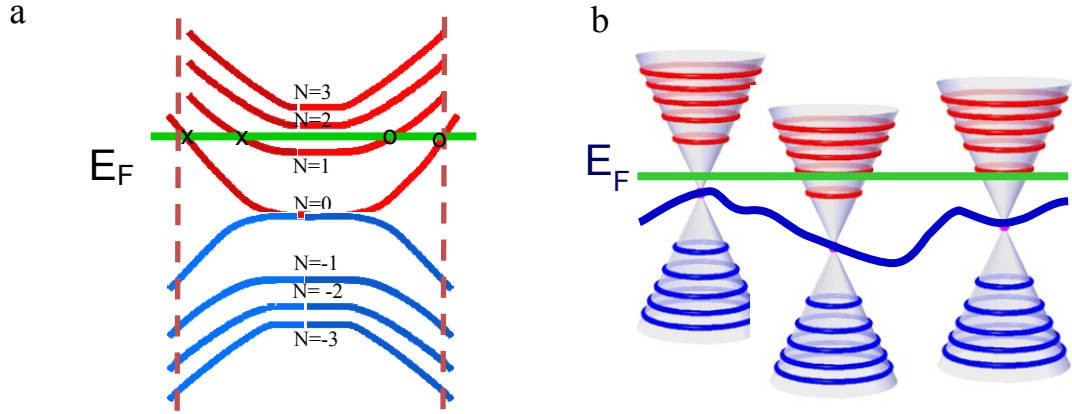

**Figure A-9. Landau levels and quantum Hall effect. a) Landau levels in the bulk showing their upward (downward for holes) bending at sample edges indicated by dashed lines. The Fermi energy (green line) lies in the gap between the N=1 and N=2 levels in the bulk and at the edges it intersects both filled LLs. The 4 intersection points define ballistic one dimensional edge channels in which the electrons move out of the page (right edge marked by circles) or into the page (left channels marked by crosses). b) In the presence of a random potential the Fermi energy cannot always be placed in a bulk gap. This may destroy the quantum Hall effect as discussed in the text.**

The condition for observing the fractional QHE [127] is more stringent. The fractional QHE occurs when as a result of strong correlations the system can lower its energy for certain filing factors by forming "composite fermions" which consist of an electron bound together with an even number of flux lines [128]. These composite fermions sense the remnant magnetic field left after having "swallowed" the flux lines, and as a result their energy spectrum breaks up into "Lambda levels" (ΛL) which are the equivalent of LLs but for the composite fermions in the smaller field. Just as the electrons display an integer QHE whenever the Fermi energy is in a gap between LLs, so do the composite fermions when the Fermi energy is in a gap between the ΛLs. The filling factors for which this occurs take fractional values $\nu = \frac{p}{2mp \pm 1}$, $p = 1,2..; m = \pm 1, \pm 2$. The characteristic spacing between the ΛLs is controlled by the Coulomb energy, and is much smaller than the spacing between LLs: $E_\Lambda \propto 0.1 \frac{e^2}{\varepsilon l_c} \sim 5 meV \frac{(B[T])^{1/2}}{\varepsilon}$ where ε is the dielectric constant of the substrate. Thus the criterion for decoupled edges in the fractional QHE case becomes $E_\Lambda > \Delta E_R \geq 30 meV \Rightarrow B > \varepsilon 50\ T$, which is larger than any dc magnetic field attainable to date. In other words, the fractional QHE is not observable in graphene deposited on SiO$_2$.

Therefore in order to access the intrinsic properties of graphene and correlation effect it is imperative to reduce the substrate-induced random potential. The remainder of this review is devoted to the exploration of ways to reduce this random potential and to access the intrinsic electronic properties of graphene.

B.   Scanning Tunneling Microscopy and Spectroscopy

In STM/STS experiments, one brings a sharp metallic tip very close to the surface of a sample, with a typical tip-sample distance of ~1nm. For positive tip-sample bias voltages, electrons tunnel from the tip into empty states in the sample; for negative voltages, electrons tunnel out of the occupied states in the sample into tip. In the Bardeen tunneling formalism [129] the tunneling current is given by

7.   $I = \frac{4\pi e}{\hbar} \int_{-\infty}^{+\infty} [f(E_F - eV + \epsilon) - f(E_F + \epsilon)] \rho_S(E_F - eV + \epsilon) \rho_T(E_F + \epsilon) |M|^2 d\epsilon$

where $-e$ is the electron charge, $f(x)$ is the Fermi function, $E_F$ the Fermi energy, $V$ the sample bias voltage, $\rho_T$ and $\rho_s$ represent the DOS in the tip and sample, respectively. The tunneling matrix M depends strongly on the tip-sample distance $z$. When the tip DOS is constant and at sufficiently low temperatures the tunneling current can be approximated by $I(r,z,V) \propto \left[ \int_{-\infty}^{eV} \rho_s(r,\varepsilon) d\varepsilon \right] \exp^{-2\kappa(r)}$ where $\kappa \sim \sqrt{2m\phi/\hbar}$ is the inverse decay length and $\phi$ is the local barrier height or average work function. The exponential dependence on height makes it possible to obtain high resolution topography of the surface at a given bias voltage. The image is obtained by scanning the sample surface while maintaining a constant tunneling current with a feedback loop which adjusts the tip height to follow the sample surface. We note that an STM image not only reflects topography but also contains information about the local DOS which can be obtained directly [130] by measuring the differential conductance:

8.   $\frac{dI}{dV}(V) \propto \rho_s(\epsilon = eV)$

Here $E_F$ is set to be zero. In STS the tip-sample distance is held fixed by turning off the feedback loop while measuring the tunneling currents as a function of bias voltage. Usually one can use a lock-in technique to measure differential conductance directly by applying a small ac modulation to the sample bias voltage.

In practice, finite temperatures introduce thermal broadening through the Femi functions in Eq.(7), leading to reduced energy resolution in STS. For example, at 4.2K the energy resolution cannot be better than 0.38meV. Correspondingly, the ac modulation of the sample bias should be comparable to this broadening in order to achieve highest resolution. The condition of flat tip DOS is usually considered satisfied for common tips, such as Pt-Ir, W or Au, as long as the sample bias voltage is not too high. Compared to a sharp tip, a blunt tip typically has a flatter DOS. In order to have reliable STS, one should make sure a good vacuum tunneling is achieved. To this end, one can check the spatial and temporal reproducibility of the spectra and ensure that they are independent of tip-sample distance [130].

### 1. Graphene on SiO$_2$

As discussed in part A, the insulating substrate of choice and the most convenient, SiO$_2$, suffers from large random potential fluctuations which make it impossible to approach the DP due to the formation of electron hole puddles [125]. STM topography on these substrates does show a honeycomb structure for single layer graphene and a triangular lattice for multi-layers [121, 131] with very few topological defects which is testimony to the structural robustness of graphene. However, in contrast to the case of graphene on graphite [65], these samples show significant corrugation on various length scales ranging from ~1-32nm due to the substrate, wrinkling during fabrication [100] and possibly intrinsic fluctuations. These corrugations can lead to broken sub-lattice symmetry affecting both transport and the STM images and can lead for example to the appearance of a triangular lattice instead of the honeycomb structure in unperturbed graphene [131, 132].

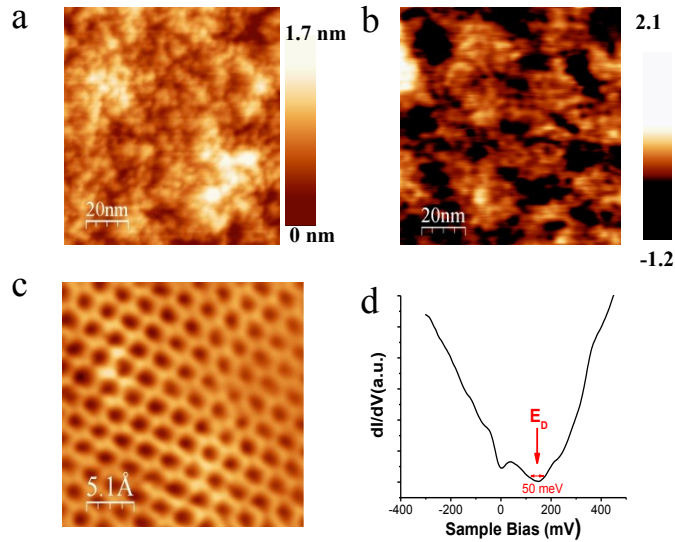

**Figure B-1. STM/STS of graphene on chlorinated SiO$_2$ . a) STM topography image of a typical 300x300 nm graphene area. Tunneling current I$_t$=20pA, and bias voltage V$_{bias}$=190mV. Legend shows height scale. b) Differential conductance map over the area in panel (a) taken close to the Dirac point (~140mV), marked E$_D$ in (d). Legend shows differential conductance scale. c) STM atomic resolution image (I$_t$=20pA, V$_{bias}$=300mV) shows honeycomb structure. d) Differential conductance averaged over the area shown in (b). . Adapted from A. Luican *et al*. Phys. Rev. B, 83 (2011)**

In the presence of scattering centers, the electronic wave functions can interfere to form standing wave patterns which can be observed by measuring the spatial dependence of dI/dV at a fixed sample bias voltage. By using these interference patterns, it was possible to discern individual scattering centers in the dI/dV maps obtained at energies far from the CNP when the electron wave length is small [133]. No correlations were found between the corrugations and the scattering centers, suggesting the latter play a more important role in the scattering process. When the sample bias voltage is close to the CNP, the electron wave length is so large that it covers many scattering centers and the dI/dV maps show coarse structures ( Figure B-1b) which are attributed to electron-hole puddles.

The Fourier transform of the interference pattern provides information about the energy and momentum distribution of quasiparticle scattering, which can be used to infer band structure [123]. While for unperturbed single layer graphene, the patterns should be absent or very weak [134], for graphene on SiO$_2$ clear interference patterns arise [133] due to strong scattering centers which are believed to be trapped charges. The dispersion E(k) obtained from the interference pattern is linear with $v_F$ = 1.5±0.2×10$^6$ and 1.4±0.2×10$^6$m/s for electron and hole states, respectively. It should be noted that these values are for states with energies significantly far from the Fermi level and the CNP. At lower energies, transport measurements yielded $v_F$=1.1×10$^6$m/s [9, 10].

## 2. *Graphene on metallic substrates*

As detailed in the introductory section epitaxial growth of graphitic layers can be achieved on a wide range of metal substrates by thermal decomposition of a hydrocarbon or by surface segregation of carbon atoms from the bulk metal[135, 136]. Graphene monolayers are relatively easy to prepare on metal surfaces and, with the right metal and growth conditions, the size of the monolayer flakes is almost unlimited. STM studies of graphitic flakes on metallic substrates have focused mostly on the structure. On Ir(111) [46], Cu(111) [46, 137, 138] and on Ru(0001) [139] (Figure B-2a) they revealed structurally high-quality monolayer graphene and continuity which is not limited by the size of terraces in the substrate, although the overall structure is often strongly modulated by the mismatch with the lattice of the underlying metal which leads to Moire super structures (Figure B-2 c). The electronic properties of these graphitic layers are strongly affected by the metallic substrates leading to significant deviations from the linear dispersion expected for free standing graphene[139](Figure B-2b). Thus, in order to access the unique electronic properties of graphene while also taking advantage of the high quality and large scales achieved on metallic substrates it is necessary to separate the graphitic layer from its metallic substrate.

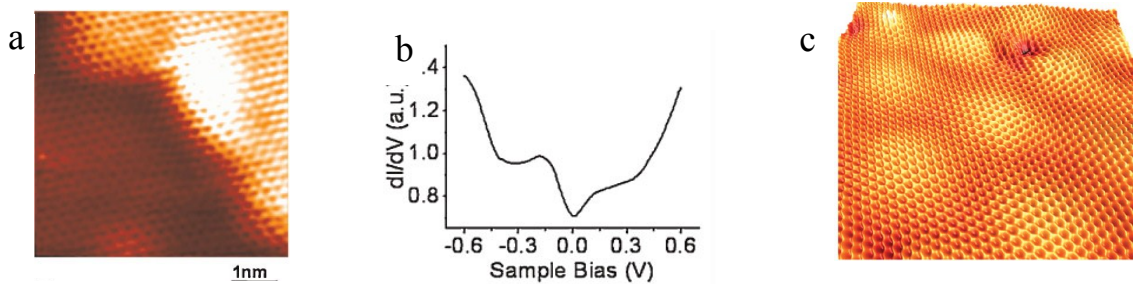

Figure B-2. STM/STS on graphene on Ru(0001) and Cu(111). a) Atomic-resolution image showing graphene overlayer across a step edge on the Ru substrate. b) Differential conductance spectrum of graphene layer on Ru substrate. Adapted from Pan et al, Adv. Mat. 21 (2009) 2777. c) Atomic resolution STM topography image of graphene on Cu showing the Moire´pattern and the honeycomb structure. Adapted from Gao et al Nano Letters, 10 (2010) 3512.

## 3. *Graphene on Graphite*

The choice of a minimally invasive substrate for gaining access to the electronic properties of graphene is guided by the following attributes: flat, uniform surface potential, and chemically pure. Going down this list, the substrate that matches the requirements is graphite, the "mother "of graphene. Because it is a conductor, potential fluctuations are screened and furthermore it is readily accessible to STM and STS studies.

**Almost ideal graphene seen by STM and STS**

During exfoliation of a layered material, cleavage takes place between the least coupled layers. Occasionally, when cleavage is partial, a region in which the layers are separated can be found adjacent to one where they are still coupled. This situation in shown in Figure B-3a where partial cleavage creates the boundary – seen as a diagonal dark ridge - between the decoupled region marked G and a less coupled region marked W. The layer separation in these regions is obtained from height profiles along lines α and β shown in the figure. In region W the layer separation, ~0.34nm, is close to the inter layer spacing 0.335nm of graphite, but in region G the larger separation, ~0.44nm, means that the top layer is lifted by ~30%. Atomic resolution STM images show a honeycomb structure in region G but a triangular one in region W. The triangular lattice in region W is consistent with the sub-lattice asymmetry expected for Bernal stacked graphite. In this stacking, which is the lowest energy configuration for graphite, the atoms belonging to sublattice A in the topmost layer are stacked above B atoms in the second layer, while B atoms in the topmost layer are above the hollow sites of the carbon hexagons of the second layer. Ab initio band structure calculations [140] show that in the presence of interlayer coupling this site asymmetry leads to a strong asymmetry in the local density of states at the Fermi level with the B atoms having the larger DOS. This leads to STM images in which the B atoms on graphite appear more prominent than the A atoms resulting in a triangular lattice[140, 141]. In the absence of interlayer coupling the DOS is symmetric between the two sublattices and one would expect to observe a honeycomb structure as seen in region G. The observation of the honeycomb structure provides an important first clue in the search for decoupled graphene flakes on graphite, but it is not sufficient to establish decoupling between the layers. This is because, even though the atomic resolution topography of the surface of HOPG was one of the first to be studied by STM, its interpretation is not unique and depends on other factors such as the bias voltage. The triangular structure discussed above is commonly seen in atomic resolution topographic images of graphite at low bias voltages, but there are also many reports of the appearance of a honeycomb structure under various circumstances which are often not reproducible [142-149].

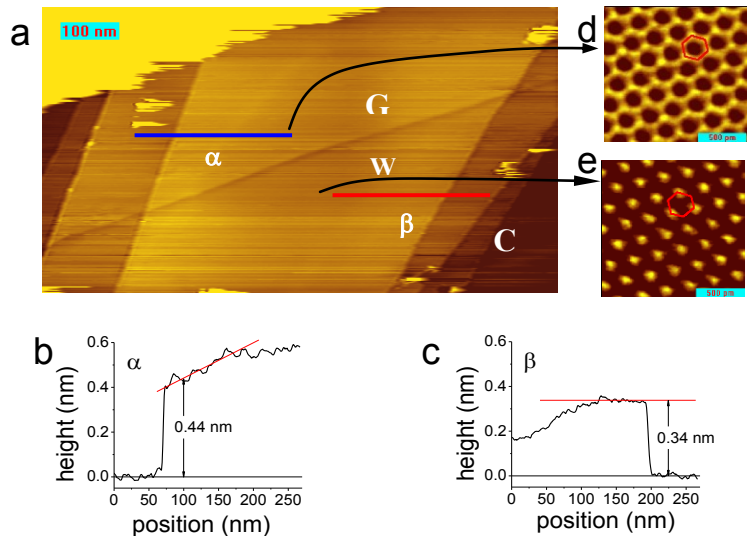

**Figure B-3. Graphene flake on the surface of graphite. a) Large area STM topography. Atomic steps are clearly visible at edges of graphene layers. A diagonal ridge separates a region with honeycomb structure (region G), from a triangular structure (region B) below. The region marked C represents the surface of graphite surrounding the flake. (b,c) Height profiles along cross sectional cuts marked α and β. (d,e) Atomic resolution images show the honeycomb structure in region G and the triangular lattice in region W.**

As we show below in order to establish the degree of coupling of the top layer to the layers underneath it is necessary to carry out spectroscopic measurements and in particular Landau level spectroscopy. In the earlier works only topographic measurements were reported [125-132] and therefore it was not possible to correlate the structure seen in STM with the degree of coupling between the layers.

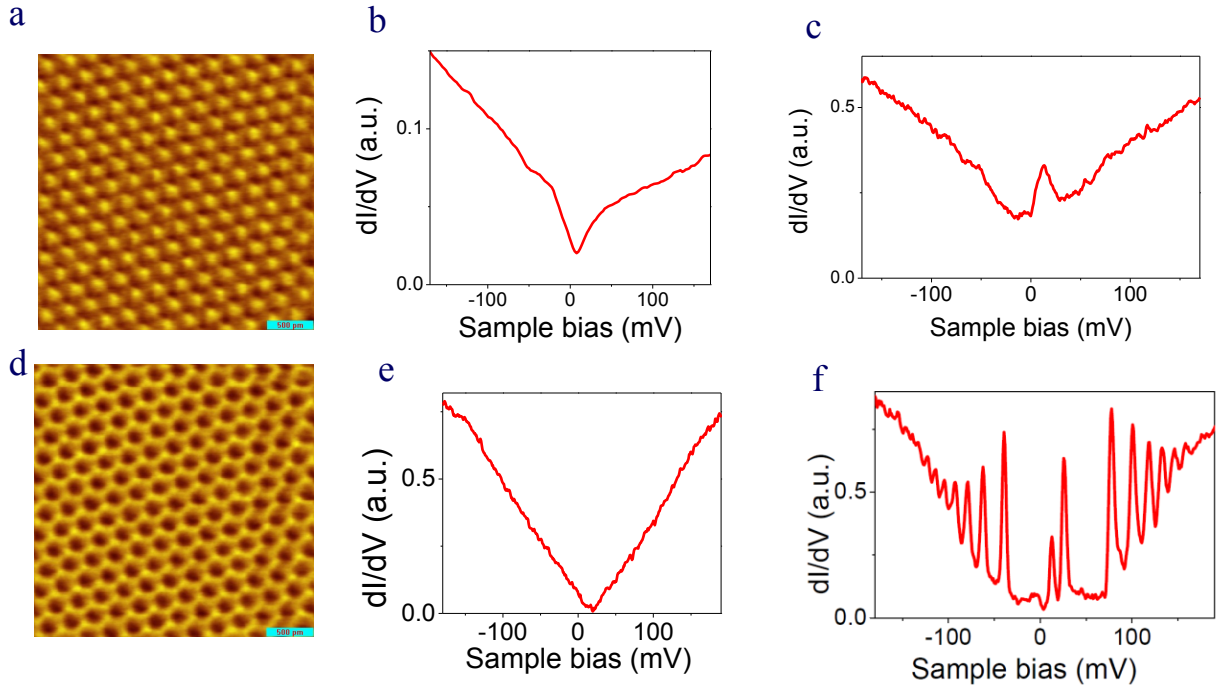

**Figure B-4. Identifying a decoupled graphene layer.** a) Atomic resolution topography in region C of Figure B-3a, shows a triangular lattice. b) STS in zero field and at T=4.2 K in region C. c) Finite field spectra ( B=3T) in region C shows no LL peak sequence. d) Atomic resolution topography in region G shows honeycomb structure. b) STS in zero field and at T=4.2 K shows the "V shaped" density of states that vanishes at the Dirac point expected for massless Dirac fermions. The Fermi energy is taken to be at zero. c) LL are clearly seen in region G.  Spectra at  T=4.2K and B=4T . (ac modulation: 2mV, junction resistance ~6GΩ).

We start in region C of Figure B-3 where atomic resolution topography images show a triangular lattice for bias voltages in the range 100mV - 800mV and for junction resistances exceeding 1GΩ as seen in Figure B-4a.  Zero field STS, Figure B-4b, shows finite differential conductance at the neutrality point, consistent with the finite DOS expected for bulk graphite. The finite field spectra shown in  Figure B-4c are again consistent with bulk graphite: no Landau level sequence is observed consistent with the energy dispersion normal to the surface. In summary the data in region C presents the characteristic features of bulk graphite.

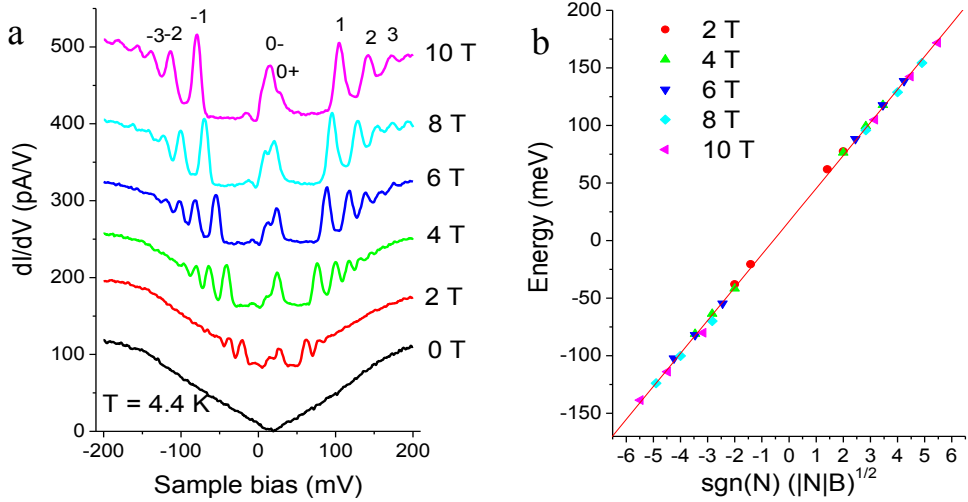

**Figure B-5.** Landau level spectroscopy of graphene. a) Evolution of Lnadau levels with field at 4.4 K and indicated values of field. b) LL energies plotted against the reduced parameter $(NB)^{1/2}$ collapse onto a straight line indicating square-root dependence on level index and field. Symbols represent the peak positions obtained from (a) and the solid line is a fit to Eq.(9).

The situation is qualitatively different in region G, where the atomic resolution spectroscopy image (Figure B-4d) shows the honeycomb structure in the entire region, which extends over ~ 400nm, with no visible vacancies or dislocations. STS in this region in zero field, Figure B-4e, shows that the DOS is V-shaped and vanishes at the DP which is ~16meV above the Femi energy ( taken as zero) corresponding to unintentional hole doping with a concentration of $n_s \sim 2 \times 10^{10} cm^2$. In the presence of a magnetic field the DOS develops sharp LL peaks (Figure B-4f). The three results in Figure B-4d,e,f are consistent with intrinsic graphene. In order to verify that the sequence of peaks in Figure B-4c does indeed correspond to massless-Dirac-fermions, Li *et al*. [65, 66] measured the dependence of the peak energies on field and level-index and compared them to the expected values (Eqn. 5):

9. $E_n = E_D \pm \sqrt{2e\hbar v_F^2 |N|B}$  N=0, ±1 …..

where $E_D$ is the energy at the DP. The N=0 level is a consequence of the chirality of the Dirac fermions and does not exist in any other known two dimensional electron system. This field-independent state at the DP together with the square-root dependence on both field and level index, are the hallmarks of massless Dirac fermions. They are the criterion that is used for identifying graphene electronically decoupled from the environment or for determining the degree of coupling between coupled layers, as discussed below.

The field dependence of the STS spectra in region G, shown in Figure B-5, exhibits an unevenly spaced sequence of peaks flanking symmetrically, in the electron and hole sectors, a peak at the DP. All the peaks, except the one at the DP, which is identified with the N = 0 LL, fan out to higher energies with increasing field. The peak heights increase with field consistent with the increasing degeneracy of the LLs. To verify that the sequence is consistent with massless Dirac fermions we plot the peak positions as a function of the reduced parameter $(|N|B)^{1/2}$ as shown in Figure B-5b. This scaling procedure collapses all the data unto a straight line. Comparing to Eqn. 9, the slope of the line gives a direct measure of the Fermi velocity, $v_F = 0.79 \times 10^6$ m/s. This value is ~ 20% below that expected from single particle calculations and, as discussed later, the reduction can be attributed to electron-phonon (e-ph) interactions.

We conclude that the flake marked as region G is electronically decoupled from the substrate

### Landau Level Spectroscopy

The technique described above, also known as LL spectroscopy, was developed by Li *et al*. [65, 66] to probe the electronic properties of graphene on graphite. They showed that LL spectroscopy can be used to obtain information about the intrinsic properties of graphene: the Fermi velocity, the quasiparticle lifetime, the e-ph coupling, and the degree of coupling to the substrate. LL spectroscopy is a powerful technique which gives access to the electronic properties of Dirac fermions when they define the surface electronic properties of a material and when it is possible to tunnel into the surface states. The technique was adopted and successfully implemented to probe massless Dirac fermions in other systems including graphene on $SiO_2$[120], epitaxial graphene on SiC [150], graphene on Pt [151] and topological insulators [152, 153].

An alternative method of accessing the LLs is to probe the allowed optical transitions between the LL by using cyclotron resonance measurements. This was demonstrated in early experiments on $SiO_2$[154, 155], epitaxial graphene[156] and more recently on graphite[157].

### Finding graphene on graphite

The flake in region G of Figure B-3, exhibits all the characteristics of intrinsic graphene – honeycomb crystal structure, V shaped DOS which vanishes at the DP, a LL sequence which displays the characteristic square root dependence on field and level index, and contains an N=0 level. One can use these criteria to develop a recipe for finding decoupled graphene flakes on graphite. For a successful search one needs the following: 1) STM with a coarse motor that allows scanning large areas in search of stacking faults or atomic steps. Decoupled graphene is usually found covering such faults as shown in Figure B-4. 2) A fine motor to zoom into subatomic length scales after having identified a region of interest. If the atomic resolution image in this region shows a honeycomb structure as in Figure B-4a one continues to the next step. 3) Scanning tunneling spectroscopy. If the region is completely decoupled from the substrate the STS will produce a V shaped spectrum as in Figure B-4b. 4) The last and crucial step is LL spectroscopy. A completely decoupled layer will exhibit the characteristic single layer sequence and scaling as shown in Figure B-4c. In the presence of coupling to the substrate the LL sequence is modified. Importantly LL spectroscopy can be used to quantify the degree of coupling to the substrate, as discussed later in the section on multi-layers.

### Landau level linewidth and electron-electron interactions.

Comparing the LL spectra in Figure B-4c with the idealized sequence of equal height delta peaks in Figure A-7, it is clear that the spectrum is strongly modified by a finite linewidth. The data in Figure B-5 is resolution limited so in order to access the intrinsic broadening of the LL high resolution spectra are obtained by decreasing the ac modulation until the spectrum becomes independent of the modulation amplitude. The peculiar V shaped lower envelope of the spectrum is a direct consequence of the square root dependence on energy as we show below. Similarly, the down-sloping of the upper envelope is a direct consequence of the linear increase in linewidth with energy.

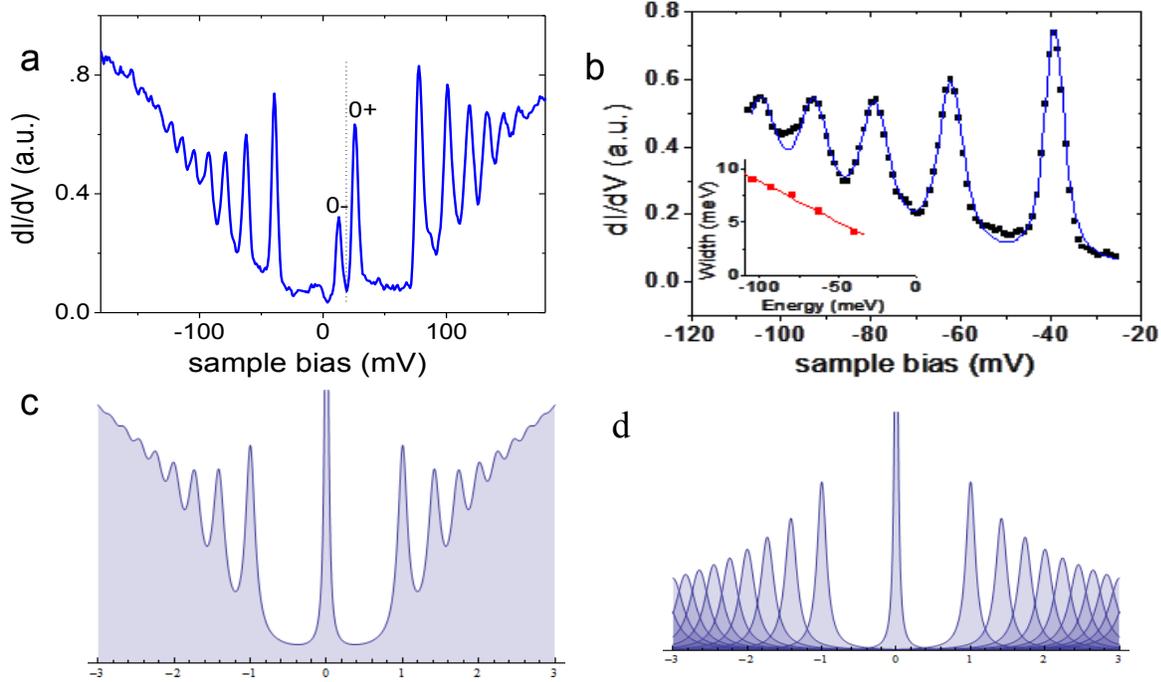

Figure B-6. Quasiparticle lifetime in graphene. a) Landau level spectrum at 4.4K and 4T. (ac modulation 2mV, setpoint 53pA at 300mV). b) High resolution spectrum on the hole side (symbols) together with a fit with sequence of Lorentzian peaks (solid line). The inset represents the energy dependence of the peak widths. c) Simulated overall density of states including the energy dependence of the linewidth. d) Individual peaks used to obtain the spectrum in c.

The sharpness and large signal-to noise ratio of the LL peaks makes it possible to extract the energy dependence of the quasiparticle lifetime from the spectrum. The level sequence can be fit to high accuracy with a sum of peak functions centered at the measured peak energies and with the line width of each peak left as free parameters. Comparing fits with various line-shapes, Li et al found that Lorentzians give far better fits than Gaussians. This suggests that the linewidth reflects the intrinsic quasiparticle lifetime rather than impurity broadening. From the measured energy dependence of the linewidth (Figure B-6b) they found that the inverse quasiparticle lifetime is:

10. $\frac{1}{\tau} = \frac{|E|}{\gamma} + \frac{1}{\tau_0}$

where E is the LL energy in units of eV, $\gamma \sim 9$fs/eV, and $\tau_0 \sim 0.5$ps at the Fermi level. The linear energy dependence of the first term is attributed to the intrinsic lifetime of the Dirac fermion quasiparticles. It was shown theoretically [158], that graphene should display marginal Fermi liquid characteristics leading to a linear energy dependence of the inverse quasiparticle lifetime arising from electron-electron interactions, as opposed to the quadratic dependence in Fermi liquids. Theoretical estimates of the life time in zero field give $\gamma \sim 20$fs/eV. Since the electron-electron interactions are enhanced in magnetic field, it is possible that the agreement would be even better if calculations were made in finite field. The energy independent term in Eqn. 10 corresponds to an extrinsic scattering mechanism with characteristic mean free path of $l_{mfp} = v_F \tau_0 \sim 400nm$. This is comparable to the sample size indicating that the extrinsic scattering is primarily due to the boundaries and that inside the sample the motion is essentially ballistic. Note that had this been a diffusive sample, with the same carrier density ( $n_s = 3 \times 10^{10}$ cm$^{-2}$ ) and mean free path, its transport mobility would be: $\mu = ev_F l_{mfp}/E_F = 220,000 \ cm^2/V \cdot \sec$.

**Line-shape and Landau level spectrum**

Several factors contribute to produce the envelope of the LL spectra Fig. B-2c: the finite lifetime of the quasiparticles which is inversely proportional to the linewidth; the uneven spacing of the LL and the energy dependence of the linewidths. Figure B-7 illustrates how the V shaped lower envelope of the spectrum builds up as the individual LLs get broader when each level has the same width and the same degeneracy (or peak area). Since the peaks are unevenly spaced ( Eqn. 9), the overlap between peaks increases at higher energies, hence the increasing background in the overall DOS with increasing energy away from the CNP (which here coincides with the Fermi energy). Comparing to the spectrum in Figure B-4c we note that that the N=1 peak is higher than the N=2 peak which is not the case in the simulated spectra. In order to simulate the down-turn of the upper envelope away from zero energy seen in the high resolution spectra of Figure B-6a one has to require the peak width to increase with energy, as shown in Figure B-6d.

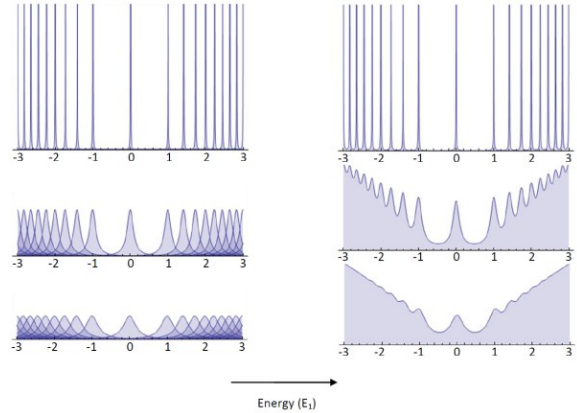

**Figure B-7.** The origin of the V-shaped background in the DOS. Left panels: illustration of the levels and their increased overlap as the linewidth is increased from the top to the bottom panel. The area under each peak is kept the constant. Right panels: overall density of states. The un-evenly spaced peaks overlap to produce the V-shaped background. Energy unit: $E_1 = \sqrt{2e\hbar v_F^2 B}$.

**Electron-phonon interaction and velocity renormalization**

The single-electron physics of the carriers in graphene is captured in a tight-binding model [11]. However, many-body effects are often not negligible. *Ab initio* density functional calculations [159] show that electron-phonon (e-ph) interactions introduce additional features in the electron self-energy, leading to a renormalized velocity at the Fermi energy $v_F = v_{F0}(1 + \lambda)^{-1}$, where $v_{F0}$ is the bare velocity and $\lambda$ is the e-ph coupling constant. Away from the Fermi energy, two dips are predicted in the velocity renormalization factor, $(v_F - v_{F0})/v_F$, at energies $E_F \pm \hbar\omega_{ph}$, where $\omega_{ph}$ is the characteristic phonon energy. Such dips give rise to shoulders in the zero field DOS at the energy of the relevant phonons, and can provide a clear signature of the e-ph interactions in STS measurements. The tunneling spectra measured on a decoupled graphene flake on graphite exhibit two shoulders that flank the Fermi energy are seen around ±150meV (Figure B-8) which are independent of tip-sample distance for tunneling junction resistances in the range 3.8-50GΩ.

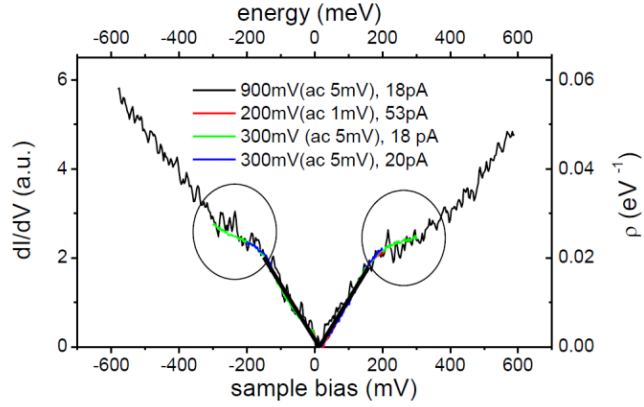

**Figure B-8. Zero field tunneling spectra at 4.4K.** Thick line is the DOS calculated according to Eq.(4). Thin lines are tunneling spectra taken with different tunneling junction settings. Circles highlight the shoulder features signaling deviations from the linear density of states.

Further analysis of these features requires a calibration of the zero field DOS. This is done by using the information obtained from LL spectroscopy to ascertain massless Dirac fermion nature of the excitations in this area, and to obtain the average value of $v_F$ for energies up to 150meV. The next step is to compare the expected DOS per unit cell with the measured spectrum in order to calibrate the arbitrary units of dI/dV. Since dI/dV is proportional to the DOS, $\rho(E)$, the linear spectra at low energies in Figure B-8 together with Eqn. 11 give:

11. $\rho_c(E) = \frac{3^{3/2}a^2}{\pi}\frac{|E-E_D|}{\hbar^2 v_F^2} = 0.123|E - E_D|$

For an isotropic band (a good approximation for the relevant energies $E <$ 150meV), the dispersion is related to the DOS by

12. $k(E) = \pm|\frac{\pi}{3^{3/2}a^2}\int_{E_D}^{E}\rho(\epsilon)d\epsilon|^{1/2}$

The result, obtained by integrating the spectrum in Figure B-8, is shown in Figure B-9a. Now the shoulders in Figure B-8 appear as kinks in the dispersion. The energy dependent velocity obtained from the dispersion:

13. $v_F = \frac{dE}{\hbar dk}$

plotted in Figure B-9b resembles that obtained by density functional theory: it exhibits two dips at the energy of the optical breathing phonon $A_1'$, $\sim \pm 150meV$, suggesting that this phonon, which couples the K and K' valleys and undergoes a Kohn anomaly, is an important player in the velocity renormalization. Incidentally, this same phonon is involved in producing the D and 2D peaks in the Raman spectra of graphene.

The $A_1'$ phonon has very large line width for single layer graphene, indicating strong e-ph coupling. However, the line width decreases significantly for bilayer graphene and decreases even more for graphite [160, 161]. Therefore e-ph coupling through the $A_1'$ phonon is suppressed by interlayer coupling and the e-ph induced velocity renormalization is only observed in single layer graphene decoupled from the substrate. Consequently and paradoxically the Fermi velocity in multilayer graphene will be closer to the bare value, as discussed in the next section.

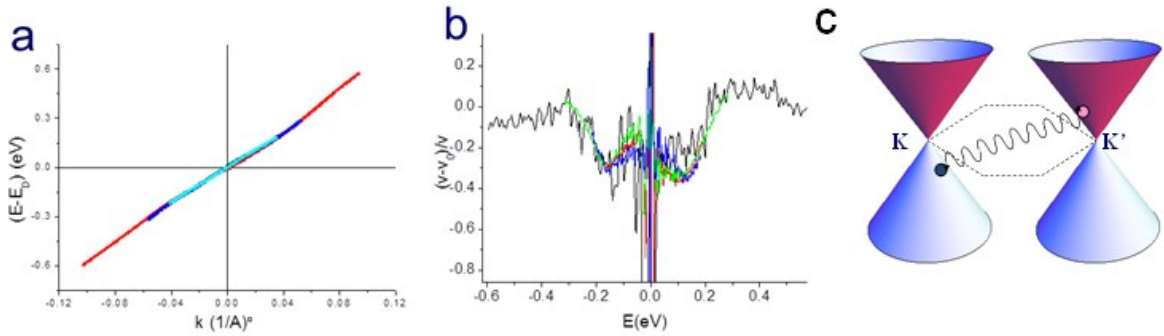

**Figure B-9.** a) Energy-momentum dispersion of graphene obtained from the data in Fig. B-8 as described in the text. b) Energy dependent Fermi velocity obtained by differentiating the dispersion in *a* . c) Schematic diagram of inter-valley scattering mediated by the $A_1'$ phonon.

## Multi-layers - from weak to strong coupling

Unlike in conventional layered materials, interlayer coupling in graphene is relatively easy to tune. For example, in Figure B-3a, the from region G to region W turns on the interlayer coupling which breaks the sublattice symmetry. Therefore the atomic resolution STM topography appears different in the two decrease in graphene-substrate spacing when crossing regions: triangular in W (Figure B-3e) and honeycomb in G (Figure B-3d). The effect of coupling on the electronic structure is illustrated by comparing STS of the two regions in

Figure B-10.

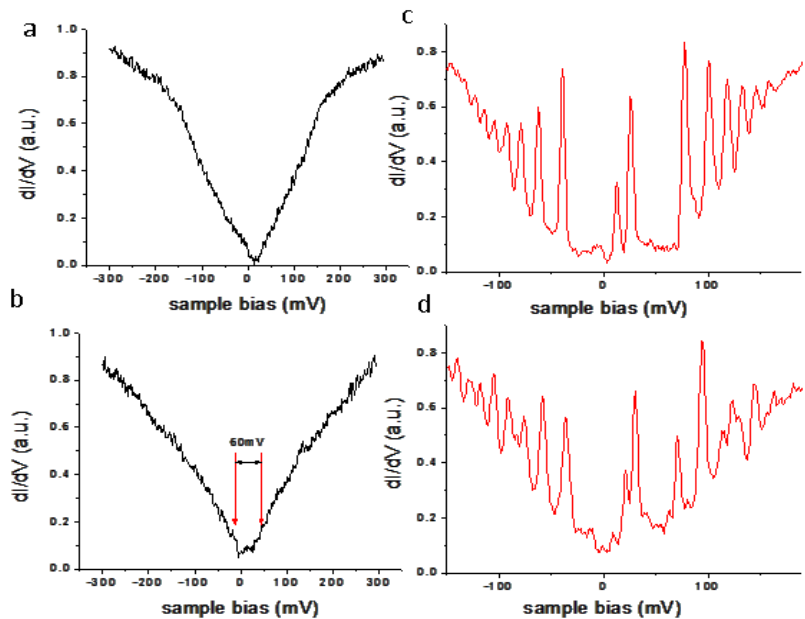

**Figure B-10.** Effect of interlayer coupling on STS spectra corresponding to the graphene flake shown in Figure B-3. a,b) Zero field STS in region G and W. c) LL spectrum at 4T in region G. d) LL spectrum at 4T in region W.

In zero field we note that the DOS vanishes at the CNP in region G, but remains finite in region W, as seen in Figure B-10 (a,b). The difference in LL spectroscopy is even more pronounced: the simple LL sequence in region G, Figure B-10c, evolves into the more complicated spectrum in region W, Figure B-10d.

Stacking faults and other defects in HOPG cause decoupling of the layers. Therefore, one often observes strong LL spectra in some regions of the surface of HOPG after cleavage, but usually more than one sequence is observed indicating coupling to the substrate [66]

For an AB stacked bilayer the interlayer coupling, $t_\perp$, the two-band dispersion of the single layer evolves into four bands [162]:

14. $$E(k) = \pm \frac{1}{2}|t_\perp \pm \sqrt{t_\perp^2 + 4(\hbar v_F k)^2}|$$

We note that the single layer linear dispersion is recovered in the limit of zero coupling. For finite interlayer coupling there are still two bands touching (Figure B-11) at the CNP, but because the bands are no longer linear the DOS does not vanish at the CNP. The other two bands are separated by an energy gap $2t_\perp$, leading to DOS jumps at $\pm t_\perp$. Such jumps are difficult to resolve in the STS. A more accurate and direct measure of the coupling between the layers can be obtained from LL spectroscopy.

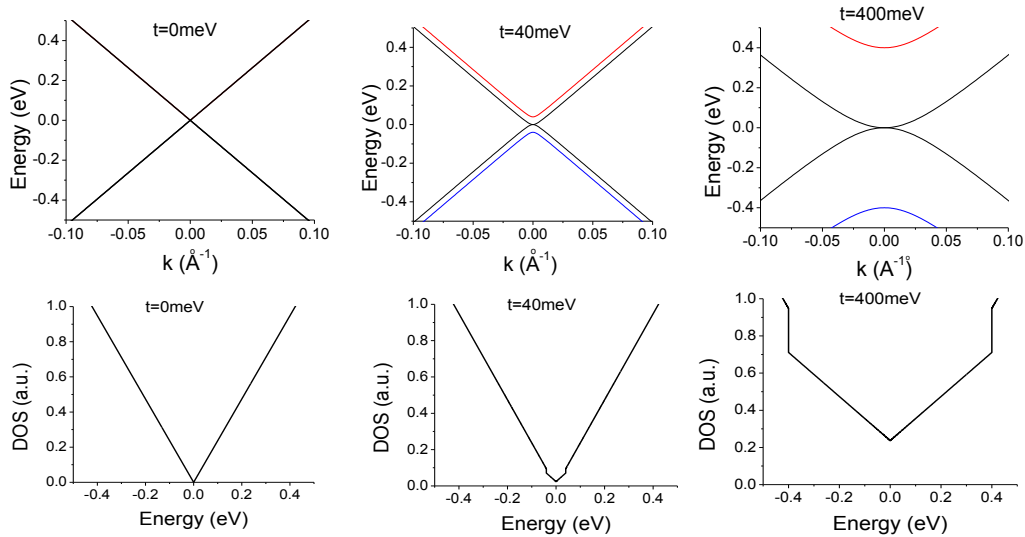

**Figure B-11. Simulated dispersion (top row) and density of states (bottom row) for graphene bilayer for indicated values of interlayer coupling strength *t*.**

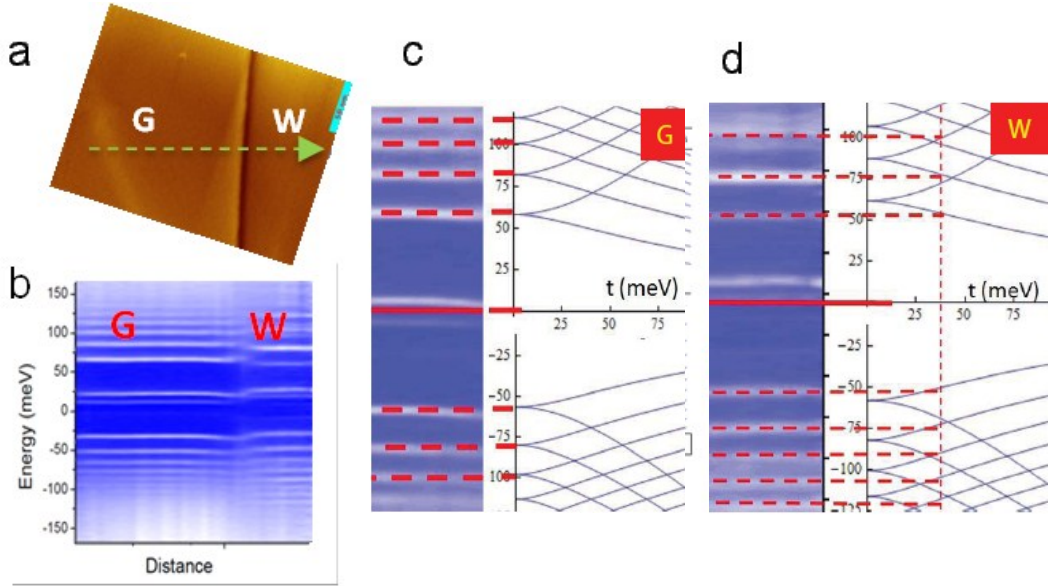

**Figure B-12.** Effect of interlayer coupling on LLs for graphene bilayer. a) Topography of flake showing the boundary between decoupled region G and weakly coupled region W. LL spectra at 4T as a function of position were recorded along the line marked "d". b) Evolution of LL spectra(4T) along the trajectory marked in panel *a* shows a qualitative change occurring across the ridge between the two regions marked by the dashed line. Intensity represents the amplitude of dI/dV. Typical tunneling spectra are shown in Figs. 9 c and d. c) Comparison of spectra in region G with calculated sequence using eqn. 15 as a function of interlayer coupling $t_\perp$ for B=4T. The sequence matches the positions of LL corresponding to zero coupling. d) Same as *c* in region W. The sequence matches the positions of LL corresponding to finite interlayer coupling of 45 meV.

In the presence of magnetic field, interlayer coupling modifies the simple sequence of massless Dirac fermions of Eq. (14) into [162]:

15. $E_N = E_D \pm \sqrt{e\hbar v_F^2 B} \frac{t'}{\sqrt{2}} \left[ 1 + \frac{2}{t'^2}(2N+1) \pm \sqrt{\left[1 + \frac{2}{t'^2}(2N+1)\right]^2 - \frac{16}{t'^4} N(N+1)} \right]^{1/2}$

where $t' = t_\perp \sqrt{\frac{eB}{\hbar v_F}}$. Once interlayer coupling is turned on, the single layer sequence splits into two, one bending toward $E_D$ and the other away from it. LL crossings occur with increasing coupling, which leads to new peaks as seen in Figure B-10d. The evolution of LLs from region G to region W is shown in Figure B-12. Comparing the LL spectra in region W to the theoretical model for a bilayer with finite interlayer coupling we obtain, as shown in Figure B-12d, an estimate of $t_\perp \sim 45$meV in this region [163] which is about one order of magnitude below the equilibrium coupling value. Although the simple model discussed above captures the main features of Figure B-12, some subtle details, e.g. electron-hole asymmetry, have not been addressed.

In the limit of equilibrium interlayer coupling, $t_\perp = 400$meV (the standard bilayer case) the spectrum consists of massive quasiparticles. These are qualitatively different from those in conventional two dimensional electron systems and are described as chiral massive fermions carrying a Berry phase of $2\pi$ [164]. The LL sequence in the bilayer is $E_n = \pm \hbar \omega_c \sqrt{N(N-1)}$ where $\omega_c = \frac{eB}{m*}$ the cyclotron frequency, m* is the effective band mass. The energy levels in this sequence are linear in field and the N=0 LL has double the degeneracy of the other LLs.

For trilayer graphene with Bernal stacking, massless Dirac fermions and massive chiral fermions coexist [66, 165]. As the number of layers increases, the band structure becomes more complex. However, for ten layers or less, the massless Dirac fermions always show up in odd number of layers [166]. Furthermore, changing the stacking sequences away from the Bernal stacking can strongly modify the band structure [167, 168]. The massive sequence can vary from sample to sample as it is controlled by interlayer coupling [65]. However, the massless sequence is quite robust, showing very weak sample dependence. For graphene multilayers, i.e. when sequences of LLs coexist, the massless sequence gives a Fermi velocity of $1.07 \times 10^6$ m/s, which is close to the un-renormalized value. This supports the theoretical expectation that e-ph coupling through $A_1'$ is suppressed by interlayer coupling as discussed in the previous section.

## 4. Twisted graphene layers

Graphite consists of stacked layers of graphene whose lattice structure contains two interpenetrating triangular sublattices, A and B. In the most common (Bernal) stacking, adjacent layers are arranged so that B atoms of layer 2 (B2) sit directly on top of A atoms of layer 1 (A1) and B1 and A2 atoms are in the center of the hexagons of the opposing layer. If two graphene layers are rotated relative to each other by an angle $\theta$ away from Bernal stacking, a commensurate superstructure, also known as Moiré pattern, is produced. The condition leading to Moiré patterns can be obtained from elementary geometry[169] $\cos(\theta_i) = (3i^2 + 3i + 1/2)/(3i^2 + 3i + 1)$, with $i$ an integer ($i$=0, $\theta$ = 60° corresponds AA stacking and $i \rightarrow \infty$, $\theta$ = 0° to AB stacking) and lattice constant of the superlattice $L = a\sqrt{3i^2 + 3i + 1}$ where $a_0 \sim 2.46$Å is the atomic lattice constant. In a continuum approximation, the period $L$

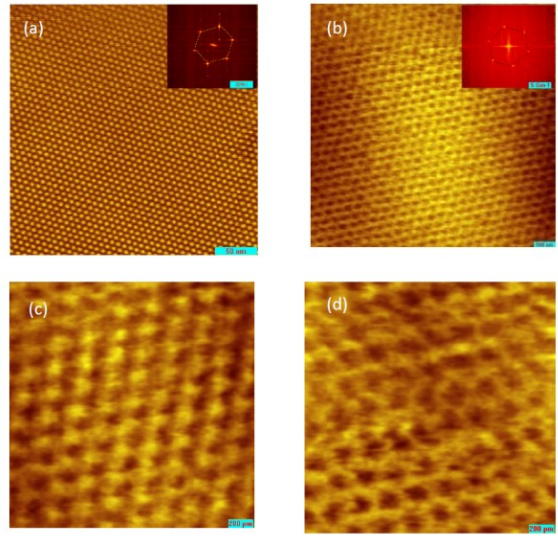

Figure B-13. Moiré pattern corresponding to a twist angle of 1.79° obtained by STM topography on a graphite surface. a) Large area image showing the super-lattice. Scale bar: 50nm. b) High resolution image showing the atomic lattice. Scale bar: 500pm. c) Zoom into a bright spot in panel *a*. Scale bar: 200pm. d) Zoom into a dark spot in panel *a*. Scale bar: 200pm. Insets: Fourier transforms of the main images.

corresponding to a twist angle $\theta$ is given by:

$$16. \quad L = \frac{a}{2 \sin\left(\frac{\theta}{2}\right)}$$

An alternative way to understand Eq.(16) is to note that when two graphene layers rotate against each other, the two hexagonal Brilouin zones also rotate (Figure B-15a) around the Γ point. As a result the K points of the two lattices separate by a displacement ΔK:

$$17. \quad \Delta K = \frac{4\pi}{3\sqrt{3}a} \sin\left(\frac{\theta}{2}\right)$$

These displacement vectors form a new hexagon, which corresponds to the Fourier transform of the Moiré pattern. Eq.(16) can be derived from Eq.(17) by using ΔK= 2π/L. The new hexagon is rotated by 30°-θ/2 relative to the original one for small angles, as seen experimentally in Figure B-13.

The freedom of stacking between graphene layers is so large that twisting away from the equilibrium Bernal stacking is possible for a wide range of rotation angles resulting in a variety of Moiré patterns. These patterns were observed very soon after STM became widely available and made it possible to explore the topography of graphite surfaces [141, 170]. An example of a Moiré pattern on the surface HOPG is shown in Figure B-13. The highly ordered triangular pattern has a period of ~ 7.7nm, much larger than the lattice constant of graphene. A better understanding of the pattern is gained by zooming into the bright and dark spots with atomic resolution, Figure B-13c,d. For the bright spots of the pattern the underlying lattice structure is triangular, indicating Bernal stacking. In between the bright spots a less ordered honeycomb-like structure is seen indicating lost registry between layers due to twisting. The connection between twist angle and the Moiré pattern period is directly revealed by comparing the pattern and its Fourier transform shown in Figure B-13. The evolution of the pattern with twist-angle (Figure B-14) illustrates the decrease in period with increasing twist angle.

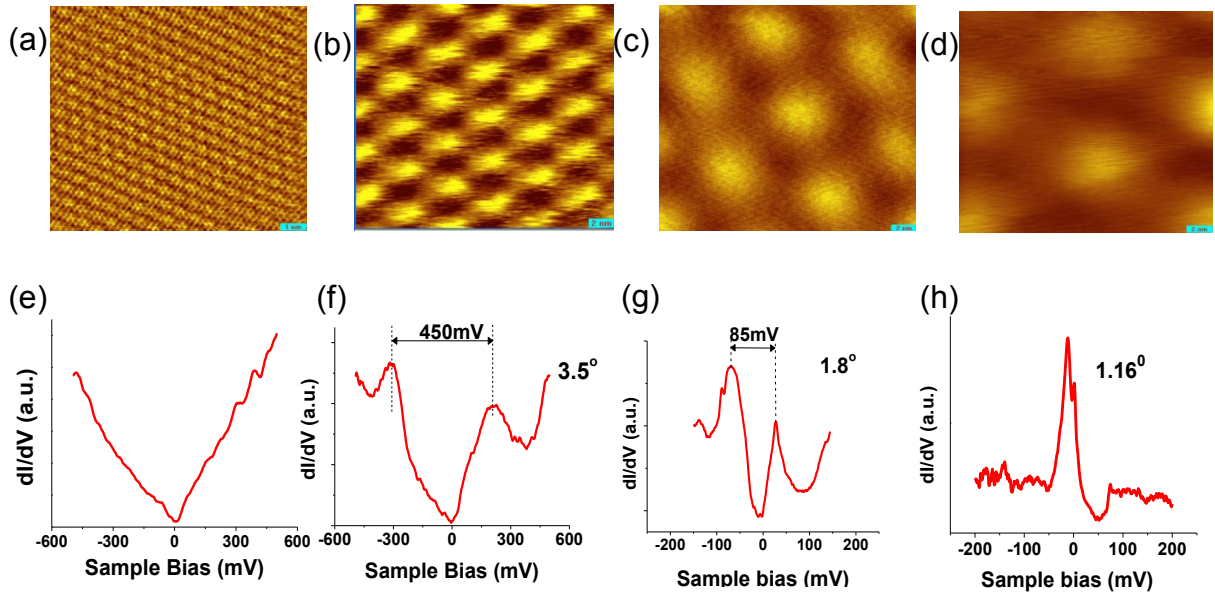

**Figure B-14. Twist angle dependence of moire patterns and van Hove singularities. Top row: Moire pattern for decreasing twist angles, a) 20.8 b) 3.48⁰, c) 1.78⁰ and d) 1.16⁰ . Scale bar: 1nm for panel *a* and 2nm for the rest. Bottom row: density of states showing van Hove singularities for indicated twist angles. e) Differential conductance for both large twist angles and untwisted regions shows no Van-Hove singularities.. (f-h) As the twist-angle decreases from 3.5⁰ to 1.16⁰ the period of the moiré pattern increases and the separation between van Hove singularities decreases.**

While twist-induced Moiré patterns have been known and understood for many years, the surprising discovery that the twist between layers also has a profound effect on the electronic band structure came only recently [171]. This realization has led to a flurry of research into the connections between interlayer twist and electronic properties [172-187]. Compared to the non-twisted case where the DOS increases monotonically with distance from the CNP, Figure B-14e, Li *et al.* [171] found that twisting away from Bernal stacking produces two pronounced peaks in the DOS which flank the CNP on both sides, Figure B-14f,g, and that their separation increases with the angle of rotation. To understand the origin of the peaks in the DOS we consider two adjacent Dirac cones belonging to the different layers in Figure B-15a. It is immediately obvious that the cones must intersect at two points at energies $\pm \hbar v_F \Delta K$ in the hole and electron sectors. At these points - and not at the DP as is the case in Bernal stacked layers - the two layers can couple to each other with coupling strength of order $t_\perp^\theta \approx 0.4 t_\perp$ [169]. Here, $t_\perp$ is the interlayer hopping for unrotated layers. At the intersections of the two Dirac cones their bands will hybridize (if the coupling between layers is finite), Figure B-15b, resulting in saddle points in the dispersion. These give rise to two Van Hove singularities which symmetrically flank the CNP and are seen as peaks in the DOS [171, 188]. It is important to realize that in the absence of interlayer coupling the Van Hove singularities will not appear. The separation between Van Hove singularities is controlled by the twist angle, θ. For angles 2°<θ<5°, the separation is

$$18. \quad \Delta E_{VHS} \approx \hbar v_f \Delta K - 2 t_\perp^\theta$$

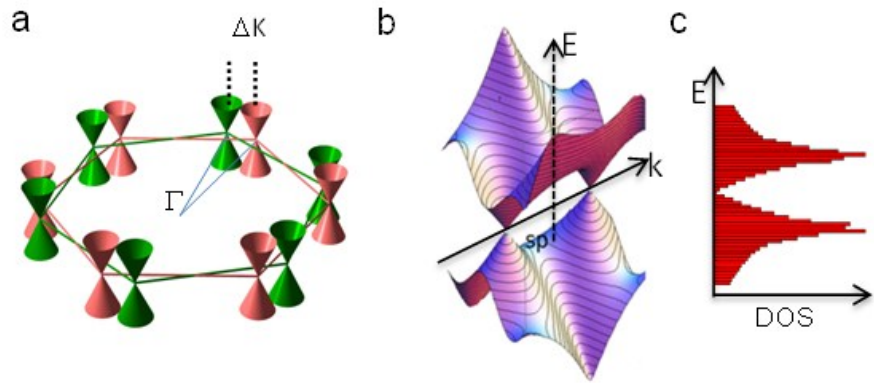

**Figure B-15. Twist angle dependence of band structure and density of states for a twisted graphene bilayer.  a) The Brillouin zones of the two layers (green and red) are rotated with respect to each other by the same angle as their relative rotation in space.  b) Saddle points in the band structure, marked sp, occur at positive and negative energies corresponding to the intersection of the Dirac cones calculated for θ=1.79°, $t_\perp \sim 0.27$eV. c)  The density of states exhibits Van Hove singularities at the saddle points.**

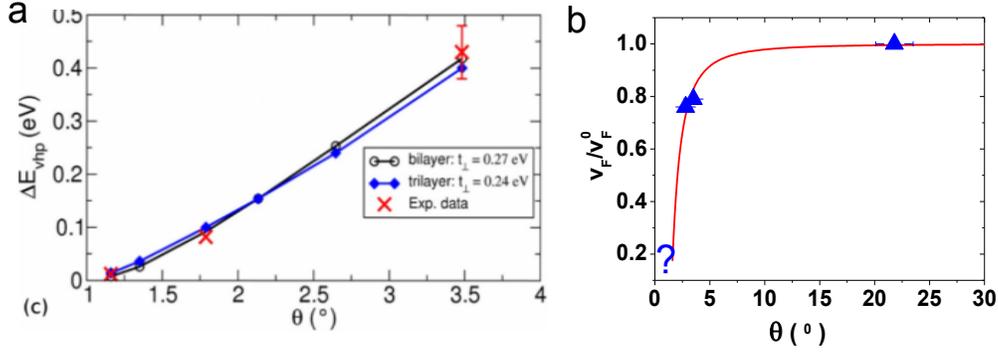

**Figure B-16.** a)Twist angle dependence of energy separation between Van Hove singularities for experimental data is compared with theory. b) Twist angle dependence of the Fermi velocity. Comparison between theory (solid line) and experimental data (symbols). The question mark at small angles corresponds to a band structure where merging of the Van Hove singularities precludes analysis based on a LL sequence.

A comparison between the measured peak separation and the theoretical calculation is shown in Figure B-16a. As the Van Hove singularities separate from each other with increasing twist angle, the low energy sector of the Dirac cones in each layer are less disturbed. Therefore, for sufficiently low energies, the electrons in twisted layers can behave like massless Dirac fermions in a single layer [169, 172-177]. However, the slope of the Dirac cone, i.e. Fermi velocity, still reflects the influence of the Van Hove singularities, leading to a renormalized Fermi velocity which depends on twist angle [169]:

$$19. \quad \frac{v_F(\theta)}{v_F^0} = 1 - 9\left(\frac{t_\perp^\theta}{\hbar v_F^0 \Delta K}\right)^2$$

The velocity renormalization can be observed experimentally by using LL spectroscopy on twisted layers in a magnetic field [188]. In Figure B-17 we illustrate these results in two adjacent regions, one of which, $M_1$, is twisted. In region $M_1$, a Moiré pattern with period 4.0nm is resolved, while in region $M_2$, the pattern is not resolved indicating an unrotated layer (or a much smaller period). In zero field, STS reveals Van Hove singularities in region $M_1$ but not in region $M_2$ even for bias voltages up to ±500meV (Figure B-17b,c). In both regions, STS in magnetic field (Figure B-17f,g) shows LLs of massless Dirac fermions with Fermi velocities of $0.87\times10^6$m/s and $1.10\times10^6$m/s for regions $M_1$ and $M_2$, respectively.

The velocity renormalization is significant only for twist angles smaller than ~10° in agreement with theory (Figure B-16b). At large angles, the Dirac cones for different graphene layers are well separated so that the low energy electronic properties and the Fermi velocity are indistinguishable from those in a single layer [120]. At very small angles less than ~2°, denoted as a question mark in Figure B-16b, the van Hove singularities become so dominant that the description of the low energy excitations in terms massless Dirac fermions no longer applies. For example at $\theta$ ~1.79° individual contributions to that spectrum from LLs and from van Hove singularities can no longer be identified, **Error! Reference source not found.**. Eventually the an the Hove singularities themselves show non-trivial field dependence [120]. Moreover, a strong spatial modulation is observed in the DOS maps at small angles, indicating the formation of a charge density wave [171, 178].

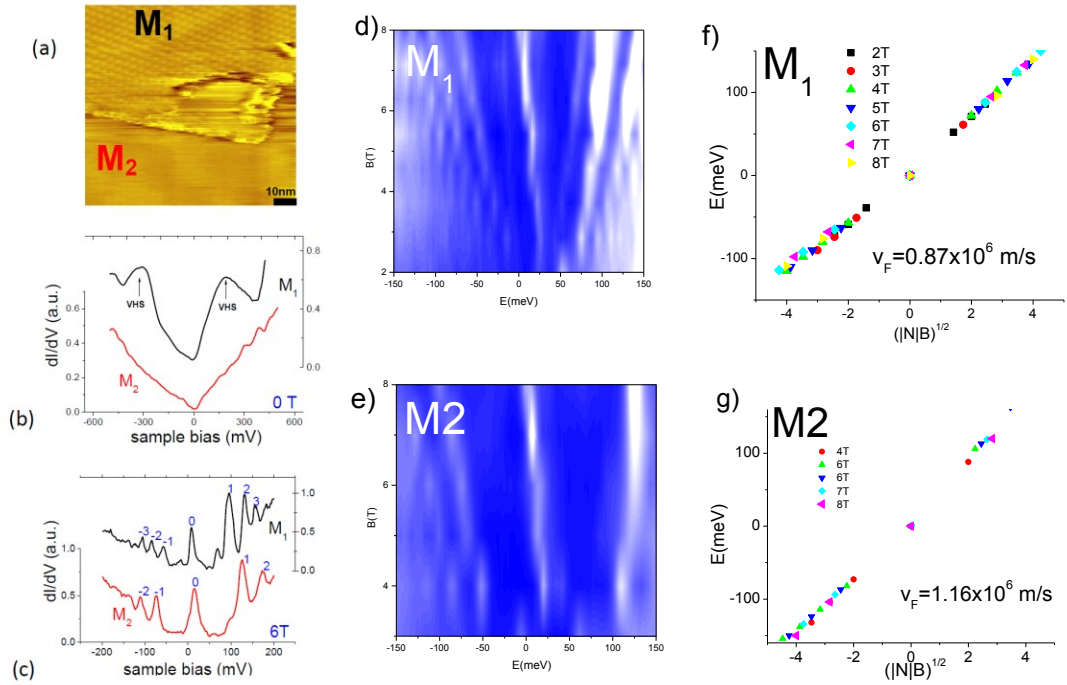

**Figure B-17. Velocity renormalization in twisted graphene. a)** STM images show region $M_1$ with a twist angle of ~3.48° and region $M_2$ with no twist. **b)** Zero field tunneling spectra show van Hove singularities, marked as "VHS", in region $M_1$. **c)** Tunneling spectra in a field of 6 T show indexed Landau levels. **d, e)** LL maps shows evolution with magnetic field in the two regions. The apparent discontinuities are the result of using discrete field points to generate the maps. **f,g)** LL peak positions plotted against reduced field show collapse of the data. Fit to Eqn. 9 gives the Fermi velocity $v_F=0.87\times10^6$ m/s in $M_1$ and $v_F=1.16\times10^6$ m/s in $M_2$.

It is important to note that the mechanism of downward velocity renormalization in twisted layers is distinctly different from that in isolated graphene layers discussed in previous sections. In the twisted layers the renormalization only occurs in the presence of coupling between layers and its magnitude is a sensitive function of the twist-angle. By contrast the velocity renormalization observed in the decoupled graphene layer supported on graphite (Figure B-5b) is due to e-ph interactions. If the 20% renormalization of the Fermi velocity seen in these data was due to coupling between twisted layers, the twist angle would have to be~$3.2^0$ according to Eqn. 19. Such a twist would result in hard-to-miss features: a Moiré pattern with a period of 2.54 nm (18 lattice spacing*)* in STM topography and two Van Hove singularity peaks in the STS ~400meV apart. The absence of these features rules out twist-induced decoupling in the partially suspended graphene layer shown in Figure B-3. In the previous section we have shown that the e-ph coupling via the $A_1^{'}$ phonon is strongest in decoupled single layer and that it becomes less important as the coupling between layers increases. As we show in Figure B-16b the twist-induced renormalization becomes negligible for angles exceeding $10^0$. For example the Fermi velocity corresponding to the $20.8^0$ twist-angle, $v_F=1.12\times10^6$ m/s is almost identical to that in multi-layers with Bernal stacking, suggesting that e-ph coupling via $A_1^{'}$ is also suppressed in twisted layers

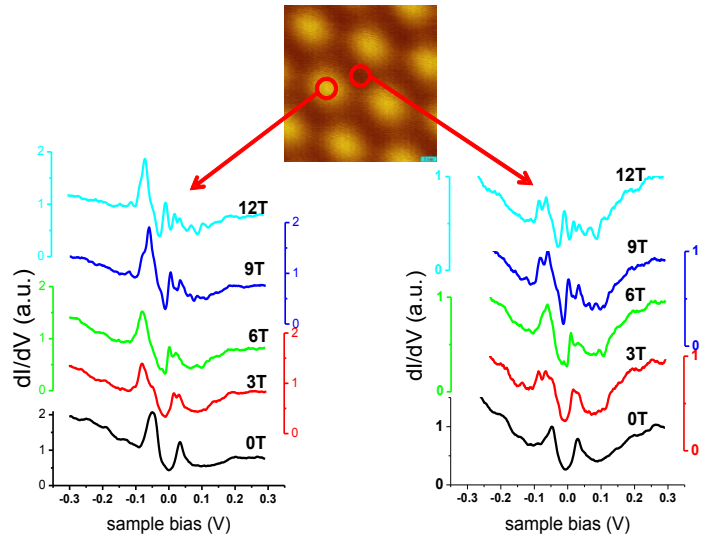

**Figure B-18** Field dependence of STS for a twist angle of 1.79°. LLs ride on the van Hove singularities. The STS show strong spatial dependence across the Moiré pattern. The positions where the spectra were taken ( indicated by arrows) correspond to a bright spot (left panel) and dark spot (right panel). The vertical scales in the right panels is magnified compared to the left panels to compensate for the lower signal intensity in the dark spots.

### 5.   *Graphene on chlorinated SiO$_2$*

The existence of electron-hole puddles strongly modifies the LLs in graphene [119, 120, 189] preventing their observation with STS [190]. One way to overcome the substrate limitation without sacrificing the ability to gate is to use suspended samples. As discussed in part C, transport measurements on suspended samples have shown that in the absence of the substrate the intrinsic DP physics including interaction effects is revealed [21, 27]. However, due to their fragility, small size and reduced range of gating the use of suspended samples is limited. Finding a minimally invasive insulating substrate on which graphene can be gated and also visualized is therefore of great interest.

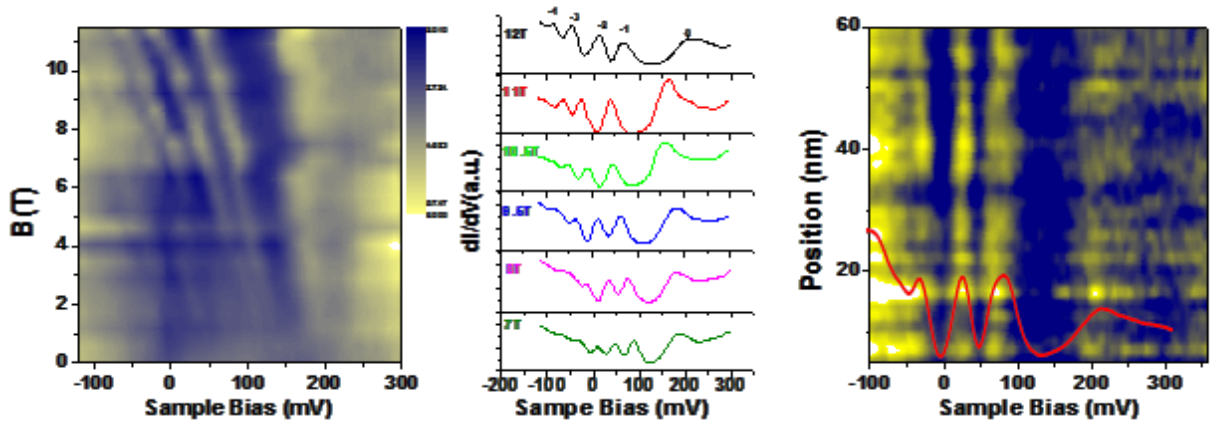

**Figure B-19.** Low temperature (4.4K) STS of graphene on a chlorinated SiO$_2$. a)  DOS map shows the evolution of LL peaks which fan away from the Dirac point and become better resolved with increasing field. b) LL spectra show well defined peaks above 7T.  Adapted from A. Luican *et al*. Phys. Rev. B, 83, 041405 (R),  (2011). Evolution of LL across the sample at 12T show well separated strips, corresponding to LL peaks (bright regions) separated by gaps (dark regions). The STS trace in red illustrates the correspondence between  LL peaks and bright regions in the map.  The spatial uniformity of the spectra indicaties that it is possible to place the Fermi energy within a gap between LLs.

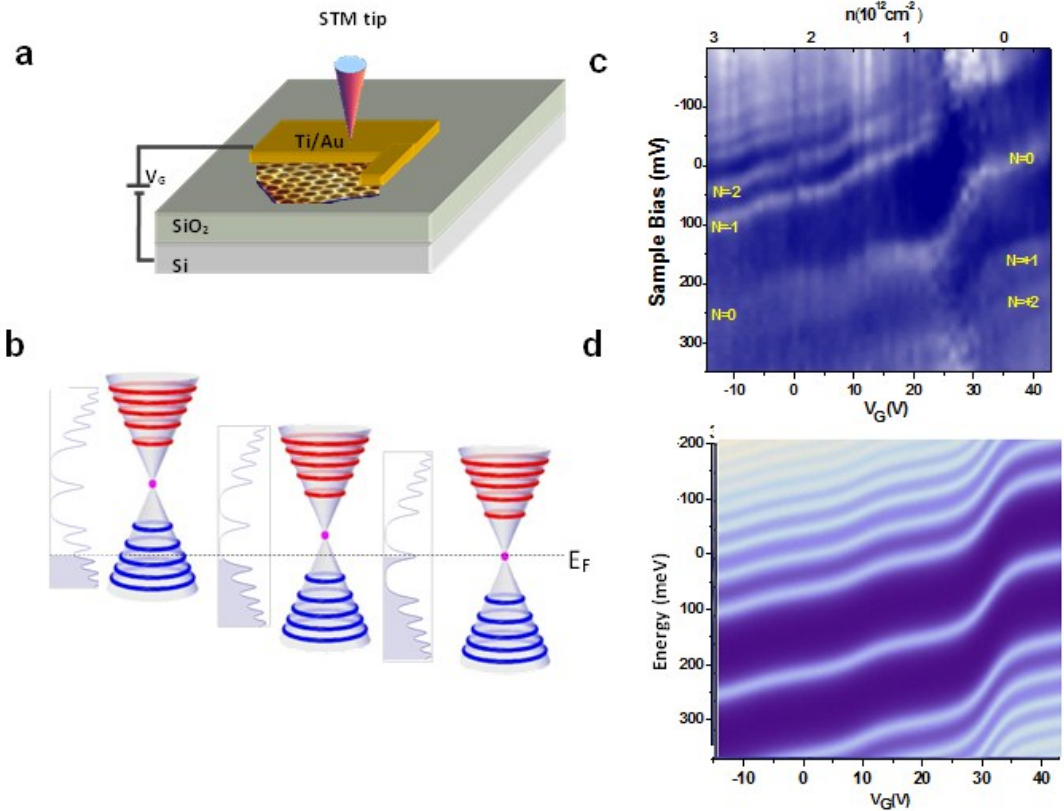

Figure B-20. Gate dependence of tunneling spectra in a magnetic field. a) Schematic diagram of measurement setup allowing simultaneous STS measurements and gating of the graphene sample on an insulating substrate with a metallic back-gate. b) Illustration of the effect of gating on the Fermi level position relative to the Dirac Point. c) Gate dependence of tunneling spectra in a magnetic field of 12 T. Bright stripes correspond to Landau levels, some of which are labeled with their indices. d) Simulation of evolution of Landau levels with gate voltages.

In the semiconductor industry it is well known that the quality of $SiO_2$ can be significantly improved by using dry oxidation in the presence of chlorine. The dry process reduces dangling bonds and the chlorine removes metal ions from the oxide which can greatly improve the uniformity and quality of the insulator [191-193]. Indeed STM and STS measurements demonstrated that for graphene supported on $SiO_2$ substrates which were treated by chlorination to minimize trapped charges and in sufficiently large magnetic fields, the LL sequence specific to single layer graphene and its dependence on carrier density can be accessed [120]. In zero field STM and STS on graphene deposited on these substrates shows a honeycomb structure and the STS displays two minima, one at the Fermi energy and the other around the CNP as shown in Figure B-1. In the presence of a sufficiently high magnetic field the spectra start resembling LLs (Figure B-19). At low fields the spectra are distorted due to the substrate-induced random potential and are strongly position dependent. Above ~ 5T a clean, although broadened, LL sequence of well separated peaks is seen [120] across the entire sample indicating that it is possible to place the Fermi energy in a gap between LLs, which is the condition for observing the QHE. The LL sequence follows that of massless Dirac fermions, Eq.(9), with a Fermi velocity of $1.07 \times 10^6$ m/s. For a direct connection between the onset of the LL sequence and the random potential we consider the dI/dV map near the CNP in Figure B-1b which reveals puddles of ~20nm in size. This imposes a length scale and a corresponding energy scale which separates between disorder-controlled and intrinsic phenomena. In finite field the intrinsic physics of the charge carriers will thus become apparent only when the magnetic length $l_B = 25.64nm/\sqrt{B}$ is smaller than the characteristic puddle size $d$. This defines a characteristic field, $B_c \approx 4\hbar/ed^2$, below which the LL spectrum is smeared out by disorder. Consistent with this picture we find no distinct LL features in the tunneling spectrum for fields below $B_c \sim 6$ T (Figure B-19), corresponding to $l_c(6T) = 10.5nm \approx d/2$.

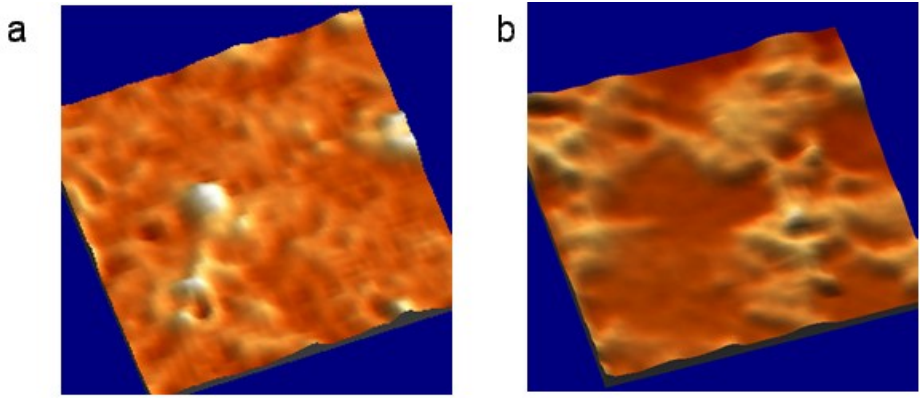

**Figure B-21. DOS maps of localized and extended states in the LL regime. a) When the Fermi energy is placed in the gap between the N=0 and N=1 LLs the electronic wavefunction is localized on impurity states (bright spots in the figure). b) When the Fermi energy is within the N=1 LL the wavefunction is extended. Image credits: A. Luican-Mayer.**

To study the effect of gating on the LLs the STS spectra are measured at a fixed magnetic field while varying the gate voltage. The data, presented as a map in Figure B-20c, shows that varying the carrier-density through gating is accompanied by pinning of Fermi energy to each LL as it is filled and followed by a jump to the next LL once a level is full. Qualitatively, one can understand the step-like features as follows: the LL spectrum consists of peaks where the DOS is large, separated by minima with low DOS. It takes a large change in the charge carrier density to fill the higher DOS region, resulting in plateaus where the Fermi level is pinned to a particular LL. The jumps in between the LLs result from the fact that filling the low density region in between the LLs does not require a large change in the carrier density [120]. To analyze the data the LL peaks are modeled as Lorenzians with width ≈30meV from which the chemical potential is numerically calculated as a function of carrier density [194]. The result of this calculation, shown in Figure B-20d, is consistent with the data. We note that as N=0 is brought closer to the Fermi level it becomes sharper and better defined. Moreover, the broader the levels, the less abrupt the jumps, indicating that in disordered samples the Fermi energy pinning effect is smeared out. A similar effect was previously observed in very high mobility GaAs samples using time domain capacitance spectroscopy [194]. Unlike the case in semiconductors where the carriers are non-relativistic, in graphene due to the fact that the LL are not equally spaced, the largest jump from N = -1 to N = 0 is followed by successively smaller jumps for higher N.

The effect of the LLs on the electronic wavefunction can be seen directly in the DOS maps shown in Figure B-21. When the Fermi energy lies in a gap between LL the wavefunction is localized on impurity states and it becomes extended when the Fermi energy is inside a LL

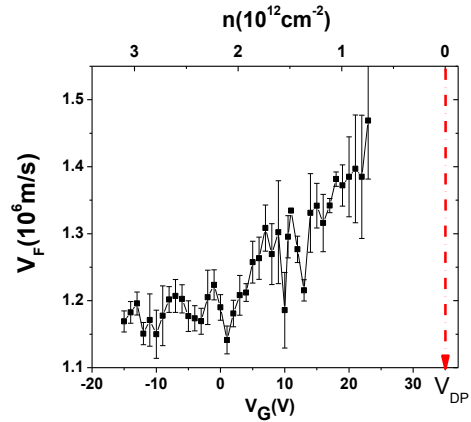

**Figure B-22.** Evolution of measured Fermi velocity with carriers density shows the velocity enhancement upon approaching the Dirac point. The Fermi velocity for each doping is obtained from LL spectroscopy at 12 T. The position of the Dirac point at 35V is indicated by the arrow (A.Luican,

The ability to measure the LL spectra while varying the carrier density makes it now possible to measure the density dependence of the Fermi velocity and to gain direct access to the effects of electron-electron interactions and screening upon approaching the DP. The Fermi velocity is obtained by plotting the LL peak energies against $\sqrt{|N|B}$ and fitting to Eqn. 9 for each carrier density. This procedure allowed to extract the density dependence of the Fermi velocity presented in Figure B-22. Significantly, the Fermi velocity increases towards lower doping level consistent with the logarithmic divergence expected due to the reduced screening ability of the two dimensional Dirac fermion carriers near the DP [195-197] [198, 199]

Unlike transport measurements that probe electrons close to the Fermi energy, STS gives access electronic states both above and below the Fermi energy and therefore provides a comprehensive view of the electronic spectrum. Particularly, careful studies of the LLs reveal that the electronic structure is not rigid [120], i.e. it varies with carrier density. Interactions between electrons can change the shape of Dirac cones and the Fermi velocity increases as the carrier density approaches zero. A similar result was obtained by measuring the cyclotron mass in suspended graphene [200].

### Fermi energy anomaly and gap-like feature

Early STS of graphene on $SiO_2$ reported a gap-like feature within ±63meV of the Fermi level [119]. The suppression of tunneling near $E_F$ and the concomitant enhancement of tunneling at higher energies were attributed to a phonon-mediated inelastic channel. Thus it was argued that phonons act as a "floodgate" that controls the flow of the tunneling electrons in graphene. In this scenario, the electrons in the STM tip were assumed to have zero lateral momentum, i.e. at the Γ point in momentum space. Momentum conservation would prevent their direct tunneling into graphene because the Dirac cones are around the K or K' points. Therefore a third momentum carrying particle, a phonon, would have to be involved to mediate the tunneling event [201]. In other words, only inelastic tunneling processes are allowed. However, such an argument is not valid for an atomically sharp tip because the electron momentum is only restricted by the uncertainty principle, $k \sim \hbar/a_T$ ($a_T$ is the size of the tip). Therefore, if the tunneling source consists of only one or few atoms, the momentum of the tunneling electron can be sufficiently large to access the K points. Indeed this large gap-like feature is not seen in STS measurements of either graphene on graphite [65] or graphene on SiC substrates [150]. However, a dip feature at the Fermi always appears for graphene on $SiO_2$, or other insulating substrates as shown in Figure B-1d. The dip feature becomes weaker in the presence of better substrate screening, and in the limit of a conducting substrate such as graphite it is barely perceptible. These observations suggest that the dip in the tunneling spectra is a manifestation of the zero bias anomaly [175, 202] rather than inelastic tunneling. The zero bias anomaly, commonly seen in mesoscopic systems and disordered semiconductors, is caused by the combined effects of disorder and interactions and leads to suppressed tunneling probabilities at the Fermi energy, regardless of the level of doping.

### *6.     Graphene on other substrates*

Many other substrates have been used to support graphene including Mica, PMMA, and SiC which are beyond the scope of this review. Below we briefly mention STM results on SiC and on h-BN.

**Graphene on SiC**

As discussed in section A-2 Epitaxial graphene grown directly on SiC is atomically flat and the sheets are continuous over macroscopic distances. Within the first few layers substantial substrate interactions cause doping, electron-phonon coupling, and distortions in the linear dispersion which are particularly pronounced in the first layer. Beyond the first few layers, the Bernal stacking of the layers in Si-face graphene causes the band structure to converge to graphite. In C-face graphene the rotational stacking of the layers, away from Bernal stacking, leads to an effective decoupling of the layers which results in exceptionally high quality and properties that resemble single layer graphene. LL spectroscopy on these rotationally stacked C-face layers yields a sequence of very sharp peaks [150], comparable to those observed in graphene on graphite [65]. Similarly cyclotron resonance measurements in this system[156, 157] reveals very sharp resonance peaks indicating long quasiparticle life time.

**Graphene on h-BN**

The use of atomically flat substrates can significantly diminish the corrugation of graphene which also reduces the magnitude of local charging fluctuations. Efforts to reduce substrate-induced perturbations, have already resulted in improved sample quality [203-205]. In this respect single crystals of hexagonal BN (h-BN) are particularly promising. Transport measurements on graphene deposited on h-BN have revealed FQHE plateaus [206] and STM measurements show that graphene on h-BN is ultra-smooth [203]. This is illustrated in Figure B-1, where we compare STM topography maps for graphene deposited on $SiO_2$ and on a single crystal of h-BN. We note that the surface corrugations is reduced by more than one order of magnitude on the surface of h-BN compared to that on the $SiO_2$ substrate.

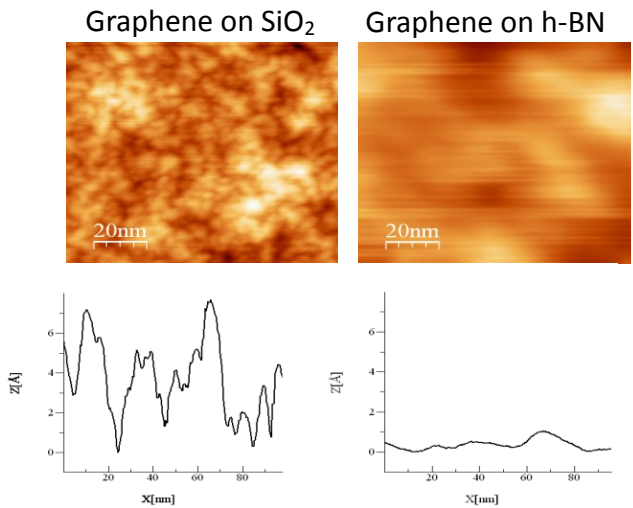

**Figure B-1.** STM topography of graphene on SiO$_2$ (left) and on h-BN (right). The bottom panels comparing the height variations for the two substrates show more than one order of magnitude decrease in corrugation on h-BN compared to SiO$_2$. ( A. Luican. *et al*. 2011 unpublished). h-BN crystals were grown by K. Watanabe and T. Taniguchi

### C. Charge Transport in Graphene

**Graphene devices for transport measurements:**

Due to the low carrier density in graphene, especially close to the CNP where the DOS vanishes, the effect of gating is substantial. In addition to tuning the carrier density this also allows tuning the Fermi energy which provides an important handle for studying and controlling the electronic properties of the material.

In the simplest electric field gating device, graphene is deposited on a Si/SiO$_2$ substrate, with the heavily doped Si as the back gate. The thickness of the SiO$_2$ is chosen (e.g., ~300nm) so that graphene shows the best contrast under an optical microscope [73]. Graphene is deposited by transferring a thin film from a source (e.g., single crystal graphite or CVD graphene on metal foil [46]) to the Si/SiO$_2$ substrate. Thus far mechanical exfoliation of graphite crystals yields devices with the highest quality. The exfoliation procedure typically starts with repeatedly peeling graphite flakes with Scotch tape, and then pressing the graphite/graphene covered tape onto a substrate[9]. For best results, the Si/SiO$_2$ substrates are baked in forming gas (Ar/H$_2$) at 200 $^0$C for one hour prior to graphene deposition to remove water and organic residue[121]. A thin foil of graphite is peeled from the bulk material using Scotch tape and transferred onto the Si/SiO$_2$ substrate. Pressure is then applied onto the graphite foil using compressed high purity nitrogen gas through a stainless steel needle, for ~5 seconds[21]. The foil is then removed and the substrate is carefully checked under an optical microscope for single layer graphene. This process is repeated until enough graphene flakes are identified. AFM is then used to confirm that a single graphene layer is present followed immediately by coating with PMMA resist. The electrical contacts and leads are then fabricated using standard e-beam lithography techniques. To remove organic and water residue, the samples are baked in forming gas at 200 $^0$C for 1 hour.

For most experiments discussed here, graphene devices were measured in a variable temperature cryostat (4.2K-300K) or dilution refrigerator (0.1 -4.2K). A small low frequency AC current (typically ~10-100nA at 17Hz) was supplied and the corresponding voltage was measured with a lock-in amplifier. The gate voltage was separately supplied by a voltage source between the source electrode and the back gate.

### Electric field gating characterization and ambipolar transport.

By varying the gate voltage, $V_g$, the carrier density in the graphene device can be tuned: $n_s = (C_g/eA)V_g = \varepsilon\varepsilon_0 V_g/ed$, where $C_g$ is the gate capacitance, $A$ the gate area, $\varepsilon$ the relative dielectric constant, and $d$ the gate dielectric thickness. For the most commonly used 300nm $SiO_2$ substrates, this yields n~ $7.4\times10^{10}$ Vg (volt) cm$^{-2}$. At very low DOS, or when the geometric capacitance of the gate electrode is very large (e.g., thin gate dielectric), a quantum capacitance in series with the geometric capacitance may become important [207]. However for most devices fabricated on 300nm $SiO_2$, within the practically obtainable carrier density, the quantum capacitance is much larger than geometric capacitance and hence its effect is negligible.

The resistivity of graphene shows a maximum when the gate voltage is tuned to bring the Fermi level to the CNP. In practice the corresponding gate voltage $V_D$ (which is zero for an ideal clean device) is device dependent, as a result of un-intentional doping from contaminants. As a function of gate voltage, graphene shows qualitative symmetric resistivity in the electron and hole branches. Such ambipolar transport behavior is a direct consequence of the near-symmetric electron/hole bands.

Table 1. The contributions of various scattering mechanisms to the carrier density dependence of the scattering time, and resulting conductivity, mobility and mean free path are compared to ballistic transport in the last row.

| Source | Scattering time | σ (conductivity) | μ (mobility) | l (mean free path) |
|---|---|---|---|---|
| Coulomb | $\sim n_s^{1/2}$ | $\sim n_s$ | $\sim const$ | $\sim n_s^{1/2}$ |
| Short range | $\sim n_s^{-1/2}$ | $\sim const$ | $\sim n_s^{-1}$ | $\sim n_s^{-1/2}$ |
| Phonons | $\sim n_s^{-1/2}$ | $\sim const$ | $\sim n_s^{-1}$ | $\sim n_s^{-1/2}$ |
| Ripples | $\sim n_s \ln^2(n_s^{1/2}a)$ | $\sim n_s^{3/2} \ln^2(n_s^{1/2}a)$ | $\sim n_s^{1/2} \ln^2(n_s^{1/2}a)$ | $\sim n_s \ln^2(n_s^{1/2}a)$ |
| Midgap states | $\sim n_s^{1/2} \ln^2(n_s^{1/2}a)$ | $\sim n_s \ln^2(n_s^{1/2}a)$ | $\sim \ln^2(n_s^{1/2}a)$ | $\sim n_s^{1/2} \ln^2(n_s^{1/2}a)$ |
| Ballistic | | $\sim n_s^{1/2}$ | $\sim n_s^{-1/2}$ | ~ L/2 |

In some short devices, though, electron-hole (e-h) resistance asymmetry is commonly observed. This asymmetry depends on the contact material. For example, while in Au-graphene junctions the asymmetry is generally weak, in Al-graphene junctions it can be so strong that an additional resistance maximum was observed at negative gate voltages [23]. Such asymmetry suggests charge transfer between the contacts and the graphene channel which leads to a local doping effect in the vicinity of the contacts. Photocurrent measurements [208] suggest that *pn* junctions may be present at the graphene-contact interface, which in transport measurements induce e-h asymmetry. In long devices, especially those measured with the standard 4-probe geometry (such as Hall bars) the transport is usually electron-hole symmetric. The ambipolar transport is a useful feature for electronics applications such as frequency multipliers[209].

**Sources of disorder and scattering mechanisms**

In the absence of disorder, the transport in graphene is ballistic and can be described by the Landauer formalism. Solving the mode dependent transmission probability of massless Dirac electrons through a graphene strip with boundary conditions restricted by the width of the strip $W$ and the separation between the contacts $L$, Tworzydło et al. [94] showed that in ballistic devices the conductance follows $\sigma \sim E_F \sim \sqrt{n_s}$, and that the corresponding mean free path is $l \approx L/2$. Consequently, the mobility in a ballistic device is: $\mu = \sigma/ne \sim n_s^{-1/2}$. This implies that for ballistic transport mobility is not a useful parameter to characterize the device quality because, unlike for diffusive transport, it depends on the carrier density.

Ballistic transport on μm length scales in graphene was demonstrated in suspended devices [21, 22]. In non-suspended devices, DC transport measurements provide information about sources of disorder and scattering mechanisms. In the presence of disorder, the transport properties of graphene can be conveniently described by Boltzmann transport yielding: $\sigma = \dfrac{e^2 v_F^2 \rho(E_F) \tau(k_F)}{2}$, where $\rho(E_F)$ is the DOS at the Fermi energy and $\tau(k_F)$ the scattering time. Here $k_F$ is the Fermi momentum. Since the energy dependence of $\tau$ is determined by the scattering mechanism, a measurement of conductivity as a function of the carrier density provides clues to the scattering mechanism.

The scattering mechanisms in graphene include Coulomb scattering (from charged impurities) [210, 211], short range scattering from point defects, phonons [212], mid gap states [213], and ripples [106, 214]. Coulomb scattering is believed to be the main limitation on device quality for graphene on $SiO_2$ substrates. Table [1] summarizes the contributions of the various mechanisms to the scattering time and resistivity. The effects on mobility $\mu$, and mean free path $l$, are deduced by using the Drude model for the conductivity, $\sigma = ne\mu$, and $l = v_F \tau$.

### 1. *Graphene-superconductor Josephson junctions*

Combining the Dirac electronic structure of graphene with superconductivity is interesting both in terms of the physics (e.g., superconducting proximity effect) and for applications (e.g., superconducting electronics and sensors). Because of the chemical inertness of graphene and the presence of its charge carriers right on the surface, achieving transparent interfaces is relatively easy and reproducible compared to other gate controllable junctions available in semiconductors. The ability to tune the junction resistance across electron and hole bands provides a wide range of accessible parameters for studying the physics of Josephson junctions. Moreover, an unusual superconducting proximity effect was theoretically predicted at graphene-superconductor (GS) interfaces: whereas a normal electron impinging on a normal-superconductor (NS) interface is "retro-reflected" as a hole due to the Andreev reflection which retraces the same trajectory, the process is specular for the Dirac quasiparticles in graphene when the Fermi energy is within the superconducting gap [63, 94]. Such "specular Andreev reflection" (SAR) is expected to leave clearly distinguishable marks in ballistic SGS junctions, where the electron mean free path exceeds the junction length, detectable through a strong and unusual gate dependence of the multiple Andreev reflections (MARs) [115, 215]. In addition, the Josephson critical current $I_c$ and the product $I_cR_N$ ($R_N$ is the normal state resistance) are expected to exhibit an unusual gate dependence in ballistic SGS junctions, which is qualitatively different from that in conventional SNS junctions[216].

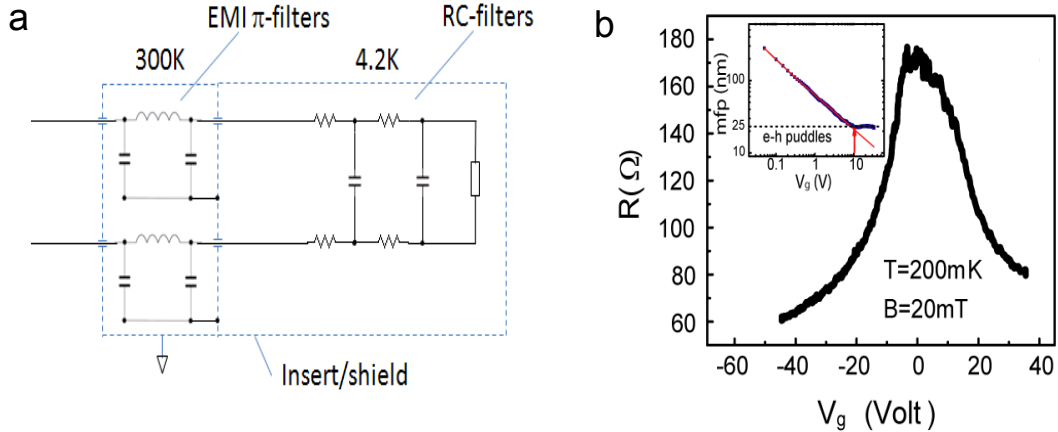

Figure C-1. a) Schematic circuit configuration for measuring the Josephson effect in graphene. b) Gate dependence of normal state resistance at T = 200mK. The superconductivity of the leads was suppressed with a small magnetic field. Inset: mean free path calculated from the transport data. The arrow indicates the onset of the puddle regime.

A GS Josephson junction was first demonstrated by Heersche et.al[116] in 2007. In this part we focus on the efforts to understand the basic properties of GS Josephson junctions fabricated on Si/SiO$_2$ substrates [117], and show that in such devices, due to substrate induced disorder, it is not possible to access the physics of the DP.

**Fabrication and measurement of graphene-superconductor junctions.**

Fabrication of S-G-S junctions is very similar to that of the typical graphene FET devices. Mechanically exfoliated single layer graphene is deposited onto Si (*p++*) /SiO$_2$ (300 nm) substrates and pre-patterned with alignment marks. Following identification of graphene with a combination of optical imaging and AFM, the superconducting leads, Ti(2 nm)/Al(30 nm), are fabricated using standard e-beam lithography and lift-off. To optimize the transparency of the G-S interface, UV ozone cleaning is carried out immediately before loading the samples into the thin film evaporation chamber. A high vacuum of ~5x10$^{-8}$ torr is reached before the deposition of the contacts. The deposition of Al is performed immediately after the Ti deposition. For the junctions discussed here the distances between the leads (junction length) are in the range $L$ ~200–400 nm, and the aspect ratios $W/L$ ~ 10–30, where $W$ is the junction width. Measurements were carried out in a dilution refrigerator with a base temperature of 100 mK using a standard four-lead technique. The Josephson effect in such micro-weak-links is delicate and vulnerable to electromagnetic interference (EMI). To filter out EM noise, a bank of two stage *RC* filters at low temperature (4 K) and a bank of EMI π-filters at room temperature were used for noise filtering as illustrated in Figure C-a.

### Superconducting proximity effect, bipolar gate-tunable supercurrent and multiple Andreev reflections

An optical micrograph of a typical S-G-S junction is shown in the upper inset of Figure C-2b. All the standard characteristics of the Josephson effect are readily observed in the S-G-S junctions[117]. A typical current-voltage characteristics (IVC) of G-S junction shows underdamped Josephson junction behavior following resistively and capacitively shunted junction (RCSJ) model, as illustrated in Figure C-2a. The sharp switching behavior in the IVC becomes smeared on approaching $T_c$. The switching current is sensitive to magnetic field, as illustrated by the Fraunhofer pattern in the lower inset of Figure C-2a. Supercurrent switching can also be induced by sweeping $V_g$ as illustrated in the inset of Figure C-2 b. As before, sharp jumps are seen between the Josephson and the normal current states, this time as a function of $V_g$. Here too hysteresis is observed. In the RCSJ model, both cases correspond to runaway of the "phase particle" moving in a tilted washboard potential $U(\varphi) = -E_J \cos(\varphi) + (I/I_C)\varphi$ with average slope $\sim I/I_c$, where $\varphi$ is the phase difference between the two superconducting banks, and $E_J = \Phi_0 I_C/2\pi$ is the Josephson energy. The slope is controlled by $I$ or by $I_C$ for the current or gate swept measurements, respectively. In both types of measurements, bipolar gate tunable supercurrent was observed which persist throughout the electron and hole branch. To obtain the value of $I_c$ in zero field, a compensating field is applied and tuned to maximize the value of $I_c$ as shown in Figure C-2c. When RF radiation is applied to the junction, the IVC shows Shapiro steps with $\Delta V = \hbar \omega_{RF}/2e$ where $\omega_{RF}$ is the angular frequency of the applied radiation.

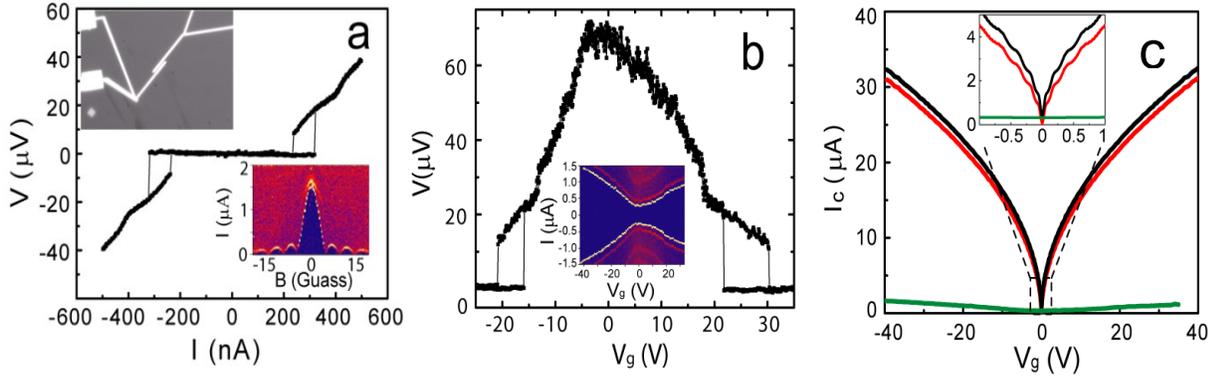

**Figure C-2. Graphene/superconductor Josephson junction. a)** Main panel: IVC showing Josephson state at T = 200 mK. Upper left inset: optical micrograph of a graphene Josephson junction. Lower right inset: magnetic field dependence of critical current exhibiting Fraunhofer pattern. **b)** Gate dependence of the voltage across a junction in Josephson state for I = 800 nA. Inset: IV curves as a function of gate voltage. The center area corresponding to the Josephson state is separated from the normal state by the switching current represented by the bright line. **c)** Gate dependence of critical current. Comparison of data (green) with theoretical predictions for ballistic SGS junction at T = 0 without (black) and with fluctuations (red) due to RF noise temperature of 300 mK. Inset: zoom into Dirac point.

The supercurrent is in general inversely related to $R_N$, the normal state junction resistance. Figure C-2b illustrates the variation of $R_N$ as $V_g$ is swept through the CNP causing the charge carriers to change from holes (negative $V_g$) to electrons (positive $V_g$). The low temperature normal state was accessed by quenching the superconductivity in the leads with a small magnetic field. Based on a comparison between the $I_C R_N$ product and the superconducting gap of the contacts, the switching currents measured here are significantly lower than theoretical predictions for ballistic junctions [94].

Outside the supercurrent regime, pronounced MAR features are observed. They develop in the bias voltage dependence of the differential resistance below $T_c$, as shown in Figure C-4a. These features consist of a series of sharp resistance minima appearing at subgap voltages $2\Delta/pe$, where $p$ is an integer and $\Delta$ is the superconducting gap of the electrodes. The first four MAR minima are indicated by dotted lines in Figure C-4a. For all the samples discussed here, the first four to six minima can be unambiguously identified, indicating high transparency of the SG interfaces. The subgap features, whose temperature dependence tracks $\Delta(T)$, are essentially independent of temperature below 500 mK. The MAR features also appear to be independent of $V_g$.

**Diffusive versus ballistic transport**

In order to determine whether the transport in these junctions is diffusive or ballistic, several aspects of the transport properties were analyzed: normal state transport, super-current values and multiple Andreev reflections. Below we show that each of these measurements indicates that the transport in these devices is strongly affected by disorder and displays all the characteristics of diffusive transport.

Above the superconducting transition temperature of the leads, the conductivity of a diffusive graphene channel $\sigma = \dfrac{e^2 v_F^2 N(E_F) \tau(k_F)}{2}$ is determined by the dependence of the DOS and the scattering time on the Fermi energy (gate voltage). At relatively large values of $V_g$ where potential fluctuations are negligible and the carrier density can be assumed uniform, the mean free path, $l$, can be estimated by measuring the normal state conductivity $\sigma$ as follows: $l = v_F \tau = \sigma h / 2 e^2 k_F$ where $k_F = \sqrt{n\pi} = \sqrt{\varepsilon \varepsilon_0 \pi / V_g e d}$. Here, $d$=300 nm is the thickness of the SiO$_2$ layer and $\varepsilon = 4$ its dielectric constant. One finds that in SGS samples on SiO$_2$ substrates the mean free path is much shorter than the lead distance, $l$~25 nm$\ll L$, indicating that these junctions are diffusive.

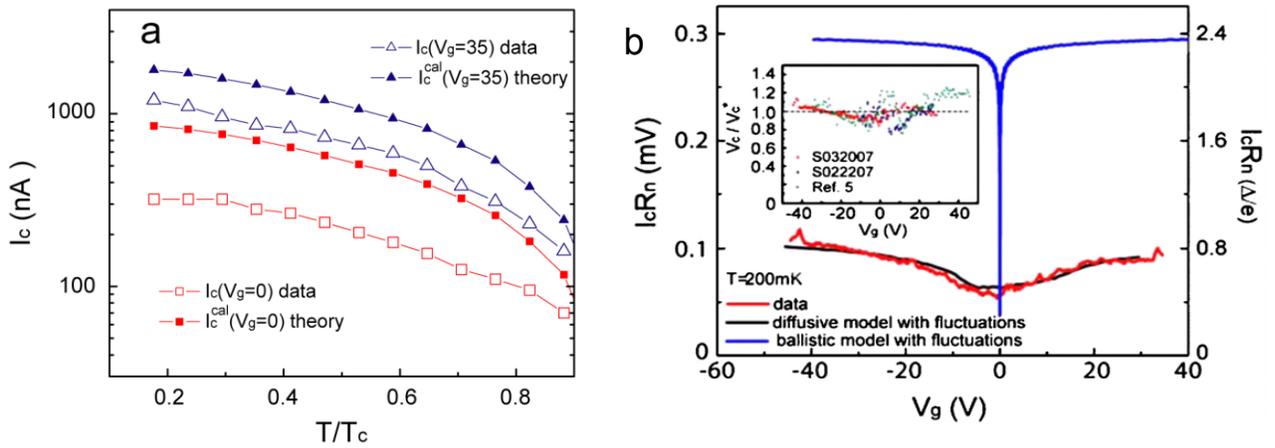

Figure C-3. Temperature and gate dependence of the Josephson effect. a) Comparison of the measured temperature dependence of the switching current, Ic, (open symbols) with calculated values for ballistic SGS junctions (solid symbols). b) Comparison of measured gate dependence of $V_c$ = I$_c$R$_n$ (red curve) with calculated values from the Likharev model including corrections for premature switching $V_c$* = I$_c$*R$_n$ (black curve). The blue curve corresponds to the ballistic model. Inset - ratio of experimental and theoretical values $V_c$/ $V_c$*, for several devices measured by different groups.

Estimating the superconducting coherence length in the diffusive limit: $\xi \sim \sqrt{\hbar D / \Delta} \sim 250 nm \sim L$, where $D = v_F l / 2$ is the diffusion coefficient, we find that these SGS devices are at the crossover between long and short diffusive Josephson junctions. The critical current through these junctions is significantly lower than the theoretical prediction for ballistic junctions [94]. The discrepancy between the measured and calculated values rapidly increases with gate voltage as shown in Figure C-3b, until at the highest gate voltages it exceeds one order of magnitude. Clearly, the gate dependence of $I_c$ in these SGS junctions cannot be accounted for in the picture of ballistic transport.

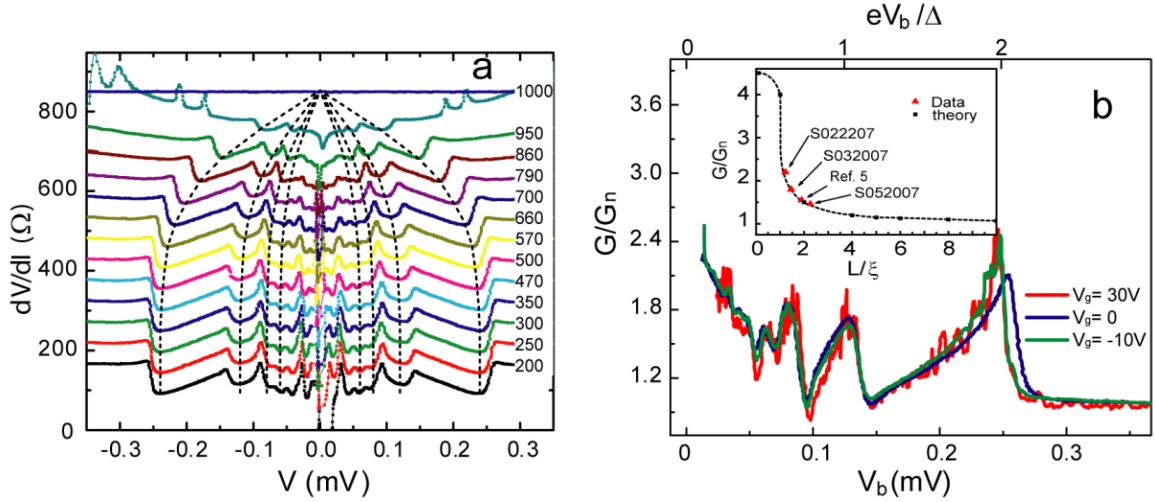

**Figure C-4. Multiple Andreev Reflections.** (a) Temperature dependence of the MAR. Curves, taken for different temperatures, are shifted vertically for clarity. The labels show temperatures in units of mK. Dotted lines indicate the first four sub-gap oscillations. The extra peaks at high bias (seen in the curve just below Tc) signal the superconducting to normal transition in the leads. b) Comparison of normalized sub-gap features at different doping levels for sample S022207. Inset: normalized differential conductance as a function of . Black squares: theoretical values for diffusive junctions from Heersche *et al.* (2007). Triangles: measured values.

Next the measured values of $I_c$ and $I_cR_N$ are compared to calculated values obtained for diffusive SNS junctions within the model proposed by Likharev [217], which treats the junction as a 1-d weak link with vanishing gap in the channel material. Using the mean free path obtained from the measured $V_g$ dependence of $R_N$, as an input parameter, the temperature dependence of $I_cR_N$ and $I_c$ as a function of gate voltage is obtained numerically by solving the Usadel equations. As illustrated in Figure C-3a, the calculated overall temperature dependence of $I_c$ qualitatively agrees with the measurement, but its magnitude is consistently larger. This discrepancy is attributed to "premature" switching induced by fluctuations due to the thermal and electromagnetic noise[218, 219]. The mean reduction in critical current can be estimated in the limit $E_J \gg E_{fl}$, as $\langle \Delta I_c \rangle \sim I_c \left[ \frac{E_{fl}}{2E_J} \ln\left( \frac{\omega_p \Delta t}{2\pi} \right) \right]^{2/3}$. Here, $E_J = \hbar I_c / 2e$ is the Josephson energy, $E_{fl}$ is a characteristic fluctuation energy, $\Delta t \sim 10^2$–$10^3$ s the measurement time, $\omega_p = \sqrt{2eI_c/\hbar C} \sim 10^{11} s^{-1}$ is the plasma frequency of the junction, and $C \sim 2 \times 10^{-13}$ F is the effective capacitance estimated from the RCSJ model. Taking into account the fluctuations, it is possible to quantitatively simulate the measured data, as shown in Figure C-3b. By contrast, in order to match ballistic junction predictions (blue curve), one would have to assume noise temperatures that are unrealistically high, which moreover would have to depend on the applied gate voltage.

The diffusive nature of the junction can also be studied using the multiple Andreev reflections features which, being relatively insensitive to the RF background, are more reliable. In the diffusive junction model[115], the MAR features are independent of carrier density but their shape is quite sensitive to the ratio $L/\xi$. On the other hand, in the ballistic SGS junction model[215], the MAR features are independent of $L/\xi$ but are quite sensitive to carrier density with the normalized conductivity at the first MAR maximum sharply dropping from 4.5 at the CNP to ~1.5 away from it.

The experimental data, shown in Figure C-3b, is in good agreement with the diffusive model and clearly indicates diffusive rather than ballistic transport in these junctions. When plotting the bias dependence of the normalized conductivity, the curves generally overlap very well with the calculations of the model. Furthermore, the shape of the MAR features, Figure C-4b in these junctions is quite sensitive to the ratio $L/\xi$. Comparing the shape of the measured MAR features with theoretical predictions [115] (Figure C-4b) it is found that they best fit diffusive junctions with $L/\xi$ ~1-2. This yields a coherence length $\xi \sim 150-300\,nm$, which corresponds to a mean free path of 10–30 nm, in agreement with the values obtained from the normal resistance of the device. For a more quantitative analysis, the normalized differential conductance was measured at the first subgap peak ($p$=1) and plotted against the ratio $L/\xi$ with $\xi$ obtained from gate dependence of resistivity in the normal state, as illustrated in the inset of Figure C-4. The data points from all the reported devices [115, 116], fall onto the theoretical curve derived for junctions in the diffusive regime.

The absence of the expected manifestations of the relativistic charge carriers in SGS junctions on Si/SiO$_2$ substrates (in particular, the gate dependence of the MARs and $I_c$ predicted by the ballistic theory), can be attributed to two limiting factors: short mean free paths and the smearing of the DP by potential fluctuations. Such limitations are not necessarily intrinsic to the material. In devices with large lead separation for example, mean free paths of 100 nm~500nm are routinely achieved. In fact if Coulomb scatterers could be eliminated leaving phonons as the only scattering mechanism, it is predicted that the intrinsic resistivity of graphene, can be as high as ~300,000 cm$^2$/Vs at room temperature[106].

## 2. *Suspended graphene*

Charged impurities are one of the prime causes of scattering and potential fluctuations. The impurities may come from various sources including polymer residues, water molecules, trapped charge centers in the substrates, etc. While contaminants on the top surface of the graphene device can be cleaned by solvents and baking, those trapped between graphene and the substrate, as well as the disorder from the substrate itself (charge trapping centers, for example) cannot be removed easily. A natural solution is to suspend graphene thus removing the influence of the substrate. Suspending graphene was first demonstrated on TEM grids [20] and above Si/SiO$_2$ substrates for electromechanical resonators [19]. Suspended graphene (SG) transport devices were fabricated and measured shortly afterwards [21, 22].

### Fabrication of suspended graphene devices.

In a typical SG device (Figure C-5), graphene is suspended from the metallic leads/contacts, which run across the sample and at the same time provide structural support. Such two-lead configuration avoids complications such as sensitivity to details of the lead geometry that arise in ballistic devices when transport measurements are carried out with a conventional Hall bar design. For the two-lead voltage configuration described here, the measured transport properties of ballistic devices depend on lead separation and doping in way that can be calculated using the Landauer [94] formalism. In an ideal ballistic graphene junction, this would yield a charge carrier mean free path equivalent of $L/2$ ($L$ being the channel length) in the Boltzmann transport

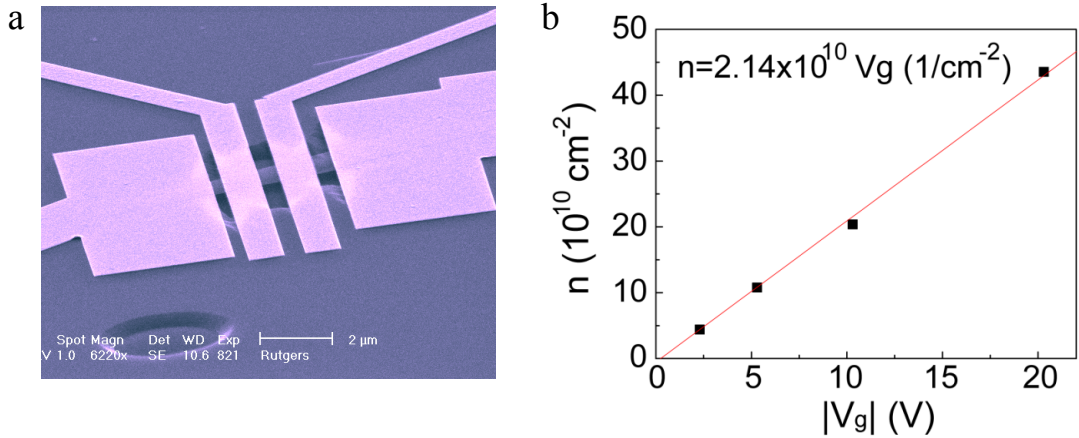

conductivity[21].

The suspended graphene (SG) devices described here are fabricated from conventional non-suspended graphene (NSG) devices with Au/Ti or Au/Cr leads deposited on Si/SiO$_2$ (300nm) substrates[21, 23]. After fabrication, the NSG devices are coated with PMMA, and an additional electron beam lithography step is carried out to open etching windows over the graphene channels. The samples are then immersed in 7:1 (NH$_4$F: HF) buffered oxide etch at 25 $^0$C for ~5 min. Due to the weak coupling of graphene to the substrate, capillary action draws the etchant underneath the whole graphene film. Hence, the etching actually starts in the entire graphene channel shortly after the sample is immersed. The isotropic etching therefore results in the suspension of the whole device, the graphene and the leads attached to it, (Figure C-5). The etchant is next replaced by deionized water, then hot acetone (to remove the PMMA) and finally hot isopropanol, with the sample remaining immersed in the liquid at all times. Finally the sample is taken out of hot isopropanol and left to dry. Due to the small surface tension of hot isopropanol and mechanical strength of graphene, devices with short channel length (~1μm) tend to survive the effects of surface tension.

A slightly different and simpler method for making SG devices is to directly apply SiO$_2$ etchant without the PMMA etching mask and hence the additional lithography for opening etching windows[22]. In this approach, the NSG devices are typically required to have relatively massive leads and to avoid completely removing the supports for the leads, typically only part of the SiO$_2$ is etched. Special attention may be required to ensure the flatness of SiO$_2$ underneath the graphene channel.

Following fabrication, the SG samples are baked in forming gas (Ar/H$_2$) at 200 $^0$C for 1 hour to remove organic residue and water right before the measurements. For further removal of the contaminants, high current annealing is performed at cryogenic temperatures after the devices are loaded and cooled [220]. This entails passing a large current through the graphene device and monitoring the voltage. Evaporation of contaminants as a result of ohmic heating can be observed through abrupt changes in the IV curves. Typical currents for effective cleaning are ~0.5mA/μm, but large sample-to-sample variation can be observed.

**Ballistic transport in suspended graphene junctions.**

For a quantitative study of the suspended graphene (SG) devices, it is important to first determine $n(V_g)$ by measuring the gate capacitance. This can be done through magneto-transport measurements which correlate the observed LLs with their corresponding filling factors $\nu = n_s h / eB$. A "fan diagram" showing LL index versus 1/B at various gate voltages (carrier densities) can be established, from which $n(V_g)$ is obtained. Typically $n(V_g) \sim \varepsilon_0 V_g / ed$ with small sample-to-sample variation due to sagging of the graphene bridge which slightly modifies the value of gap to the Si substrate, $d$, around the nominal $SiO_2$ thickness.

Electron–hole asymmetry is generally observed in most two terminal devices, including the SG devices. The asymmetry becomes more significant with decreasing channel length [23], suggesting that it is due to the contacts. We limit the discussion to the hole branch where well-defined Shubnikov-de-Haas (ShdH) oscillations were observed in the SG devices.

In the absence of magnetic field, we focus on understanding the potential fluctuations and scattering in the SG devices. Comparing the $R(V_g)$ curve before and after suspension (Figure C-6a) the reduction in potential fluctuation is evident from the significant sharpening of the curve. On the hole branch the half-width at half-maximum (HWHM) is almost an order of magnitude smaller than that of the best graphene-on-$SiO_2$ samples reported. The residual carrier density, which also determines the amplitude of the random potential fluctuations, is obtained from the density dependence of the resistivity or conductivity ( Figure C-6b, Figure A-8f ) as the density for which the resistivity saturates ~$10^{11}$ cm$^{-2}$. This corresponds to a Fermi energy fluctuation of ~30meV which imposes an energy scale below which the electronic properties are controlled by electron-hole puddles, consistent with the fact that the resistivity curve are independent of temperature below 200K.

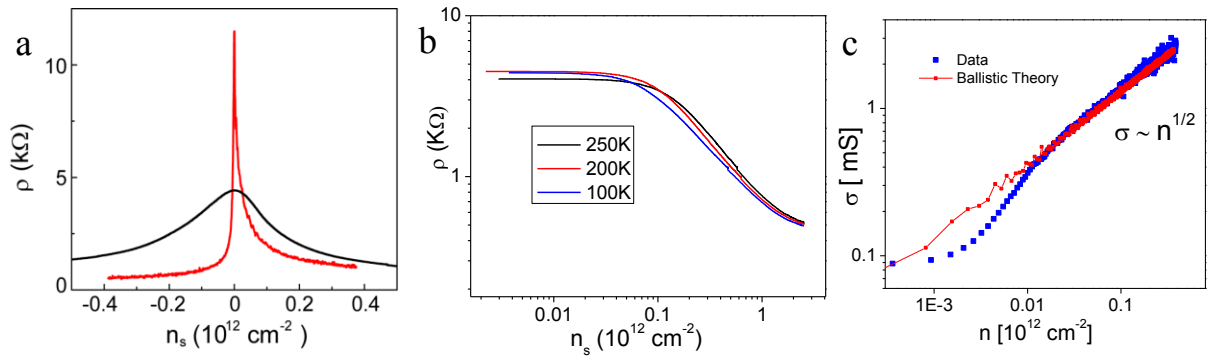

The residual carrier density in the SG sample, obtained as the value at which the data deviates from the theoretical ballistic curve (Figure C-6c), is one order of magnitude smaller than for the NSG sample. As a result of the strongly reduced residual carrier density, the resistance near the CNP shows strong temperature dependence in the SG device, in marked contrast to NSG samples, as illustrated in Figure C-7b. Whereas in NSG the maximum resistivity saturates below ~200 K, in SG it continues to grow down to much lower temperatures consistent with reduced random potential fluctuations.

In Figure C-7b we compare the temperature dependence of the Fermi energy fluctuations (random potential) for two SG samples and one NSG sample. At high temperatures, the slopes of the SG curves approach $k_BT$ as expected for thermally excited carriers and in agreement with theoretical calculations[221]. At low temperatures, where the fluctuation energy is controlled by charge inhomogeneity, its value in the SG samples, $\Delta E_R \sim 3$ meV, is much smaller than in the best NSG samples reported ($\Delta E_R \sim 30$ meV[222]). A direct consequence of the low level of potential fluctuations in the SG samples is that one can follow the intrinsic transport properties of Dirac fermions much closer to the CNP than is possible with NSG samples.

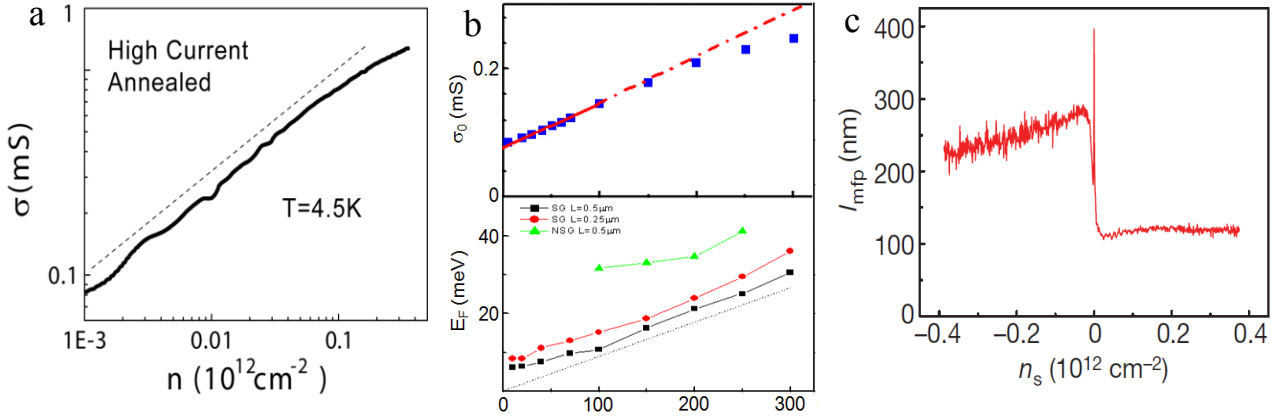

Figure C-7. a) Conductivity of suspended graphene shows $n^{1/2}$ dependence in near-ballistic suspended graphene devices. b) Temperature dependence of minimum conductivity of suspended graphene (top) and effective Fermi energy fluctuation for suspended and non-suspended graphene (bottom). c) Mean free path in suspended graphene. On the hole side it's value is ~L/2, where L is the distance between leads.

As a result of the strong reduction of scattering the $R(V_g)$ curve in the SG samples at low temperatures approximately follows the dependence expected for ballistic transport as shown in Figure C-6c and the mean free path is roughly $\sim L/2$ as seen in Figure C-7c. The gate dependence of the measured conductivity can be fit by assuming that midgap states contribute a resistivity in series with the ballistic junction:[21, 213], $\sigma = \sigma_{bal}^{-1} + \sigma_{mg}^{-1}$, where $\sigma_{bal} = \frac{L}{W}\frac{4e^2}{h}\sum_n T_n$ is the ballistic contribution to the conductivity and $\sigma_{mg} = \frac{e^2}{n_i h}\frac{2}{\pi}k_F^2(\ln k_F R_0)^2$ describes scattering by midgap states produced by topological defects of characteristic size $R_0$ and density $n_i$. The fit at 4 K gives $R_0 \sim 3.4$ nm and $n_i \sim 1 \times 10^{10}$ cm$^{-2}$. We note that for a ballistic junction the mobility, $\mu = \frac{\sigma}{ne} \propto n_s^{-1/2}$ depends on carrier density, so it is meaningless to assign a mobility unless one specifies the density at which it is measured. This is in contrast to the usual case of diffusive transport where the mobility is constant. The maximum mobility observed just outside the potential fluctuation regime exceeds 200,000 cm$^2$/Vs in the SG devices. At high carrier densities the mobility in SG and NSG becomes comparable (~10,000 cm$^2$/Vs), suggesting effects from short range scattering and contact resistance.

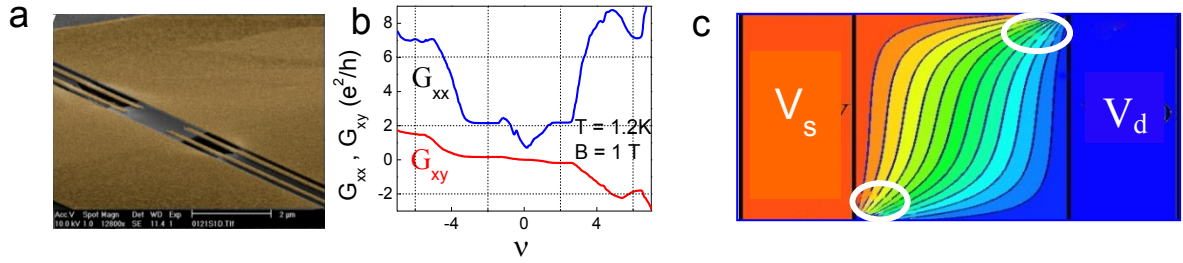

**Figure C-8.** Absence of QHE in SG devices with Hall bar lead configuration. a) SEM micrograph of SG sample with Hall bar lead geometry. Scale bar 2µm. b) Longitudinal and Hall conductance for the sample in (a). The values of $G_{xy}$ are reduced below the expected $\nu e^2/h$, while $G_{xx}$, instead of vanishing for $\nu$ corresponding to the QHE, develops plateau-like features. This indicates that the Hall voltage probes are short-circuited due to hot-spots. c) Equipotential line distribution in a sample with $W/L$=1.78, for a large Hall angle ($\sigma_{xy}/\sigma_{xx} = 20$), illustrating the "hot spots" at opposite corners of the sample marked by white circles. Adapted from Skachko et. al, Ref. 30.

### 3. *Hot spots and the fractional QHE.*

The fractional QHE, as the hallmark of strong correlations, is an important stepping stone for establishing the presence and extent of electron-electron correlations in graphene. SG devices where the Fermi energy fluctuations, $\Delta E_R \sim$ 3meV, are one order of magnitude smaller compared to the best NSG samples, were thus considered ideal candidates for studying the fractional QHE in graphene. However, even though the condition for observing the fractional QHE in SG devices (see A6), $5meV \dfrac{(B[T])^{1/2}}{\varepsilon} > \Delta E_R$ is already satisfied in fields as low as a few Tesla, all efforts to observe the fractional QHE in SG devices using the standard Hall bar lead geometry failed. Surprisingly, the SG devices even failed to exhibit precise integer QHE plateaus [22].

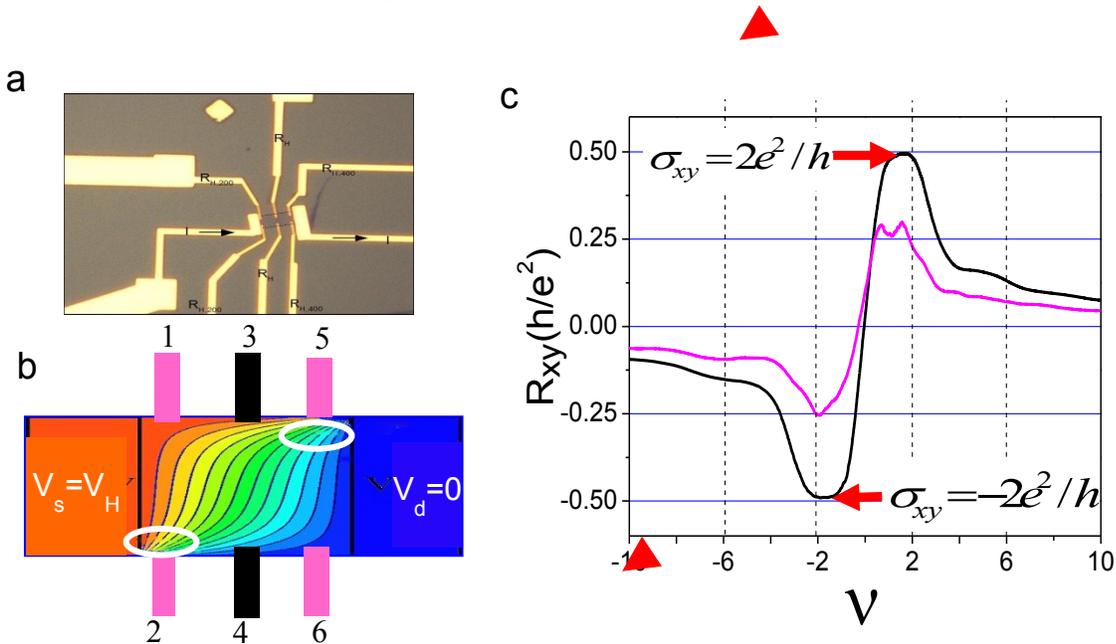

**Figure C-9.** Effect of hot spots on the QHE. a) SEM micrograph of large NSG device 10 x 4 µm$^2$ with 6 voltage leads, one pair placed in the center of the sample and the other two within 100nm of the current leads on both sides. B) Equipotential lines for conditions corresponding to a QHE plateau together with position of voltage leads for sample shown in panel *a*. c) Comparison of Hall resistance obtained from voltage measurement for center leads (3,4) black line, and for leads within the hot spot region (5,6) magenta line, clearly shows the shorting of the Hall voltage for the leads that cover the hot spot. Data taken at B=9T and T= 20 K. Data from Skachko *et al*. 2011, unpublished.

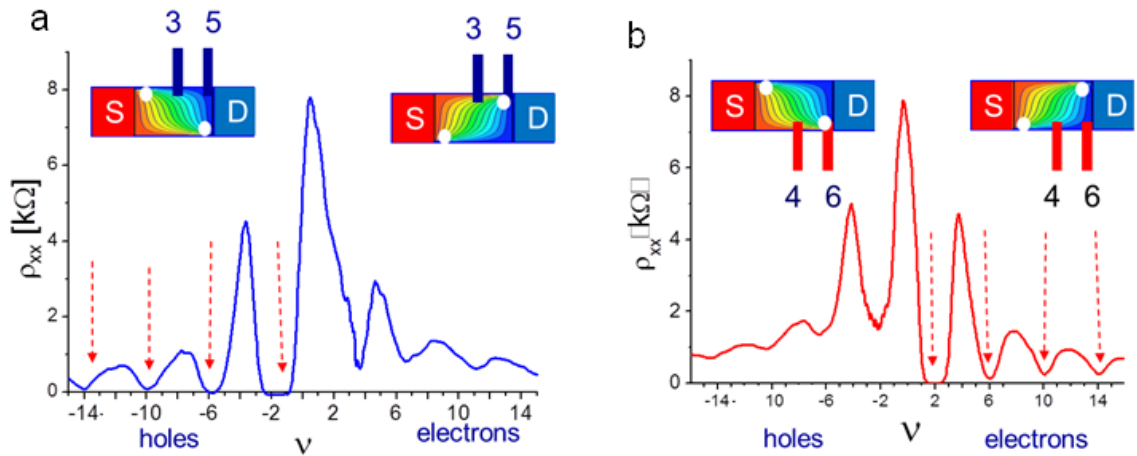

Figure C-10. Filling factor dependence of longitudinal voltage at 9T and 20K. a) Longitudinal resistivity measured between leads 3,5 of device shown in Figure C-9a. The hot spot positions (white dots) are along opposite diagonals for the electron and hole sectors. In the hole sector, where the leads are outside the hot spots the longitudinal resistance vanishes on the Hall plateaus. In the electron sector where the lead 5 is in the hot spot the longitudinal resistance does not vanish indicating shorting of the Hall voltage. b) Same as panel *a* but longitudinal voltage is measured on the opposite side of the sample between leads 4 and 6. Now the shorting occurs in the hole sector where lead 6 is in the hot spot. Data from Skachko *et al* . 2011, unpublished.

The cause for this failure was explained by Du *et al*. [27, 29, 30] who showed that the Hall bar measurement geometry shorts out the Hall voltage in the small SG samples (Figure C-9) and demonstrated that problem can be circumvented in a two terminal geometry. Using a two terminal measurement configuration they were able to observe both the integer QHE and the fractional QHE in SG devices. Subsequently, using similar two terminal measurement geometries, other groups also reported the fractional QHE in SG devices [28, 223].

To understand the limitation on measurement geometry in SG devices one needs to consider the distribution of electrical potential and current in the QHE regime (Figure C-9c). Since the Hall angle is $90^0$ in this regime, the lines represent both lines of current and lines of constant potential. Markedly all equipotential lines converge in two diagonally opposite corners of the device where all the dissipation takes place. These points of convergence, sometimes referred to as hot spots[224-229] are a consequence of the peculiar potential distribution at large Hall angles (the case of plateaus in the QHE) where most of the potential drop, roughly equal to the Hall voltage, occurs at opposite corners of the sample close to the current leads [230, 231]. The position of the hot spots can be obtained by using the right hand rule to determine the direction of the force on the moving carrier. Thus if the drain is at the left current lead, the carriers are holes and the field pointing out of the page the hot spots will be along the diagonal running from bottom left to top right. The position of the hot spots shifts to sides if one of the following is reversed: sign of the carriers, direction of the field or of the current. This is indeed what is observed experimentally as discussed below and illustrated in Figure C-9c and Figure C-10.

In order to elucidate the role of hot spots in shorting out the QHE Skachko *et al.*[232] [30] carried out measurements in a large NSG device that had leads placed both inside and outside the hot spots (Figure C-9a). Figure C-9c shows a comparison between the Hall resistance measured with leads outside (3,4) and inside (5,6) the hot spot region. When the leads are outside the hot spot region the Hall resistance shows the precise quantization in units of $e^2/2h$, as expected for the integer QHE. In contrast the value measured with leads inside a hot spot is significantly reduced due to shorting of the QHE. A similar effect is seen for the longitudinal resistance as illustrated in Figure C-10. Measuring the longitudinal resistance between two leads that are outside the hot spot region reveals the expected vanishing longitudinal resistance on Hall plateaus. However if one of the leads is placed inside a hot spot this is no longer the case. Notably, reversing the current direction or position of leads reverses the position of the hot spots and consequently the sign of the carrier charge for which shorting occurs. A similar reversal is observed if the field direction is reversed.

### QHE with two terminal measurements

Due to the small sample size of SG devices (typical length L<0.5–1 µm and width W<1.5–3 µm), which is necessary to ensure mechanical and structural integrity of the sample, it is practically impossible to avoid placing the voltage leads outside the hot spot regions. As a result shorting the Hall voltage in SG devices measured using the standard Hall bar lead configuration is essentially unavoidable[30].

Shorting the Hall voltage can be circumvented in the two-terminal lead configuration shown in Figure C-5a. One drawback of this configuration is that the two-terminal devices do not simultaneously provide information on $R_{xx}$ and $R_{xy}$. Instead the measured two-terminal conductance is a combination of both longitudinal and transverse magneto transport. The relation between magneto-resistance oscillations and the QHE measured in two-terminal devices calculated by Abanin and Levitov [233]. It was shown theoretically that, for clean samples and low temperatures, the two-terminal conductance displays plateaus at values that are precisely the same as those obtained on QHE plateaus. In between the plateaus the conductance is non-monotonic and depends on the aspect ratio W/L as shown in Figure C-11a. Figure C-11b shows the two-terminal conductance versus filling factor for the SG sample shown in Figure C-5a. In this device W>L, and the conductance is expected to overshoot between plateaus, as is indeed observed. The two-terminal measurements reveal well-defined plateaus associated with the anomalous QHE that appear already in fields below 1 T. Above 2 T additional plateaus develop

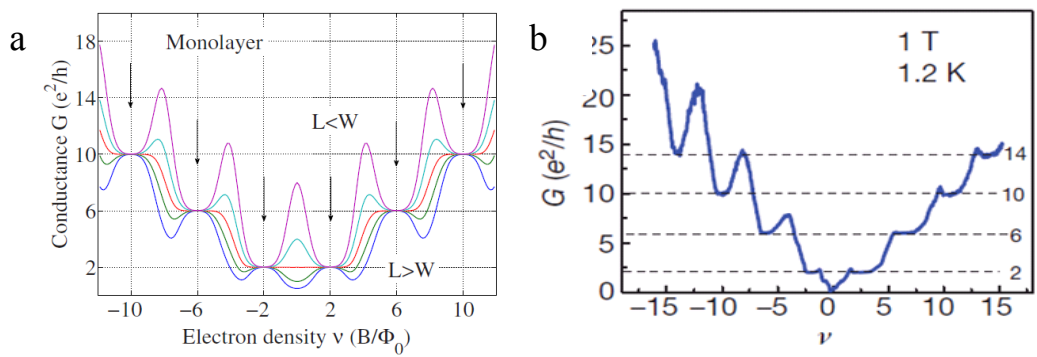

Figure C-11. a) Theoretical predictions for the two terminal conductance in the QH regime for rectangular channels with different aspect ratios [197]. b. Measured two terminal conductance in a suspended graphene device with *W/L>1*.

at ν= -1 and at ν=3, reflecting interaction-induced lifting of the spin and valley degeneracy.

**Fractional QHE**

The experimental evidence supporting the massless Dirac fermion picture of the charge carriers in graphene was quick and compelling appearing already in the very first magneto-transport measurements. Most of the initial work supported the single particle picture of the charge carriers in graphene, but no evidence for collective effects and interactions could be found. Magneto-transport experiments on graphene-on-$SiO_2$ samples showed no evidence of interactions or correlations for magnetic fields below 25 Tesla. In higher fields, the appearance of QHE plateaus at ν=0,1,4 suggested that interaction effects do exist in graphene [234]. However, experimental observation of such effects was strongly suppressed by substrate-induced potential fluctuations.

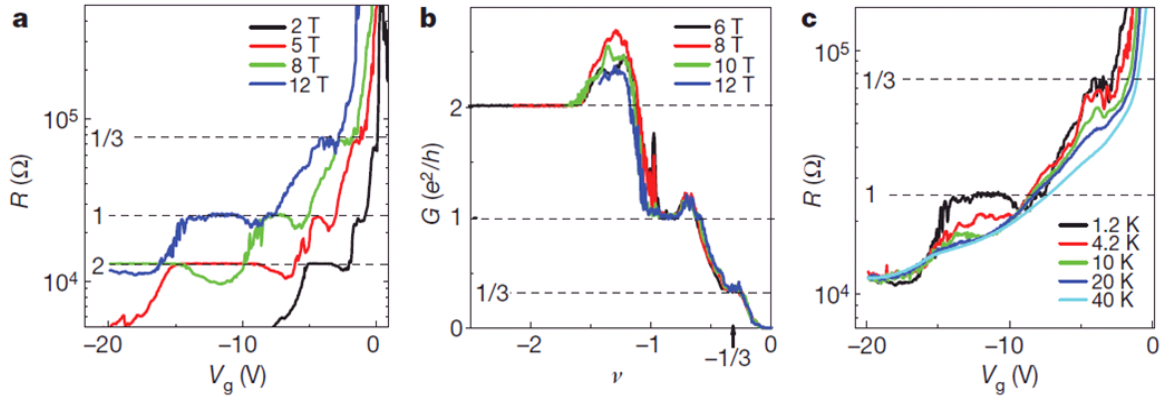

**Figure C-12. FQHE in suspended graphene.** a) Gate voltage dependence of resistance for a SG device at 1.2 K, showing QH plateau of ν=1 at B>2 Tesla and ν=1/3 at B>5Tesla. b) Conductance as a function of filling factor in relatively strong magnetic fields. C) Temperature dependence of the quantum Hall plateau features. The plateaus at ν=1/3 become smeared out with increasing T and disappear for T~20 K.

SG devices, which exhibit ballistic transport in zero field, are well suited to study interaction effects and magnetically induced phases provided that he Hall voltage is not shorted out by the voltage leads as demonstrated by Du *et al.*[27]. Using a two-terminal geometry they showed that integer plateaus arising from interaction effect (ν=1, 3, etc.) can be clearly observed in SG in magnetic field as low as ~2 Tesla. At low temperatures and above ~2 T, they observed a fractional QHE plateau at ν=1/3 which becomes better defined with increasing field (Figure C-12a). When plotting G versus ν, the curves for all values of B collapse together (Figure C-12b), and the plateaus at ν=1/3, 1 and 2 show accurate values of the QH conductance. The FQHE reflects the formation of an incompressible condensate, which can be described by a Laughlin wavefunction [127]. In the composite-fermion generalization of the FQHE [128], the FQHE state can be mapped onto the integer QHE of composite fermions, giving rise to the filling- factor sequence $\nu = \dfrac{p}{2sp \pm 1}$ (with s and p integers) which corresponds to the formation of weakly interacting composite particles consisting of an electron and an even number of captured magnetic flux lines. In this picture, the FQHE with $\nu = 1/3$ corresponds to the integer QHE with $\nu = 1$ for the composite particles consisting of one electron and two flux lines. Excitations out of this state would produce fractionally charged quasiparticles $q^* = e/3$, at an energy cost of the excitation gap, $\Delta_{1/3}$, which provides a measure of the state's robustness. Despite the qualitative difference in LL spectra between Dirac fermions in graphene and the non-relativistic electrons in semiconductors, the $\nu = 1/3$ state is formally expected to be the same in both cases [235] but with the pseudospin in graphene playing the role of the traditional electron spin in the non-relativistic case.

An order of magnitude estimate of $\Delta_{1/3}$ can be obtained from the temperature at which the $\nu = 1/3$ plateau disappears (Figure C-12c): $\Delta_{1/3} \approx 20K \approx 0.008 E_C (12T)$, where $E_C(K) = e^2/4\pi\varepsilon_0 l_B = 650 B^{1/2}/\varepsilon$ (with $B$ in units of tesla) is the Coulomb energy, $\varepsilon_0$ the permittivity of free space and $l_B = \sqrt{\dfrac{\hbar}{eB}}$ is the magnetic length and $\varepsilon$ the effective dielectric constant. A more precise value can be obtained by using the method described below.

**Activation gap obtained from two terminal measurements**

Obtaining the activation gap from a two terminal measurement is technically challenging due to several reasons: the presence of contact resistance, the mixing of longitudinal and transverse components ($\sigma_{xx}$ and $\sigma_{xy}$), and the shape dependence of the two-terminal magneto-resistance At low fillings, the resistance $\dfrac{1}{\nu}\dfrac{h}{e^2}$ is large compared to the typical contact resistance ($\sim$ 100 Ohm for Au/Ti leads), hence the latter has a negligible effect on the analysis. The dependence of the magneto transport on sample geometry and the connection between longitudinal and transverse components which can be obtained using a conformal mapping approach [233, 236] makes it possible to extract the QH activation gap from a two terminal measurement. The first step is to de-convolute $\sigma_{xx}$ and $\sigma_{xy}$ from the measured two terminal resistance by using the conformal invariance of the magneto-transport problem [233]. Because of the 2D nature of the problem, $\sigma_{xx}$ and $\sigma_{xy}$ can be interpreted as the real and imaginary parts of a complex number $\sigma = \sigma_{xx} + i\sigma_{xy}$, and thereupon the transport equations become conformally invariant. Applied to a rectangular two-lead geometry, the conformal mapping yields a specific dependence of the two-terminal conductance on $\sigma_{xx}$, $\sigma_{xy}$ (or more directly $\rho_{xx}, \rho_{xy}$) and the sample aspect ratio $A=L/W$: $R = |\rho_{xy}| + \rho_{xx} g(l)$, where $g(A)$ is a function of the aspect ratio[29]. The value of $g(A)$ is positive for $A>1$ and negative for $A<1$. It vanishes at $A=1$. For $A>>1$ the function $g(A) \sim A$; while for $A<<1$ it is $g(A) \sim -1/A$. Deviation from a quantized conductance value obtained from the measured two terminal resistances therefore offers a way to extract $\rho_{xx}$ from two terminal measurements.

Figure C-13 shows the simulated $\sigma_{xx}$ obtained by this method from the measured two-terminal resistance for an SG device[29], at 1.2K and 12 Tesla as input parameters. Conformal mapping analysis thus allows to obtain $\sigma_{xx}$ and its temperature dependence from the two-terminal measurement. An Ahrenius plot of the temperature dependence of $\sigma_{xx}$ at the minima is then used to extract the activation gap $\Delta$: $\sigma_{xx} \sim \exp(-\Delta/2k_B T)$ (Figure C-13b). The best fit values obtained are $\Delta/k_B = 10.4K$ for $\nu = 1$ and $\Delta/k_B = 4.4K$ for $\nu = 1/3$. Fits to the variable-range

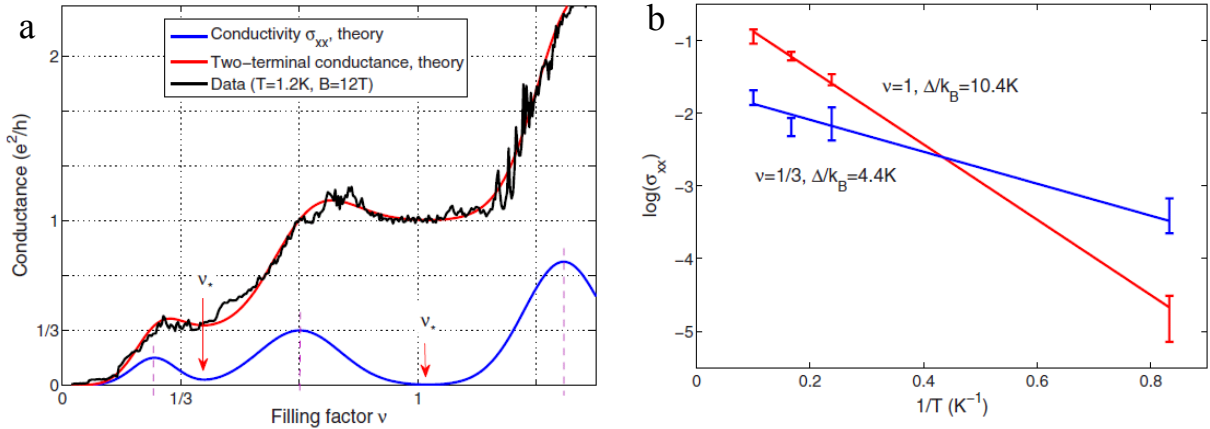

**Figure C-13.** a) Fit (red curve) of the two terminal resistance measured at 1.2K in 12Tesla (black curve), and extraction of longitudinal conductivity (blue curve). b) Arrhenius plot of the longitudinal conductivity obtained in (a), from which the activation energy was obtained.

-hopping dependence, $\sigma_{xx} \sim \exp\left[-(T_*/T)^{1/2}\right]$, were also attempted but no discernible statistical advantage over the activation dependence was observed.

The theoretical prediction for the $\nu = 1/3$ gap in an ideal 2D electron system [105, 235] is $\Delta_{1/3} \approx 0.1 E_C$, where $E_C(K) = e^2/4\pi\varepsilon_0 l_B = 650 B^{1/2}/\varepsilon$. In order to compare the measured and theoretical values one needs to know the dielectric constant. For SG devices, if self-screening is ignored, $\varepsilon = 1$, in which case the measured value is only 8% of the predicted one. However, at low carrier densities self-screening effects which renormalize the dielectric constant cannot be ignored. Its value which is determined by the screening properties of the graphene layer is at present not well established and subject to debate. It ranges from ~5 obtained with random phase approximation (RPA)[196] or GW methods, to 15 obtained from inelastic X-ray scattering on graphite[237]. Using the RPA value, $\varepsilon \sim 5$, the measured values of the gap is within a factor of 2 of the predicted value at 12T.

We note that the value of $\Delta_{1/3}$ in SG is much larger than the corresponding gap in the 2DES in semiconductors. This is because of large deviation of the latter from an ideal 2D system, due to the finite thickness of the quantum wells (10nm to 30nm) which weaken the Coulomb interactions, leading to an almost order of magnitude reduction in the energy gap [238]. The gap is further reduced due to the larger dielectric constant ($\varepsilon \sim 12.9$ in GaAs/GaAlAs heterostructures).

### 4. *Magnetically induced Insulating Phase*

Next we discuss transport near the CNP ($\nu = 0$). Models for lifting of the 4-fold spin and valley degeneracy fall into two categories depending on whether the spin or valley degeneracy is lifted first[239] [240-243]. Both predict insulating bulk, but the former supports counter-propagating edge states and thus is a conductor while the latter with no edge states is an insulator. In the spin-first scenario, where both spin and valley degeneracy can be lifted for all LLs, plateaus occur at all integer values of $\nu$. In contrast the valley-first scenario does not permit plateaus at odd filling-factors other than $\nu = \pm 1$. Experiments addressing this issue in non-suspended graphene are inconclusive [118, 244-246]. While tilted field experiments support the spin-first scenario [118], the absence of plateaus at $\pm 3, \pm 5$ is consistent with the valley-first scenario. The fact that both insulating and conducting behavior were reported further contributes to the uncertainty.

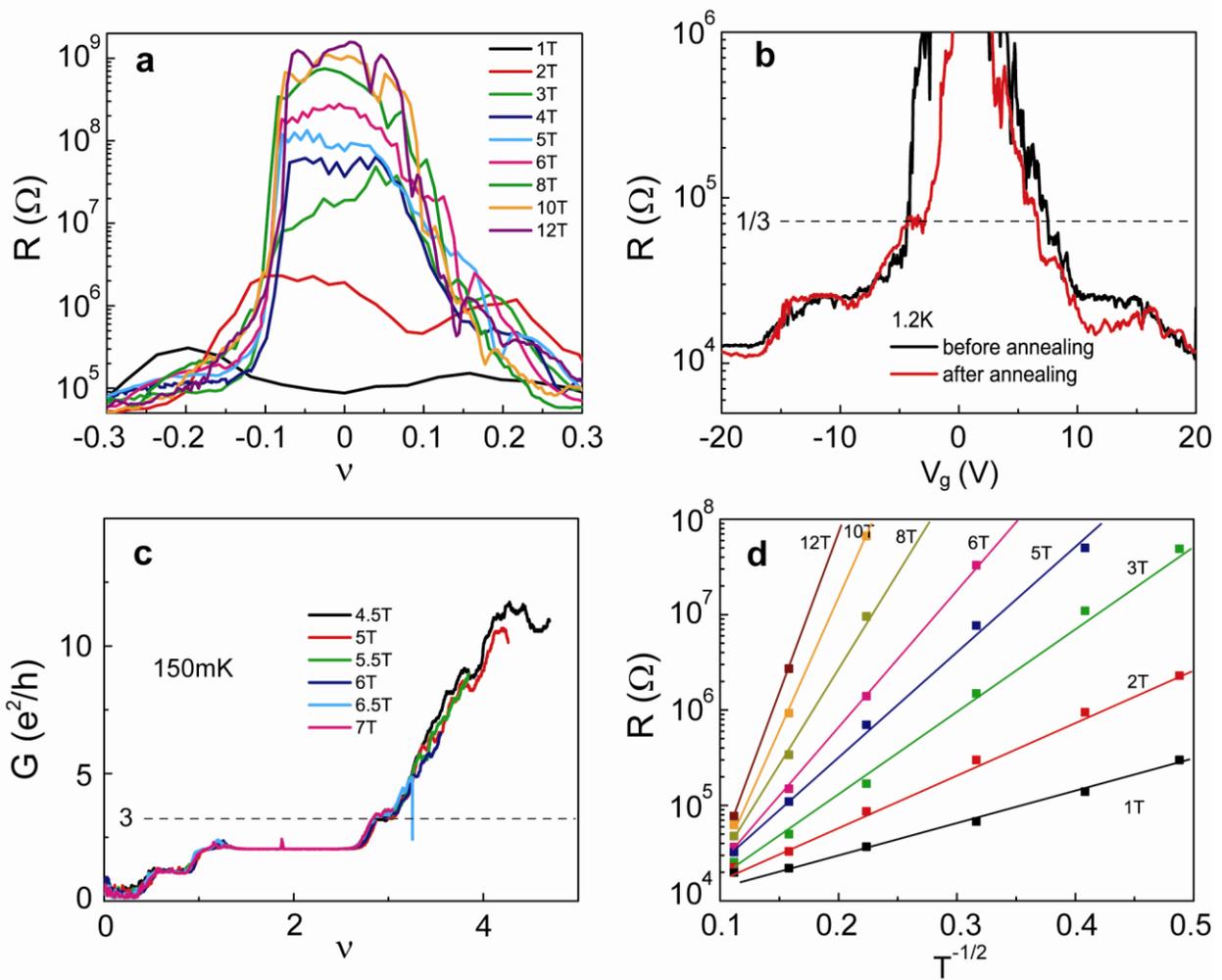

Figure C-14. Insulating behavior at ν=0. a) Resistance as a function of filling factor at indicated magnetic fields. For $|\nu|<0.1$, the resistance increases sharply with field. The maximum resistance value measured above 8 Tesla is instrument-limited. b) Competition between FQHE and insulating behavior. The sample was warmed up to room temperature and re-cooled to 1.2K. Due to the condensation of contaminants on the graphene channel, the insulating regime became broader swallowing the FQHE plateau at ν= −1/3 . Upon current annealing the sample was re-cleaned almost to its pristine condition causing the insulating regime to recede and the plateau at ν= −1/3 to reappear. c) QHE plateaus of a SG sample which showed ν =3 indicating the lifting of the degeneracy in the N=1 LL. d) Log plot of maximum resistance for ν=0 as a function of $T^{-1/2}$ for the field values shown. The solid lines are guides to the eye.

To address this question in SG samples Du *et al*.[27, 30] studied four samples in fields up to 12T and temperatures ranging from 1K to 80K. All samples were insulating at $\nu = 0$ for high fields and low temperature. Consistently they found that the higher the sample quality, as measured by the saturation carrier-density, the sharper the transition, the narrower the region of filling factor where it is observed and the earlier its onset (lower fields and higher temperatures). In the highest quality sample the onset of insulating behavior scaled linearly with field. This is clearly seen in Figure C-14a where the sharp onset of insulating behavior at $|v| \sim 0.1$ is marked by a dramatic increase in resistance. In these samples the maximum resistance value is instrument limited to $\sim 1 G\Omega$. In lower quality samples the insulating region is broader, the onset less sharp and the resistance lower. Interestingly the FQHE states were only observed in samples with narrow insulating regions, suggesting a competition between the two ground states. This is illustrated in Figure C-14b where the insulating phase, having become broader after contamination, "swallowed" the 1/3 plateau. Current annealing the sample brought it back almost to its pristine condition again revealing the 1/3 plateau.

Can the SG data shed light on the nature of the insulating phase? The appearance of a plateau at $v = 3$, shown in Figure C-14c, favors the spin-first splitting scenario. However, since the spin splitting scenario supports counter-propagating edge states, it is inconsistent with an insulating $v = 0$ state. A possible solution would entail a gap opening in the edge states and thus a mechanism to admix them. This would require a mechanism to flip spins and valleys such as magnetic impurities or segments of zigzag edges [247]. An alternative explanation is that the system undergoes a transition to a new broken symmetry phase such as a Wigner or a more exotic skyrme crystal [239, 248, 249]. In this case pinning would naturally lead to insulating behavior. The temperature dependence of the $v = 0$ state is summarized in Figure C-14d. The details of the temperature dependence of the maximum resistance show strong sample-to-sample variation, but all curves fit a generalized activated form: $R_{max} = R_0 \exp(-T_0/T)^{\alpha}$ with $\alpha \sim 1/3 - 1$. In the best sample (Figure C-14d) $\alpha \sim 1/2$ for all fields with $T_0 \sim B^2$. This may provide a hint to the nature of the insulating state, but more work is needed to resolve this question.

## Acknowledgements

We thank A. Luican-Mayer, I. Skachko and A.M.B. Goncalves for help with data and figures. Funding provided by NSF-DMR-090671, DOE DE-FG02-99ER45742, and Lucent